\documentclass[aps, prd, twocolumn,superscriptaddress,nofootinbib,10pt]{revtex4-2}
\usepackage{graphicx}
\usepackage{dcolumn}
\usepackage{multirow}
\usepackage[table]{xcolor}
\usepackage{colortbl,hhline}
\usepackage{amssymb,amsmath,amsthm,mathrsfs}
\usepackage{epstopdf}
\usepackage{lipsum}
\usepackage{hyperref}
\usepackage{empheq}
\usepackage{color}
\usepackage[normalem]{ulem} 
\usepackage{physics}
\usepackage[normalem]{ulem}
\usepackage[ruled,vlined]{algorithm2e}
\usepackage{enumitem}
\usepackage{empheq}
\usepackage{comment}
\usepackage{mathtools}
\usepackage{stackengine}
\usepackage{tikz}
\usepackage[caption=false]{subfig}
\usepackage{amsfonts} 
\DeclareMathSymbol{\shortminus}{\mathbin}{AMSa}{"39}

\newcommand{\be}{\begin{equation}}
\newcommand{\ee}{\end{equation}}
 \newcommand{\bea}{\begin{eqnarray}}
\newcommand{\eea}{\end{eqnarray}}

\newcommand{\ie}{\textit{i.e.}\ }
\newcommand{\eg}{\textit{e.g.}\ }

\newcommand{\vk}{{\mathbf k}}
\newcommand{\vx}{{\mathbf x}}

\newcommand{\CLns}{{\tt ${\mathcal C}$osmo${\mathcal L}$attice}}

\def \FFdual {F_{\mu\nu}\tilde{F}^{\mu\nu}}

\begin{document}

\vspace*{0.5cm}

\title{The non-linear dynamics of axion inflation: a detailed lattice study}

\newcommand{\addressIFIC}{Instituto de F\'isica Corpuscular (IFIC), Consejo Superior de Investigaciones Cient\'ificas (CSIC) and Universitat de Val\`{e}ncia, 46980, Valencia, Spain}
\newcommand{\addressEHU}{Department of Physics, University of the Basque Country, UPV/EHU, 48080, Bilbao, Spain}
\newcommand{\addressEHUQC}{EHU Quantum Center, University of the Basque Country UPV/EHU, Leioa, 48940 Biscay, Spain}

\author{Daniel G. Figueroa} \email{daniel.figueroa@ific.uv.es} \affiliation{\addressIFIC} 
\author{Joanes Lizarraga}\email{joanes.lizarraga@ehu.eus} \affiliation{\addressEHU}\affiliation{\addressEHUQC} 
\author{Nicol\'as Loayza}\email{nicolas.loayza@ific.uv.es} 
\affiliation{\addressIFIC} 
\author{Ander Urio}\email{ander.urio@ehu.eus} \affiliation{\addressEHU}\affiliation{\addressEHUQC} 
\author{Jon Urrestilla}\email{jon.urrestilla@ehu.eus} \affiliation{\addressEHU}\affiliation{\addressEHUQC} 

\date{\today}

\begin{abstract}
We study in detail the fully inhomogeneous non-linear dynamics of axion inflation, identifying three regimes: weak-, mild-, and strong-backreaction,
depending on the duration of inflation. We use lattice techniques that explicitly preserve gauge invariance and shift symmetry, and which we validate against other computational methods of the linear dynamics and of the homogeneous backreaction regime. Notably, we demonstrate that the latter fails to accurately describe the truly local dynamics of strong backreaction. We investigate the convergence of simulations of local backreaction, determining the requirements to achieve an accurate description of the dynamics, and providing useful parametrizations of the delay of the end of inflation. Additionally, we identify key features emerging from a proper local treatment of strong backreaction: the dominance of magnetic energy against the electric counterpart, the excitation of the longitudinal mode, and the generation of a scale-dependent chiral (im)balance. Our results underscore the necessity to accurately capture the local nature of the non-linear dynamics of the system, in order to correctly assess phenomenological predictions, such as \eg the production of gravitational waves and primordial black holes. 
\end{abstract}

\keywords{cosmology, early Universe, inflation, ultra-slow-roll, primordial black holes}

\maketitle

\section{Introduction}

Inflation remains to date as the leading framework to explain the large scale properties of the Universe, including the observed anisotropies of the cosmic microwave background (CMB)~\cite{Akrami:2018odb}. An inflationary phase can be realised by an {\it inflaton} scalar field $\phi$ `slowly' rolling down a sufficiently `flat' potential $V(\phi)$, so that enough number of e-folds are obtained. As inflationary constructions can be very sensitive to unknown ultraviolet (UV) physics, an appealing mechanism to protect $V(\phi)$ from radiative corrections is to promote the inflaton to an {\it axion-like particle} (ALP), invoking a {\it shift} symmetry $\phi \rightarrow \phi + c$, with $c$ a constant. The inflaton's shift symmetry is only broken by non-perturbative
effects and hence the flatness of the potential can be ensured `naturally', as originally proposed in~\cite{Freese:1990rb, Adams:1992bn}. Possible interactions of an ALP inflaton with other fields become then very restricted. 

While several implementations of axion-driven inflation scenarios have been proposed~\cite{Freese:1990rb, Adams:1992bn,Dimopoulos:2005ac, Easther:2005zr, Bachlechner:2014hsa, McAllister:2008hb, Silverstein:2008sg}, we focus on scenarios where the operator with lowest dimensionality that couples the inflaton to an Abelian gauge sector is present,
\begin{eqnarray}
 {\cal L}_\text{int} = - \frac{1}{4 \Lambda} \phi F_{\mu \nu} \tilde F^{\mu \nu} \,,
 \label{eq:phiFFtilde}
\end{eqnarray}
with $\Lambda$ some energy scale, and $F_{\mu \nu}$ ($\tilde F_{\mu \nu}$) the (dual) field strength of a gauge field $A_\mu$. These scenarios, referred to as {\it axion inflation}, are particularly interesting from an observational point of view, given the wide range of phenomena they can exhibit. A distinctive feature of them is the exponential production of a helical gauge field during inflation~\cite{Turner:1987vd,Garretson:1992vt,Anber:2006xt,Anber:2009ua,Barnaby:2010vf,Adshead:2013qp,Cheng:2015oqa}, which can have many interesting cosmological consequences. In the present work we assume $\phi$ is coupled only to a {\it hidden} Abelian gauge sector, \ie $A_\mu$ is a
{\it dark photon}. For given $V(\phi)$, the excitation of the gauge field is therefore determined solely by the strength of $\Lambda$. In practice the excitation of the gauge field is fully controlled by the dimensionless parameter $\xi \equiv |\dot{\phi}|/(2 \Lambda H)$, where $\dot{\phi}$ is the velocity of the inflaton field and $H$ is the Hubble rate. 

During inflation, the excited gauge field sources curvature perturbations, via inflaton fluctuations produced through the inverse decay process $A+A \rightarrow \delta\phi$. The sourced scalar perturbations are non-Gaussian of the equilateral type, with non-linear parameter $f_{\rm NL}^{\rm {eq}}\simeq7\cdot 10^5\times(H^6/|\dot\phi|^3)(e^{6\pi\xi}/\xi^9)$~\cite{Barnaby:2010vf}. Current CMB constraints on this type of non-gaussianities~\cite{Planck:2019kim}, $f^{\rm eq}_{\rm NL} = -26 \pm 47$, translate into a strong constraint on $\xi$ at CMB scales. For $V(\phi) = {1\over 2}m^2\phi^2$ this reads $\xi_{\rm CMB}\lesssim 2.5$ at $95\%$ C.L.~\cite{Barnaby:2010vf,Barnaby:2011qe,Meerburg:2012id,Pajer:2013fsa}, or equivalently $\Lambda^{-1} \lesssim 35m_p^{-1}$, with $m_p \simeq 2.435 \cdot 10^{18}$ GeV the reduced Planck mass.

The quantity $\xi$ that controls the amplification of $A_\mu$ typically increases as inflation carries on, since $H$ decreases and $|\dot\phi|$ increases during standard slow-roll inflation. Thus, an initially small value $\xi \lesssim 2.5$ at CMB scales is expected to grow to larger values at much shorter scales. For a specific inflaton potential $V(\phi)$, the evolution of $\xi$, specially towards the end of inflation, will depend on the amount of backreaction of $A_\mu$, both onto the inflationary expansion and the inflaton dynamics.

The growth of $\xi$ can result in large overdensities, which might form primordial black holes (PBH) when re-entering the horizon during radiation domination (RD)~\cite{Linde:2012bt}. In order to compute this effect one needs to follow the excitation of large curvature perturbations in the strong backreaction regimen~\cite{Barnaby:2011qe}. Taking this into consideration, Ref.~\cite{Linde:2012bt} showed that to prevent an excess of dark matter in the form of PBH's, a bound $\xi_{\rm CMB} \lesssim 1.5$ must be imposed at CMB scales. For a quadratic inflaton potential, this implies an upper bound on the axion-gauge coupling as $\Lambda^{-1} \lesssim 23m_p^{-1}$~\cite{Linde:2012bt}. While the PBH limit is stronger than the non-Gaussianity limit, it relies however on an approximation of the strong backreaction regime which, in light of the results we present in this paper, will very likely need re-evaluation.

Tensor metric perturbations are also effectively sourced during inflation by the helical gauge field~\cite{Barnaby:2010vf}. Once tensor modes cross back inside the horizon after the end of inflation, they behave as a stochastic GW background (GWB) with a key phenomenological signature: one of the GW chiralities is much larger than the other. The amplitude of such parity violating background could be potentially observable by the Laser Interferometer Space Antenna (LISA) or the Einstein Telescope (ET) observatories~\cite{Sorbo:2011rz,Barnaby:2011qe,Cook:2013xea,Adshead:2013qp,Bartolo:2016ami,Maggiore:2019uih,LISACosmologyWorkingGroup:2022jok}. Realistic assessments of the direct detection of such background require, however, a revision of the impact of the strong backreaction on the GWB spectrum, similarly as for scalar perturbations. 

Furthermore, a high frequency GWB is also expected from the preheating dynamics after axion inflation~\cite{Adshead:2019igv,Adshead:2019lbr,Adshead:2023mvt}. Current limits on the effective number of relativistic degrees of freedom 
translate into a bound on the axion-gauge coupling, which for $V(\phi) = {1\over2}m^2\phi^2$ reads $\Lambda^{-1} \lesssim 14m_p^{-1}$~\cite{Adshead:2019igv,Adshead:2019lbr}. While the GW-preheating limit is stronger than the non-Gaussianity and PBH limits, it depends on the details of the last stages of inflation and of a potential early PBH dominated phase ensued after inflation~\cite{Domcke:2020zez}. As the strong backreaction inflationary phenomenology uncovered in our present work might affect this limit, we keep open the exploration of couplings beyond current preheating bounds, in order to fully understand observational constraints of axion inflation.

Increasingly accurate analysis of the gauge field backreaction during inflation have been developed during the last years. After initial backreaction-less studies solving the linear dynamics of $A_\mu$~\cite{Martin:2007ue,Demozzi:2009fu}, it was soon understood that backreaction of the gauge field is simply unavoidable~\cite{Cheng:2015oqa,Notari:2016npn,DallAgata:2019yrr,Domcke:2020zez,Gorbar:2021rlt,Peloso:2022ovc,Galanti:2024jhw}, for (sufficiently) large couplings. Under the assumption that the inflaton remains homogeneous, the state of the art to address the backreaction dynamics in axion inflation is based on the {\it gradient expansion formalism} (GEF), which considers a tower of infinite coupled partial differential equations over the expectation value of gauge field correlators. Phenomenological results based on the mentioned formalism have been analyzed in~\cite{Durrer:2023rhc,vonEckardstein:2023gwk,Durrer:2024ibi}. While this formalism is computationally very efficient, it ignores the local nature of the term $F\tilde F$ and couplings between inflaton gradients  and $A_\mu$. A perturbative approach extending the GEF to account for inflaton gradients has also been recently proposed~\cite{Domcke:2023tnn}.

Lattice approaches, on the other hand, capture the full locality of the problem, and have been used to re-assess various phenomenological effects~\cite{Adshead:2019lbr,Adshead:2019igv,Caravano:2021bfn,Caravano:2022epk,Adshead:2023mvt,Figueroa:2023oxc,Caravano:2024xsb,Sharma:2024nfu}.Ref.~\cite{Figueroa:2023oxc} in particular, which, from now on,  will be referred to as {\tt Paper~I}, captured for the first time the full dynamical range of the {\it local} strong backreaction regime till the end of inflation.
It showed that a proper lattice implementation can reproduce previous homogeneous backreaction results when switching off axion gradients, and further and most importantly, it demonstrated the importance of capturing consistently the inhomogeneity of the problem, as significant departures in the dynamics were found when including the axion gradients. While accurately dealing with the full non-linear problem, including the local features of strong backreaction, lattice simulations are however computationally very demanding. 

In summary, axion inflation can have many interesting phenomenological consequences. Together with the production of non-Gaussian curvature perturbations~\cite{Barnaby:2010vf,Barnaby:2011qe,Barnaby:2011vw,Cook:2011hg, Barnaby:2011qe,Pajer:2013fsa} and chiral GWs~\cite{Sorbo:2011rz, Barnaby:2011qe, Cook:2013xea,Adshead:2013qp,Bastero-Gil:2022fme,Garcia-Bellido:2023ser}, it might also lead to successful magnetogenesis~\cite{Garretson:1992vt, Anber:2006xt, Adshead:2016iae, Durrer:2023rhc}, baryon asymmetry mechanism~\cite{Giovannini:1997eg,Anber:2015yca,Fujita:2016igl,Kamada:2016eeb,Jimenez:2017cdr,Cado:2022evn}, efficient (p)reheating~\cite{Adshead:2015pva,Cuissa:2018oiw,Adshead:2023mvt} and sizeable post-inflationary GW production~\cite{Adshead:2018doq,Adshead:2019igv,Adshead:2019lbr}.  The details depend on the choice of $V(\phi)$, the scale $\Lambda$, and on the field content assumed, and can be rather complex, possibly requiring to consider fermion production~\cite{Adshead:2015kza,Adshead:2018oaa,Domcke:2018eki,Domcke:2019qmm,Cado:2022pxk} and thermal effects~\cite{Ferreira:2017lnd,Ferreira:2017wlx}.  It has also been considered the possibility that the gauge sector is represented by the non-abelian gauge group SU(2)~\cite{Maleknejad:2011sq,Maleknejad:2011jw,Maeda:2013daa,Maleknejad:2016qjz,Dimastrogiovanni:2016fuu,Maleknejad:2016dci,Dimastrogiovanni:2018xnn,Maleknejad:2020yys,Wolfson:2020fqz,Mirzagholi:2020irt,Watanabe:2020ctz,Wolfson:2021fya,Fujita:2022jkc,Iarygina:2023mtj,Fujita:2023axo,Dimastrogiovanni:2024lzj,Brandenburg:2024awd,Badger:2024ekb}. A major goal of the present work is to show that, even for the simplest case where the inflaton is only coupled to a U(1) dark photon, the details of strong backreaction are extremely complicated, and hence one should not trust phenomenological consequences which depend on such non-linear regime, unless the latter is fully under control. 

In this work we present the details of a consistent lattice formulation of the interaction of a shift-symmetric field with a gauge field, including the case when the expansion of the universe is dictated by such fields. This formalism is a natural generalization of the formulations introduced in~\cite{Figueroa:2017qmv,Cuissa:2018oiw}, and has been already used in {\tt Paper I}. In this manuscript we elaborate on the technical difficulties one finds when dealing with such a lattice formulation of the problem. We discuss the numerical challenges encountered when applying the formalism to an axion inflation scenario where the inflaton is just coupled to a dark photon, and has potential $V(\phi) = {1\over2}m^2\phi^2$. While a quadratic potential is in conflict with current CMB constraints~\cite{BICEP:2021xfz,Tristram:2021tvh}, we stick to this choice in order to compare and extend previous results from the literature. Our
methodology can be applied to arbitrary potentials, and we plan to present those results elsewhere.

\section{Axion inflation}\label{sec: basic}

In this section we provide an overview of the axion inflation model. Using $(-,+,+,+)$ metric signature and the reduced Planck mass $m_p \simeq 2.435 \cdot 10^{18}$ GeV, we
consider an action $S_{\rm tot} = S_{\rm g} + S_{\rm m}$, with $S_{\rm g}~\equiv~\int {\rm d}x^4 \sqrt{-g}\frac{1}{2}m_p^2 R$ the standard Hilbert-Einstein term for gravity, and a matter action given by
\begin{eqnarray}
S_{\rm m} &=& -\int {\rm d}x^4 \sqrt{-g}\left[\frac{1}{2}\partial_\mu \phi\partial^\mu\phi+V(\phi)\right. \\
&&\hspace{2.2cm}\left. +\, \frac{1}{4}F_{\mu\nu}F^{\mu\nu} - \frac{\alpha_{\Lambda}}{4}\frac{\phi}{m_p} \FFdual \right],\nonumber
\end{eqnarray}
with $\phi$ the inflaton, and $A_{\mu}$ the gauge field of a hidden $U(1)$ gauge sector. We indicate the coupling strength of the axion-gauge interaction by $\alpha_{\Lambda} \equiv m_p/\Lambda$, and define the field strength of $A_\mu$ and its dual, as usual, by
\bea 
F_{\mu\nu}\equiv\partial_\mu A_\nu-\partial_\nu A_\mu\,,~~~
\tilde F_{\mu \nu}\equiv \frac{1}{2}\epsilon_{\mu\nu\rho\sigma}F^{\rho\sigma}\,,
\eea
where  $\epsilon_{\mu\nu\rho\sigma}$ is the completely antisymmetric Levi-Civita tensor in a curved space-time, with $\epsilon_{0123}= 1/\sqrt{-g}$.

The action is invariant under local $U(1)$ transformations, $A_\mu \rightarrow A_\mu + \partial_\mu\alpha$, with $\alpha(x)$ an arbitrary real function. Except for the potential term $V(\phi)$, the action displays a shift symmetry $\phi\to\phi +c$, with $c\in \mathbb{R}$. The potential $V(\phi)$ is assumed to be generated by an external mechanism and breaks the shift symmetry explicitly. The axion-inflation model includes the lowest dimensional interaction compatible with a shift-symmetric inflaton, \ie the five-dimensional axial coupling $\phi F\tilde F$ in Eq.~(\ref{eq:phiFFtilde}). While our approach is applicable to any potential, we choose a quadratic one 
\be
V(\phi) = \frac{1}{2}m^2\phi^2\;,
\ee
to compare with previous results from the literature. The  inflaton mass is set to $m \simeq 6.16\cdot10^{-6}m_p$, as dictated by fitting the observed temperature anisotropies of the CMB~\cite{Planck:2018jri}.

Assuming a spatially flat Friedmann-Lema\^itre-Robertson-Walker (FLRW) background, varying the action leads to the equations of motion (EOM)
\begin{empheq}[]{alignat=2}
&\ddot{\phi} = -3H\dot{\phi}+\frac{1}{a^2}\nabla^2\phi-m^2\phi+\frac{\alpha_\Lambda}{a^3 m_p}\vec{E}\cdot\vec{B}\,,\label{eqn:eom1}\\
&\dot{\vec{E}} = -H\vec{E}-\frac{1}{a^2}\vec{\nabla}\times\vec{B}-\frac{\alpha_\Lambda}{a m_p}\left(\dot{\phi}\vec{B}-\vec{\nabla}\phi\times\vec{E}\right),\label{eqn:eom2}\\
&\vec{\nabla}\cdot\vec{E} \,\, = -\frac{\alpha_{\Lambda}}{am_p}\vec{\nabla}\phi\cdot\vec{B}\,,\label{eqn:Gauss}
\end{empheq}
where $\dot{} \equiv \partial/\partial t$ are derivatives with respect to cosmic time $t$, $a(t)$ is the scale factor, and $H=\dot a/a$ is the Hubble rate. We define the electric field $\vec E$ in the temporal gauge $A_0 = 0$, and the magnetic field $\vec B$, as
\be
E_i\equiv\dot A_i\,,\quad B_i\equiv\epsilon_{ijk}\partial_j A_k\,.
\ee
We note that Eq.~(\ref{eqn:Gauss}) represents the Gauss constraint. 

The expansion of the Universe is governed by the Friedmann equations
\begin{eqnarray}
\ddot{a}&=&-\frac{a}{3m_p^2}\big( 2\rho_{\rm K}-\rho_{\rm V}+\rho_{\rm EM} \big)\,,\label{eqn:eom3}\\ 
H^2&=&\frac{1}{3m_p^2}\big(\rho_{\rm K}+\rho_{\rm G}+\rho_{\rm V}+\rho_{\rm EM}\big)\,,\label{eqn:Hubble}
\end{eqnarray}
where the different homogeneous energy density contributions are given by  
\begin{equation}
\begin{split}
&\rho_{\rm K} \equiv \frac{1}{2}\langle\dot{\phi}^2\rangle\; , \quad \rho_{\rm G} \equiv \frac{1}{2a^2}\langle(\vec\nabla\phi)^2\rangle\; ,
\\
&\rho_{\rm V} \equiv \frac{1}{2}m^2\langle \phi^2 \rangle\;, \quad\rho_{\rm EM} \equiv \frac{1}{2a^4}\langle a^2\vec E^2+\vec B^2\rangle \;,
\end{split}
\label{eqn:energyDensityTerms}
\end{equation}
with $\langle ... \rangle$ indicating volume averaging. Labels K, G, V denote the kinetic, gradient and potential energy densities of the inflaton, respectively, while EM denotes the electromagnetic energy density associated to $A_\mu$. We note that Eq.~(\ref{eqn:Hubble}) represents the Hubble constraint.

The treatment on the dynamics of the system of Eqs.~(\ref{eqn:eom1})-(\ref{eqn:Gauss}) and (\ref{eqn:eom3})-(\ref{eqn:Hubble}) can be done at various levels of approximation, which we review in the following.
 
\subsection{Linear regime: chiral instability}\label{subsec:linear} 
 
In the linear regime the expansion of the universe is just dictated by the inflaton, which is considered to be homogeneous, \ie $\vec{\nabla} \phi = 0$. The contributions of the gauge field to the expanding background or to the inflaton dynamics are neglected. In this regime the EOM read 
\begin{eqnarray}
\ddot{\phi}&=&-3H\dot{\phi} -m^2\phi \,,\label{eq:eoml1}\\
\dot{\vec{E}}&=&-H\vec{E}-\frac{1}{a^2}\vec{\nabla}\times\vec{B}-\frac{\alpha_{\Lambda}}{am_{p}}\dot{\phi}\vec{B}\,,\label{eq:eoml2}\\
\ddot a &=& -\frac{a}{3m_p^2}\left( 2\rho_{\rm K}-\rho_{\rm V} \right)\,,\label{eq:eoml3}
\end{eqnarray}
while the constraint equations are reduced to
\begin{eqnarray}
\vec{\nabla}\cdot\vec{E}&=& 0\,\label{eq:gauss},\\
H^2&=&\frac{1}{3m_p^2}\left(\rho_{\rm K}+\rho_{\rm V}\right)\,\label{eq:ene}.
\end{eqnarray}
In order to understand the gauge field dynamics we first write Eq.~(\ref{eq:eoml2}) in terms of $\vec A(t,{\bf x})$ and using the conformal time $d\tau \equiv dt/a(t)$,  
\begin{eqnarray}
\left( \frac{d^2}{d\tau^2} - \nabla^2 + \frac{\alpha_{\Lambda}}{m_p} \; \frac{d \phi}{d\tau}  \;\vec{\nabla} \times \right) \vec{A}(\tau,{\bf x}) = 0 \; .
\end{eqnarray}
Next, we Fourier transform $\vec A$, decomposing its Fourier modes into a polarization vector basis, 
\begin{eqnarray}
\vec A(\tau,\vx) = \sum_{\lambda} \int \frac{{\rm d}^3k}{(2\pi)^3}  A^\lambda(\tau,{\bf k})\vec{\varepsilon}^{\,\lambda}(\hat\vk)  e^{i\vk\cdot\vx}\, .
\label{eq:FTransform}
\end{eqnarray}
We choose a chiral vector basis $\lbrace \vec{\varepsilon}^{\,+}(\hat\vk)\, ,\vec{\varepsilon}^{\,-}(\hat\vk)\rbrace$, which satisfies the properties
\bea
\begin{split}
\hat \vk\cdot\vec\varepsilon^{\,\lambda}(\hat\vk)=0\,,\quad
\hat \vk \times \vec \varepsilon^{\,\lambda}(\hat\vk)=-i\lambda \vec\varepsilon^{\,\lambda}(\hat\vk)\,,\\
\varepsilon_i^{\,\lambda}(\hat\vk)^*=\varepsilon_i^{\,\lambda}(-\hat\vk)\,,\quad \vec \varepsilon^{\,\lambda'}\hspace*{-1mm}(\hat\vk)\cdot\vec\varepsilon^{\,\lambda}(\hat\vk)^* = \delta_{\lambda\lambda'}\,.
\end{split}
\label{eqn:polarisationvectors}
\eea
We promote the Fourier amplitude into a quantum operator $A^\lambda(\tau,{\bf k}) \rightarrow \hat A^\lambda(\tau,{\bf k}) \equiv \hat a_{{\bf k}}\mathcal{A}^\lambda(\tau,{\bf k})$ + $\hat a_{-{\bf k}}^\dag\mathcal{A}^{\lambda}(\tau,-{\bf k})^{*}$ by means of standard annihilation and creation operators that verify  $[\hat a_{{\bf k}}, \hat a_{{\bf k}'}^{\dagger}] \equiv \delta^{(3)}({\bf k}-{\bf k}').$ 
Using the slow-roll condition $\tau \simeq -1/(aH)$, the mode functions $\mathcal{A}^{\pm}(\tau,{\bf k})$ follow the equation 
\bea
&&\left(\partial_\tau^2+k^2\pm sign(\dot\phi)\frac{2 k \xi}{|\tau|}\right) \mathcal{A}^\pm(\tau,{\bf k})=0\,,\label{eq:linA}\\
\vspace*{2mm}
\hspace{-1.5cm}{\rm with} && \hspace{1.8cm}
\xi \equiv \frac{|\dot{\phi}|}{2 H \Lambda }\,.
\eea 
For values $\xi \gtrsim \mathcal{O}(1)$, one of the chiral modes develops an exponential growth, while the other remains in vacuum. 
For $\dot{\phi} < 0$ ($>0$), it is $A^{+}$ ($A^{-}$) the one that experiences the instability. Considering $\xi$ to be approximately constant, as expected deep inside the slow-roll dynamics, a general solution for Eq.~\eqref{eq:linA} is given by
\bea\label{eq:Coulomb}
\mathcal{A}^\pm(\tau,{\bf k}) = \frac{1}{\sqrt{2k}}[iF_0(\pm \xi, -k\tau) + G_0(\pm \xi, -k\tau)]\,,
\eea
with $F_0$ and $G_0$ the regular and irregular Coulomb wave functions~\cite{Anber:2006xt,Anber:2009ua}. The coefficients have been chosen so that for scales deep inside the Hubble radius, $- k\tau \gg 1$, both chiralities are in the so-called Bunch-Davies (BD) vacuum,    
\bea
\mathcal{A}^\pm(\tau,{\bf k}) ~~\xrightarrow[~~-k\tau \gg 1~~]{} ~~
\frac{1}{\sqrt{2 k}}e^{-i k \tau}\,. 
\label{eq:BD}
\eea
Without loss of generality, we choose $\dot\phi < 0$, so that the excited mode is $\mathcal{A}^+$. In the limit $-k\tau\ll 2\xi$, the solution~(\ref{eq:Coulomb}) for this mode  can be approximated by
\begin{equation}
\mathcal{A}^+(\tau,{\bf k}) \underset{k|\tau| \ll 2\xi}{\simeq} \frac{1}{\sqrt{2k}}\left(\frac{k}{2\xi aH}\right)^{1/4} e^{\pi\xi - 2\sqrt{2\xi k/(aH)}}\,,\label{eqn:amplifiedAplus}
\end{equation}
which exhibits very clearly the exponential nature of the chiral instability.

Depending on the evolution of $\xi$, the exponential growth of the unstable mode may become sufficiently large so that its contribution to the expansion of the universe (in the form of electromagnetic energy density) may become gradually more and more relevant. Furthermore, the term $\,\vec{E}\cdot\vec{B}\,$ in the $rhs$ of Eq.~(\ref{eqn:eom1}) may also become relevant, affecting the inflaton dynamics. In other words, for sufficiently large coupling $\alpha_\Lambda$, as inflation proceeds, there is always a point at which the backreaction-less approximation becomes unreliable. We must, therefore, refine the treatment of the dynamics, as we do next, in order to incorporate the backreaction of the gauge field. 

\subsection{Homogeneous backreaction}
\label{sec:BRhomo}

In order to account for the gauge field backreaction in the system, it is essential to keep simultaneously both the electromagnetic energy in Eq.~\eqref{eqn:eom3} as well as the term $\propto~ \vec{E}\cdot\vec{B}\,$ in Eq.~\eqref{eqn:eom1}. The homogeneous backreaction approach incorporates the latter as the expectation value $\langle \vec{E}\cdot\vec{B} \rangle$, so that the inflaton dynamics remains homogeneous. This leads to the following set of EOM
\begin{eqnarray}
\ddot{\phi}&=&-3H\dot{\phi} -m^2\phi+\frac{\alpha_\Lambda}{a^3 m_p}\langle\vec{E}\cdot\vec{B}\rangle \,,\label{eq:eomh1}\\
\dot{\vec{E}}&=&-H\vec{E}-\frac{1}{a^2}\vec{\nabla}\times\vec{B}-\frac{\alpha_{\Lambda}}{am_{p}}\dot{\phi}\vec{B}\,,\label{eq:eomh2}\\
\ddot a &=&-\frac{a}{3m_p^2}\left( 2\rho_{\rm K}-\rho_{\rm V}+\rho_{\rm EM} \right)\,,\label{eq:eomh3}
\end{eqnarray}
while the constraint equations read now
\begin{eqnarray}
\vec{\nabla}\cdot\vec{E}&=&0\,\label{eq:gaussl},\\
H^2&=&\frac{1}{3m_p^2}\left(\rho_{\rm K}+\rho_{\rm V}+\rho_{\rm EM}\right)\,\label{eq:enel}.
\end{eqnarray}
Different approaches have been developed to solve these equations. A straightforward method is to evaluate the expectation value $\langle \vec{E}\cdot\vec{B} \rangle$ every time step, as an integral over the gauge field mode functions, and later plug this back into the EOM, repeating this procedure iteratively ~\cite{Cheng:2015oqa,Notari:2016npn,DallAgata:2019yrr,Domcke:2020zez}. An alternative method is to solve the EOM by rewriting them in an entirely different fashion, in terms of expectation values. The so-called {\it Gradient Expansion Formalism} (GEF) makes use of this idea, obtaining an infinite tower of coupled ordinary differential equations~\cite{Sobol:2019xls,Gorbar:2021rlt,Durrer:2023rhc,Durrer:2024ibi,vonEckardstein:2023gwk}. The new equations are solved iteratively up to a truncated order that ensures the convergence of the solutions as compared to the mode-by-mode solution. While both formalisms lead to the same results, the GEF is computationally much more efficient than the iterative procedure. 

To a first approximation, the most significant impact of the homogeneous backreaction is on the dynamics of the inflaton. The gauge field excitation acts as a `friction' that reduces the inflaton velocity, delaying in this way the end of inflation. Furthermore, an oscillatory behaviour is displayed by $\dot\phi$, and hence by $\xi$, as a consequence of the delayed response of the inflaton's velocity to the excitation of the gauge field~\cite{Cheng:2015oqa,Notari:2016npn,DallAgata:2019yrr,Domcke:2020zez}. Remarkably, in the homogeneous backreaction regime the system remains fully chiral. In other words, only the exponentially enhanced mode is excited ($A^+$ in our case), while the other ($A^-$ in our convention) remains in vacuum (or at most it is marginally excited if the inflaton velocity changes sign due to a strong oscillation). 

A more refined approach has been developed recently in order to accommodate the gradients of the inflaton field in the GEF methodology~\cite{Domcke:2023tnn}. This formalism goes a step further than standard GEF by considering not just 2-point functions of the gauge field, but also 3-point functions including the gradient of the inflaton. The results reproduce previous findings from standard GEF, namely the delay of the end of inflation and the oscillatory behaviour of $\xi$. Although it effectively captures the onset of the truly local backreaction, it soon fails when the inflaton gradients exceed a certain threshold, beyond which full lattice approaches are required.

\subsection{Local backreaction}\label{eqn:NonLinear}

In order to capture accurately the dynamics of the backreaction, we need to allow for inhomogeneities of the inflaton and of the gauge field, consistently and without restrictions. The truly local dynamics of the system are characterised by Eqs.~(\ref{eqn:eom1})-(\ref{eqn:eom2}) and (\ref{eqn:eom3}), for which there is no possible analytical solution. Because of the non-linear nature of the equations, approximations such as the previously discussed homogenous approaches fail to follow correctly the backreaction dynamics~\cite{Figueroa:2023oxc}. Therefore, a lattice approach must be used in order to capture accurately the local dynamics. Before we move on into the more technical details of our lattice formulation, we introduce here 
the form of the exact local EOM in the continuum, adapted to the actual dynamics of inflation. In order to do this, we make use of the number of e-foldings $\mathcal{N}$, defined by 
\begin{equation}
    d\mathcal{N}=Hdt\,.
\end{equation}
Keeping the previous definition of the electric field, $E_i \equiv {\dot A}_i \equiv H A_i '$, with $ ' \, \equiv \, d/d\mathcal{N}$ and $H \equiv \dot a/ a$, and introducing the inflaton conjugate momentum $\pi_\phi \equiv \dot\phi = H\phi'$, we rewrite the EOM of the system as
\begin{eqnarray}
\hspace*{-0.7cm}H({\pi}_{\phi}' + 3\pi_{\phi}) \hspace*{-1mm}&=&\hspace*{-1mm} \frac{1}{a^2}\nabla^2\phi-m^2\phi+\frac{\alpha_\Lambda}{a^3 m_p}\vec{E}\cdot\vec{B}\,, \label{eq:eomef1}\\
\hspace*{-0.7cm}H\big(\vec{E}' + \vec{E}\,\big) \hspace*{-1mm}&=&\hspace*{-1mm} \frac{1}{a^2}\vec{\nabla}\times\vec{B} -\frac{\alpha_\Lambda}{a m_p}\big(\pi_\phi\vec{B}-\vec{\nabla}\phi\times\vec{E}\,\big), \label{eq:eomef2}\\
\hspace*{-0.7cm}-3HH' &=&{1\over m_p^2}(3\rho_{\rm K}+\rho_{\rm G}+2\rho_{\rm EM})\,,
\label{eq:eomef3}
\end{eqnarray}
while the constraint equations 
\begin{eqnarray}
\vec{\nabla}\cdot\vec{E} &=& -{\alpha_{\Lambda}\over am_p}\vec{\nabla}\phi\cdot\vec{B}\,,\label{eq:cons1eFolding}\\
H^2 &=& {1\over 3m_p^2}(\rho_{\rm K}+\rho_{\rm G}+\rho_{\rm V}+\rho_{\rm EM})\,,
\label{eq:cons2eFolding}
\end{eqnarray}
remain intact. We note that the scale factor can be written now explicitly as $a \equiv a_* e^{(\mathcal{N}-\mathcal{N}_*)}$, with $a_*$ and $\mathcal{N}_*$ of arbitrary choice, whereas the homogeneous energy densities $\rho_{\rm K}$, $\rho_{\rm G}$, $\rho_{\rm V}$ and $\rho_{\rm EM}$ are still given by Eq.~(\ref{eqn:energyDensityTerms}).

Aside from the time variable change, the fundamental difference of this system of equations with respect to the homogeneous approaches is the inclusion of inhomogeneities of the inflaton. This implies, first of all, that the backreaction of the gauge field on the inflaton is local, via $\vec E \cdot \vec B$, and not through the expectation value $\langle \vec E \cdot \vec B \rangle$. It also implies that, for consistency, we need to maintain the laplacian term $\propto \nabla^2\phi$ in the inflaton's EOM. Furthermore, the term $\pi_\phi \vec B$ in the gauge field's EOM, will no longer be interpreted as the magnetic field times a homogeneous inflaton velocity, as the latter will also become spatially dependent.

\subsection{Chiral Projection}

We note that whilst the helicity decomposition of the gauge field is set in Fourier space, we aim to solve for the non-linear dynamics of the system in real space, \eg Eqs.~(\ref{eq:eomef1})-(\ref{eq:eomef3}), where the decomposition into the two linearly independent helicities is not explicit. It is therefore convenient to lay out a procedure to go back and forth between the Cartesian and the chiral basis.

We start by noticing the property $\epsilon_{ijl}\hat k_j \varepsilon^\lambda_l = -i\lambda \varepsilon^\lambda_i$ of the helicity vectors $\lbrace \vec\varepsilon^{\,+},\vec \varepsilon^{\,-}\rbrace$, {\it c.f.}~Eq.~(\ref{eqn:polarisationvectors}), which we can conveniently rewrite as $(-i\epsilon_{ijl}\hat k_l)\varepsilon_j^\lambda = \lambda\varepsilon_i^\lambda$. We recognise there the helicity operator~\cite{altaisky2005wavelets}
\bea
    \Sigma_{ij}(\hat\vk)\equiv -i\epsilon_{ijl}\hat k_l\,,
\eea
which, by construction, is antisymmetric, $\Sigma_{ij} = -\Sigma_{ji}$, and has $\vec \varepsilon^{\,\pm}$ as eigenvectors, with eigenvalues $\pm1$, \ie $\Sigma_{ij}\varepsilon_j^\lambda = \lambda \varepsilon_i^\lambda$. 
Extending the basis to a full orthonormal triad $\lbrace \vec\varepsilon^{\,+},\vec \varepsilon^{\,-}, \hat k \rbrace$, and relabelling the vectors as $\vec\varepsilon^{\,(-1)}\equiv \vec\varepsilon^{\,-},\vec\varepsilon^{\,(0)}\equiv \hat k,
\vec\varepsilon^{\,(+1)} \equiv \vec \varepsilon^{\,+}$, we see that $\vec\varepsilon^{\,(0)}$ is also an eigenvector of $\Sigma_{ij}$ with vanishing eigenvalue, \ie $\Sigma_{ij}\varepsilon_j^{\,(0)} = 0$. We can therefore write 
\begin{eqnarray}
\Sigma_{ij}\varepsilon_j^{\,(\sigma)} = \sigma \varepsilon_i^{\,(\sigma)}\,,~~~\sigma = -1,0,+1.
\end{eqnarray}

Any vector $\vec A({\bf k})$ that lives in Fourier space can always be expressed as a linear combination of a (fixed) Cartesian basis, 
$\vec A \equiv$ $A_1\hat e_1$ + $A_2\hat e_2$ + $A_3\hat e_3$, or as a linear combination of the chiral basis as $\vec A \equiv$ $A^{(-1)}\vec\varepsilon^{\,(-1)}$ + $A^{(0)}\vec\varepsilon^{\,(0)}$ + $A^{(+1)}\vec\varepsilon^{\,(+1)}$, with $A^{(+1)} \equiv A^+$ the positive helicity component, $A^{(-1)} \equiv A^-$ the negative helicity component, and  $A^{(0)} \equiv A^{\parallel}$ the longitudinal component. We can think of $\vec A({\bf k})$ as the Fourier transform of the gauge field $\vec A(\mathcal{N},{\bf x})$ at an arbitrary time $\mathcal{N}$. Then, it holds that
\bea
\Sigma_{ij}(\hat\vk)A^{(\sigma)}_j(\vk) = \sigma A^{(\sigma)}_i(\vk)\,,~~~ \sigma = -1,0,+1\,,
\eea
where $A_i^{(\sigma)} \equiv A^{(\sigma)}\varepsilon_i^{(\sigma)}$ are the chiral components of $\vec A$. This property can be used to construct a proper helicity projector as
\bea
\Pi^{\pm}_{ij}(\hat\vk) &\equiv& \frac{1}{2}\left[(\Sigma^2(\hat\vk))_{ij} \pm \Sigma_{ij}(\hat\vk)\right]\nonumber\\
  &=& \frac{1}{2}\left(P_{ij}(\hat\vk) \pm  \Sigma_{ij}(\hat\vk)\right)\,,\label{eq:ChiralProjector}
\eea
where in the second line we have used $(\Sigma^2(\hat\vk))_{ij} \equiv \Sigma_{il}(\hat\vk)\Sigma_{lj}(\hat\vk) = P_{ij}(\hat\vk)$, with $P_{ij}(\hat\vk) \equiv \delta_{ij} - \hat k_i\hat k_j$ the transverse projector.It then follows that $\Pi^{\lambda}_{ij}(\hat\vk)A_j^{\parallel}(\vk) = 0$ and $\Pi^{\lambda}_{ij}(\hat\vk)A_j^{\lambda'}(\vk) = {1\over2}(1+\lambda\lambda')A_i^{\lambda'}(\vk) \equiv A_i^{\lambda}(\vk)\delta_{\lambda\lambda'}$. From here we arrive at the desired property for a chiral projector: given the Cartesian components of a vector, say the gauge field $\vec A = (A_1,A_2,A_3)$ or the electric field $\vec E = (E_1,E_2,E_3)$, we can obtain their chiral components by means of the projection operation
\begin{eqnarray}
&& \hspace{-1cm} A^{\pm}_i(\vk) \equiv \Pi^{\pm}_{ij}(\hat\vk)A_j(\vk)\,,~~  E^{\pm}_i(\vk) \equiv \Pi^{\pm}_{ij}(\hat\vk)E_j(\vk)\,. 
\label{eqn:helicityPro} 
\end{eqnarray}

Conversely, we can also obtain the Cartesian components of a vector, given its chiral components. This requires to write explicitly the Cartesian form of the chiral vectors, which can be constructed as $\vec\varepsilon^{\,\pm}(\hat\vk) \equiv [\vec u(\hat\vk) \pm i \vec v(\hat\vk)]/\sqrt{2}$, with $\vec v(\hat\vk) \equiv (\hat e_3 \times \hat {\bf k})/|\hat e_3 \times \hat {\bf k}| = (-\sin\varphi,\cos\varphi,0)$ and $\vec u(\hat\vk) \equiv  \vec v \times \hat {\bf k} = (\cos\theta\cos\varphi,\cos\theta\sin\varphi,-\sin\theta)$, given that $\hat {\bf k}~=~(\sin\theta\cos\varphi,\sin\theta\sin\varphi,\cos\theta)$, where $\theta$ and $\varphi$ are the spherical polar and azimuthal angles. We then obtain
\bea\label{eq:Chiral2Cartesian}
A_i = \varepsilon_i^{+}A^+ + \varepsilon_i^{-}A^- + \hat k_iA^{\parallel}\,, ~~i = 1,2,3 
\eea
where $\varepsilon_i^\pm(\theta,\varphi) \equiv (u_i(\theta,\varphi) \pm iv_i(\theta,\varphi))/\sqrt{2}$.

\section{Lattice formulation}
\label{sec:lattice_approach}

We present a consistent lattice formulation of the problem at hand, consisting of an appropriate spatial discretization scheme for the EOM~(\ref{eq:eomef1})-(\ref{eq:cons2eFolding}) that respects all relevant symmetries involved. We also introduce our implementation of a lattice version of the chiral projector operator in Sec.~\ref{subsec:Lattice_toolkit}. 

A consistent lattice formulation of the EOM of a system with axion-gauge interaction $\phi F\tilde{F}$ was originally introduced in~\cite{Figueroa:2017qmv} for Abelian gauge sectors (see also~\cite{Moore:1996qs,Moore:1996wn}), and later on adapted in~\cite{Cuissa:2018oiw} to account for self-consistent evolution in an expanding background. There, an appropriate lattice gauge invariant version of $\phi F\tilde F$ was derived, so that the Bianchi identities and the shift symmetry of the system are preserved exactly at the lattice level, see~\cite{Cuissa:2018oiw,Figueroa:2017qmv} for details. The lattice equations presented in those works required, however, an implicit evolution scheme for symplectic time integrators, due to the fact that the evolution of the electric field depends on itself at other lattice sites (see~\cite{Figueroa:2020rrl} for a discussion on different time integrators). Here, we present a modified version of such lattice equations -- already used in {\tt Paper I} --, suitable for non-symplectic time integrators like, \eg Runge-Kutta 2. This evades the necessity to use implicit schemes, while still reproducing the continuum equations to order $\mathcal{O}(\delta t^2,\delta x^2)$, just as in the original implicit integrators from~\cite{Cuissa:2018oiw,Figueroa:2017qmv}. 

We consider that $\phi$ lives on the lattice {\it sites} $\textbf{n}$, $A_i$'s on the {\it links} (\ie halfway in between lattice sites, at $\textbf{n}+\hat{\imath}/2$), and we define
spatial lattice derivatives as usual, using {\it forward/backward} operators $\Delta^{\pm}_{i}\varphi \equiv \pm \frac{1}{\delta x}(\varphi_{\pm i}-\varphi)$, where $\delta x$ is the {\it lattice spacing}, with the $\pm {i}$ subscripts indicating spatial displacements in one lattice unit in the direction $\hat{\imath}$. We define electric and magnetic fields on the lattice as
\be\label{eq:EandBinLattice}
E_i(\textbf{n}+\hat{\imath}/2) \equiv 
\dot A_i\,,~~
B_i(\textbf{n}+\hat{\jmath}/2+\hat{k}/2) \equiv \sum_{j,k}\epsilon_{ijk}\Delta^+_jA_k\,, 
\ee
and introduce \textit{improved} versions that live on the lattice sites ${\bf n}$, as
\begin{equation}
\begin{gathered}
 E_{i}^{(2)} (\textbf{n}) \equiv  \frac{1}{2}\left(E_{i}+E_{i,-i}\right)\,, 
 \\
 B_{i}^{(4)} (\textbf{n})\equiv \frac{1}{4}\left(B_{i}+B_{i,-j}+B_{i,-k}+B_{i,-j-k}\right)\, .
\end{gathered}
\label{eq:E2B4definition}
\end{equation}
Following~\cite{Figueroa:2017qmv,Cuissa:2018oiw}, we use the above ingredients to build an action which preserves gauge invariance and the shift symmetry of $\phi$, at the lattice level (see Appendix~\ref{App:LatticeFirstPrinciples}). Contrary to~\cite{Figueroa:2017qmv,Cuissa:2018oiw}, however, we do not discretize the time derivatives in the lattice action, but instead leave them as continuum derivatives. The EOM follow from variation of the lattice action Eq.~(\ref{eq:latticeAction}), which expressed directly in terms of the number of e-foldings $\mathcal{N}$, with $'\equiv d/d\mathcal{N}$, read as
\begin{widetext}
\begin{eqnarray}
\pi'_{\phi} &=& - 3\pi_{\phi} + \frac{1}{H}\left(\frac{1}{a^2}\sum_{i} \Delta_{i}^{-}\Delta_{i}^{+}\phi - m^2\phi+ \frac{\alpha_{\Lambda}}{a^{3}m_{p}}\sum_{i}E_{i}^{(2)}B_{i}^{(4)}\right)
\, , \label{eqn:eom1LatProgram}\\
E'_{i} &=& - E_{i} - {1\over H}\left(\frac{1}{a^2}\sum_{j,k}\epsilon_{ijk}\Delta_{j}^{-}B_{k} + \frac{\alpha_{\Lambda}}{2am_{p}}\left(\pi_{\phi}B_{i}^{(4)} + \pi_{\phi, +\hat{\imath}}B_{i,+\hat{\imath}}^{(4)}\right) \right. \nonumber\\
&& \hspace{5cm} - \left.\frac{\alpha_{\Lambda}}{4am_{p}}\sum_{\pm}\sum_{j,k}\epsilon_{ijk}\left\lbrace \left[(\Delta_{j}^{\pm}\phi)E^{(2)}_{k,\pm \hat{\jmath}}\right]_{+\hat{\imath}}+\left[(\Delta_{j}^{\pm}\phi)E^{(2)}_{k,\pm \hat{\jmath}}\right\rbrace\right]\right)\,, \label{eqn:eom2LatProgram}
\end{eqnarray}
\end{widetext}
while the Gauss constraint reads
\begin{eqnarray}
\sum_{i}\Delta_{i}^{-}E_{i} = -\frac{\alpha_{\Lambda}}{2am_{p}}\sum_{\pm}\sum_{i}\left(\Delta_{i}^{\pm}\phi \right)B_{i,\pm \hat{\imath}}^{(4)} \; . \label{eq:GaussLatProgram} 
\end{eqnarray}
The lattice counterparts of the Friedmann equations, Eq.~({\ref{eqn:eom3}}) and ({\ref{eqn:Hubble}}), are
\begin{eqnarray}
 - 3HH' &=& \frac{1}{m^2_{p}}\left(3\rho^{\rm L}_{\rm K}+\rho^{\rm L}_{\rm G}+2\rho^{\rm L}_{\rm EM}\right)\, ,\label{eqn:eom3LatProgram}\\
 3H^2 &=& \frac{1}{m^2_{p}}(\rho^{\rm L}_{\rm K}+\rho^{\rm L}_{\rm G}+\rho^{\rm L}_{\rm V}+\rho^{\rm L}_{\rm EM})\; , \label{eq:HubbleLatProgram} 
\end{eqnarray}
where the different energy density contributions (\ref{eqn:energyDensityTerms}) are computed in the lattice as
\begin{equation}
\begin{gathered}
\rho^{\rm L}_{\rm K} \equiv \frac{1}{2}\left\langle\pi^2_{\phi}\right\rangle_{\rm V}\; , \quad \rho^{\rm L}_{\rm G} \equiv \frac{1}{2a^2}\Big\langle(\vec\Delta^{\hspace{-0.5mm}+}\hspace{-0.5mm}\phi)^2\Big\rangle_{\rm V}\; ,
\\
\rho^{\rm L}_{\rm V} \equiv \frac{1}{2}m^2\left\langle \phi^2 \right\rangle_{\rm V}\;, \quad\rho^{\rm L}_{\rm EM} \equiv \frac{1}{2a^4} \Big\langle a^2 \vec E^2 + \vec B^2\Big\rangle_{\rm V} \;,
\end{gathered}
\label{eqn:energyDensityTermsl}
\end{equation}
with $\langle ... \rangle_{\rm V} \equiv \frac{1}{N^3}\sum_{\mathbf{n}}(...)$ denoting volume averaging over a lattice of $N$ points per dimension. As long as $N$ is sufficiently large to guarantee that -- given a lattice spacing $\delta x$ -- the (comoving) lattice length $L = N\delta x$ is much bigger than the (comoving) scales $\lambda_* \sim 2\pi/k_*$ of the excited fields (say $k_*$ is the typical peak scale of the fields' power spectra), this procedure leads to a well-defined notion of a homogeneous and
isotropic expanding background, within the given (comoving) volume $V = L^3$.

We have used the public package \CLns~\cite{Figueroa:2020rrl,Figueroa:2021yhd,Figueroa:2023xmq} as the computational environment to implement our lattice formulation. 

\subsection{The reciprocal lattice}
\label{subsec:Lattice_toolkit}

\noindent We use a discrete Fourier transform given by
\be
 f(\textbf{n}) \equiv \frac{1}{N^3}\sum_{\tilde{\textbf{n}}}e^{-i\frac{2\pi}{N}\tilde{\textbf{n}}\textbf{n}}f(\tilde{\textbf{n}})\; , \; f(\tilde{\textbf{n}}) \equiv \sum_{\textbf{n}}e^{+i\frac{2\pi}{N}\textbf{n}\tilde{\textbf{n}}}f(\textbf{n})\; ,
\ee 
where $\tilde{\textbf{n}} = (\tilde{n}_1,\tilde{n}_2,\tilde{n}_3)$ refers to the sites in the reciprocal lattice, with $\tilde{n}_i=-N/2 + 1,...,-1,0,1,...,N/2$. We choose periodic boundary conditions in the position lattice, so that $f({\bf n}+\hat{\jmath} N) = f({\bf n})$, for $j = 1,2,3$. This implies the existence of a minimum infrared (IR) momentum $k_{\text{IR}}=2\pi/L$ on the lattice, so that the continuum (comoving) momenta are identified as $\textbf{k}=k_{\text{IR}}\tilde{\textbf{n}}$. There is also a maximum ultraviolet (UV) momentum that we can represent in the lattice, $k_{\text{UV}}=\sqrt{3}{N\over 2}k_{\text{IR}}$. 

While different choices of lattice operators are possible to represent a given continuum derivative operation, each choice implies a different {\it lattice momentum} $ {\bf k}_{\rm L}(\tilde{\bf n})$, that will depart differently from the continuum momenta ${\bf k}(\tilde{\bf n})$ in the ultraviolet (UV) scales ($|\tilde{\bf n}| > N/2$) of the lattice. The Fourier transform of any lattice derivative operator over a function $f$  satisfies the relation
\be
   (\nabla_i f)(\tilde{\textbf{n}}) = -i k_{\text{L},i} (\tilde{\textbf{n}}) f (\tilde{\textbf{n}}) \; ,
\ee
which defines the lattice momentum $k_{\text{L}, i}(\tilde{\textbf{n}})$ ascribed to the lattice representation of $\nabla_i$. In our case, we use backward/forward operators which, if interpreted as centred halfway of the lattice sites, at ${\bf n} + \hat{\imath} / 2$,  lead to the lattice momentum (both for forward and backward operations)
\be\label{eq:LatticeMementum}
k_{\text{L},i} = 2\frac{\sin(\pi \tilde{n}_i/N)}{\delta x}\; .
\ee
As expected, Eq.~(\ref{eq:LatticeMementum}) tends to the linear momentum $k_{\text{L},i}(\tilde{\textbf{n}}) \simeq 2\pi\tilde{n}_i/\delta xN \equiv k_{\text{IR}} \tilde{n}_i$, at the IR scales of the lattice ($|\tilde{\textbf{n}}|\ll N/2$). We refer the reader to~\cite{Figueroa:2011ye,Figueroa:2020rrl} for further discussion on lattice derivative operators and their lattice momenta. 

We use Eq.~(\ref{eq:LatticeMementum}) to build the lattice version of the chiral  projector~(\ref{eq:ChiralProjector}) as 
\be
\Pi^{{\rm L},\lambda}_{ij}(\tilde{\textbf{n}}) = \frac{1}{2}\left( \delta_{ij} - \frac{k_{\text{L},i}k_{\text{L},j}}{k_{\text{L}}^2} -\lambda \frac{i}{k_{\text{L}}}\epsilon_{ijk}k_{\text{L},k}\right)\; ,
\label{eq:projector_lattice}
\ee
where $\lambda = \pm$ and $k_{\text{L}} = (k_{\text{L},1}^2+k_{\text{L},2}^2+k_{\text{L},3}^2)^{1/2}$. Given the Cartesian components of a vector on the lattice, $A_i(\tilde {\bf n})$, we will obtain its chiral components $\lambda = \pm$ as
\begin{eqnarray}
A_j^\lambda(\tilde{\textbf{n}}) = \Pi^{{\rm L},\lambda}_{ij}(\tilde{\textbf{n}})A_j(\tilde{\textbf{n}})\,.    
\end{eqnarray}
The Fourier transform of these $A_i^\lambda(\tilde{\textbf{n}}) \longrightarrow A_i^\lambda({\textbf{n}})$ will be transverse only with respect to the forward/background derivative operators, \ie $\Delta_i^\pm A_i^\lambda({\bf n}) = 0$\footnote{Note that $\pm$ in $\Delta_i^\pm$ refers to forward (+) / backward (-) derivatives, unrelated to the chiral component $\lambda$, which are also either $+$ or $-$ (independently of the choice of derivative).}. 

The lattice version of Eq.~(\ref{eq:Chiral2Cartesian}),
\begin{eqnarray}\label{eq:Chiral2CartesianLattice}
A_i = \varepsilon_{{\rm L},i}^{+}A^+ + \varepsilon_{{\rm L},i}^{-}A^- + \hat k_{{\rm L},i}A^{\parallel}\,, ~~i = 1,2,3
\end{eqnarray}
allow us to obtain the Cartesian components of a vector, given its chiral components. The difference between Eq.~(\ref{eq:Chiral2Cartesian}) and Eq.~(\ref{eq:Chiral2CartesianLattice})  is that, in the latter, we use a lattice version of the chiral vectors, which we construct using 
the azimuthal and polar angles $\theta_{\rm L},\varphi_{\rm L}$ of the lattice momentum $\hat {\bf k}_{\rm L} = (\sin\theta_{\rm L}\cos\varphi_{\rm L},\sin\theta_{\rm L}\sin\varphi_{\rm L},\cos\theta_{\rm L})$, with
\begin{eqnarray}
\hspace{-4mm}\vec\varepsilon_{{\rm L}}^{\,\pm}(\tilde{\bf n}) &\equiv& [ \vec u(\hat\vk_{\rm L}) \pm i  \vec v(\hat\vk_{\rm L})]/\sqrt{2}\,,\\
\vec v(\hat\vk_{\rm L}) &=& (-\sin\varphi_{\rm L},\cos\varphi_{\rm L},0)\,,\\ 
\vec u(\hat\vk_{\rm L}) &=& (\cos\theta_{\rm L}\cos\varphi_{\rm L},\cos\theta_{\rm L}\sin\varphi_{\rm L},-\sin\theta_{\rm L})\,.
\end{eqnarray}
We refer the reader to App.~\ref{app:chiralBaseProj} for an explicit version of the polar angles in terms of the lattice momentum.

Finally, we discuss briefly the lattice definition of power spectrum. Let $f(t,\vx)$ be a generic field, with ensemble average in the continuum, say at a fixed time, given by 
\begin{eqnarray}
\langle {f}^2 \rangle = \int \hspace{-1mm}d\log k~\Delta_{\tt f}(k)\,,  ~~~\Delta_{f}(k)=\frac{k^3}{2\pi^2}\mathcal{P}_{f}(k)\,,
\nonumber\\
\langle f(\vk)f^{*}(\vk') \rangle = (2\pi)^3\mathcal{P}_{f}(k)\delta(\vk-\vk')\,.~~~~~
\label{eqn:powerSpectrum}
\end{eqnarray}
As explained in~\cite{CLPS}, there are different approaches to define the power spectrum on a lattice that mimic $\Delta_f({\bf k})$. In this work we opt to use \CLns's {\tt Type~I-Version~1} power spectra and canonical binning. This means that in a lattice with $N$ sites per dimension and lattice spacing $\delta x$, there are $l_{\rm max} = [\sqrt{3}N/2]$ bins, labelled as $l = 1, 2, 3, ..., l_{\rm max}$, with representative momentum of each bin given by $k_l =
k_{\rm IR}l$, $k_{\rm IR} \equiv 2\pi/(N\delta x)$. The lattice power spectrum is then
\begin{eqnarray}\label{eq:PSI.v1}
\Delta_{f}(k_l) = {k_l d x\over 2\pi N^5}\#_{l}\left\langle\left|f (\tilde{\bf n}')\right|^2\right\rangle_{R(l)}\,,
\end{eqnarray}
with $\#_l$ the exact multiplicity of modes within the $l$-th bin, and $\langle\left|f (\tilde{\bf n}')\right|^2\rangle_{R(l)} \equiv {1\over\#_l} \sum_{\tilde{\bf n}' \in R(l)} \left|f (\tilde{\bf n}')\right|^2$ an angular average over all modes $\tilde{\bf n}'$ inside the $l$-th bin, represented by $R(l) \equiv \lbrace \tilde{\bf n}' ~/~ |\tilde{\bf n}'| \in [l-1/2,l+1/2)\rbrace$. We note that within the IR region ($l < N/2$) of the lattice, Eq.~(\ref{eq:PSI.v1}) is well approximated by $\Delta_{f}(k_l) \simeq (k_l^3/2\pi^2) (\delta x^3/ N^3)\large\langle| f (\tilde{\bf n}')|^2\large\rangle_{R(l)}$, resembling closely the continuum form $(k^3/2\pi^2)\mathcal{P}_{f}(k)$. Furthermore, the expectation value obtained with the lattice power spectrum just defined, $\left\langle f^2 \right\rangle = \sum_{l=1}^{l_{\rm max}} {1\over l}\Delta_{f}(k_l)$, is exactly identical to the lattice volume average $\left\langle f^2 \right\rangle_V \equiv \frac{1}{N^3}\sum_{\bf n} f^2({\bf n})$, and mimics the continuum counterpart $\int d\log k~\Delta_{f}(k)$ as desired, since ${1/l} \equiv {k_{\rm IR}/k_l} \simeq dk/k \equiv d\log k$. In practice, we compute power spectra in our simulations for $\phi,\vec A^\pm,\vec A,\vec E^\pm,\vec E,\vec B$ as $\Delta_\phi,\Delta_A^{(\pm)},\Delta_A,\Delta_E^{(\pm)},\Delta_E,\Delta_B$ via Eq.~(\ref{eq:PSI.v1}) with $|f(\tilde{\bf n})|^2 =$ $|\phi(\tilde{\bf n})|^2$, $|\vec A^\pm(\tilde{\bf n})|^2$, $|\vec A(\tilde{\bf n})|^2$, $|\vec E^\pm(\tilde{\bf n})|^2$, $|\vec E(\tilde{\bf n})|^2$, and $|\vec B(\tilde{\bf n})|^2$, respectively, with the electric and magnetic fields given by Eq.~(\ref{eq:EandBinLattice}). 

For a discussion on other choices of power spectra representations on a lattice, we point the reader to \CLns's {\it technical note I}~\cite{CLPS}. 

\section{Lattice Simulations, Part I. Linear regime, onset of non-linearities and homogeneous backreaction}

Our simulations need to capture the evolution of all relevant modes involved in the dynamics, starting from the initial vacuum configuration of the gauge field, passing through the initial linear evolution, and eventually reaching the non-linear regime. As we will see, the range of modes $k_{\rm IR} \leq k \leq k_{\rm UV}$ that are truly relevant for the dynamics, depends strongly on the strength of the axion-gauge interaction. The range of modes becomes wider as we increase the coupling $\alpha_\Lambda$, making the simulations with the largest couplings computationally highly demanding. Furthermore, in simulations requiring such a large range of modes, the integrated contribution from the initial vacuum tail of the gauge field may dominate over the inflaton contribution. This would fake and affect the initial dynamics, which should be only inflaton-driven during the linear regime. 

In this section, we introduce various procedures to start running our simulations, which represent different strategies on how to minimize the computational cost of our runs, and/or how to prevent the gauge field vacuum configuration to distort the physical dynamics. In section~\ref{subsec:IC_Lattice} we present our method to set up the initial condition on the lattice when the gauge field is still in vacuum. In Section~\ref{subsec:LinRegimeLatticeTH} we introduce a simplified version of the lattice EOM to simulate only the linear regime, where we also present tests on the robustness of our lattice simulations in such regime. We then introduce a criterium to switch from the linear evolution to the fully non-linear dynamics in Section~ \ref{subsec:switch}, and we also propose a method to start simulations with the gauge field already excited (way above the BD vacuum solution). We finalize with Section~\ref{subsec:resultsHomo}, where we show our ability to reproduce the homogeneous backreaction regime on the lattice, which serves a twofold purpose: as a consistency check on the robustness of our lattice formulation, and as a reference for comparison  against the truly physical local backreaction regime of the system, which we discuss thoroughly in Section~\ref{sec:localBR}.

\subsection{Initial condition on the lattice}
\label{subsec:IC_Lattice}

In order to run any simulation, we need first to initialize all the fields in the lattice. In the case of the gauge field, we know that the Bunch-Davies (BD) vacuum~(\ref{eq:BD}) is the solution that sets the initial condition for each helicity component, for wavelengths well inside the Hubble radius. Such condition characterizes an initial quantum vacuum state $|0\rangle$ (for which $\hat a_{\bf k}|0\rangle = 0$), whereas solving for Eqs.~(\ref{eq:eomef1})-(\ref{eq:eomef3}) [or more precisely for their lattice counterparts Eqs.~(\ref{eqn:eom1LatProgram})-(\ref{eqn:eom2LatProgram}) and  (\ref{eqn:eom3LatProgram})] assumes implicitly the use of classical fields. Given the quantum expectation value $s_{A}^2(\tau,\vk) \equiv \langle 0| \hat A^\lambda(\tau,{\bf k}) \hat A^{\lambda^\dag}\hspace{-0.7mm}(\tau,{\bf k})| 0\rangle \equiv |\mathcal{A}^\lambda(\tau,{\bf k})|^2$ for the chirality $\lambda$, the BD condition implies $s_{A}^2 \longrightarrow {1\over2k}$. This corresponds to initially vanishing occupation numbers, $n_{\bf k} = 0$, however, we know that one of the gauge field chiralities ($\lambda = +$ in our convention) will grow exponentially fast during the linear regime. This leads, in turn, to exponentially growing occupation numbers, $n_{\bf k} \gg 1$, and as a result, we can actually approximate the vacuum expectation values of products of field operators, by ensemble averages of random fields, see \eg \cite{Garcia-Bellido:2002fsq}. This corresponds to the classical limit of the excited field(s). 

In practice, at some initial time when wavelengths are still deep inside the Hubble radius, we create a random realization of the corresponding Fourier amplitudes of each chirality. Initially, such classical stochastic configuration fails, of course, to capture well the truly quantum nature of the gauge field, as the commutator of the field amplitude and its conjugate momentum vanishes. However, as the unstable gauge field chirality grows exponentially during the linear regime, the gauge field becomes classical, and quantum expectation values become well approximated by ensemble averages over classical field realizations on the lattice. In that moment, the quantum nature of the field is no longer relevant, as the quantum field commutator becomes negligible compared to the field amplitudes, which satisfy $|A(\tau,{\bf k})|^2\approx|A^+(\tau,{\bf k})|^2 \gg 2\times{1\over 2k}$ within the finite range of modes where the excitation is supported. 

In order to configure the gauge field initial values, we create a random realization of the helicity components as described above. In practice, we first re-write the BD condition~(\ref{eq:BD}) as a function of the e-folding time variable $\mathcal{N}$. 
Making use of the conformal time relation $\tau\simeq -1/(aH)$ deep inside inflation, we write the gauge and the electric field chiral amplitudes as
\bea
A^{\pm}(\mathcal{N},\tilde{\bf n}) &\simeq& f^{\pm}_ke^{ik/(aH)}\,,
\label{eqn:ABDcosmic}\\
E^{\pm}(\mathcal{N},\tilde{\bf n}) &\simeq& -\frac{i}{a}kg_k^{\pm}e^{ik/(aH)}\,,\label{eqn:EBDcosmic}
\eea
where $f^+_k, f^-_k, g^+_k, g^-_k$ are the independent amplitude realizations of a Gaussian random field with zero mean and root-mean-square ${1/\sqrt{2k}}$. We note that $\mathcal{N}$ appears only implicitly throughout $a=a(\mathcal{N})$ and $H=H(\mathcal{N})$, and we highlight that $k$ in Eqs.~(\ref{eqn:ABDcosmic})-(\ref{eqn:EBDcosmic}) corresponds to the modulus of the linear momentum, $k = |\vk| = k_{\rm IR}|\tilde{\bf n}|$, to mimic correctly the continuum spectra.

Using Eq.~(\ref{eq:Chiral2CartesianLattice}), we finally convert the chiral fluctuations~(\ref{eqn:ABDcosmic})-(\ref{eqn:EBDcosmic}) into Cartesian components of the gauge and the electric field, $A_i(\mathcal{N},{\bf n})$ and $E_i(\mathcal{N},{\bf n})$, where we highlight that the angles $\theta_{\rm L}, \varphi_{\rm L}$ in Eq.~(\ref{eq:Chiral2CartesianLattice}) correspond to the polar and azimuthal angles of the lattice momentum $ {\bf k}_{\rm L}(\tilde{\bf n})$, the Cartesian components of which are given in Eq.~(\ref{eq:LatticeMementum}). A more detailed discussion of the implementation of the Bunch-Davies vacuum in the lattice can be found in Appendices~\ref{app:chiralBaseProj} and~\ref{App:BD}.

The initialization of the inflaton and of the expansion rate, on the other hand, are more straightforward compared to the gauge field. We start our simulations when the inflationary dynamics are still well dominated by the homogeneous inflaton, so we set up a spatially independent initial amplitude $\phi$ and conjugate momentum according to the slow-roll condition $\pi_\phi \simeq - \phi/3 H$. The initial Hubble rate in that moment is dictated by the kinetic and potential contributions, $\rho_{\rm K}^{\rm L}$ and $\rho_{\rm V}^{\rm L}$, respectively, to the Hubble law, {\it c.f.}~Eq.~(\ref{eq:HubbleLatProgram}). Technically, we should also add fluctuations on top of the homogeneous field, say with a spectrum of quantum vacuum fluctuations in Bunch-Davies. However, in practice, we can set the inflaton fluctuations to zero during the linear regime, as they would not be well balanced classically in Eq.~(\ref{eq:GaussLatProgram}), given the gauge field initialization described before. Furthermore, and most relevantly, when the dynamics become non-linear, the inflaton develops classical inhomogeneities (due to the local gauge field backreaction), which actually become rapidly much larger than those expected from vacuum fluctuations. 

\subsection{Linear regime on the lattice}
\label{subsec:LinRegimeLatticeTH}

If we start with all modes deep inside the Hubble radius, the initial dynamics captured in the lattice should only correspond to the linear regime. We therefore start our simulations evolving first a lattice version of the EOM that capture only the linear physics,
\begingroup
\allowdisplaybreaks
\begin{eqnarray}
\pi'_{\phi} &=& - 3\pi_{\phi} - \frac{1}{H}m^2\phi\,,
\label{eqn:eom1LatProgramLin}\\ \label{eqn:eom2LatProgramLin}
E'_{i} &=& - E_{i} - {1\over H}\Big[\frac{1}{a^2}\sum_{j,k}\epsilon_{ijk}\Delta_{j}^{-}B_{k} \\ 
&& \hspace{1.45cm} +~ \frac{\alpha_{\Lambda}}{2am_{p}}\Big(\pi_{\phi}B_{i}^{(4)} + \pi_{\phi, +\hat{\imath}}B_{i,+\hat{\imath}}^{(4)}\Big) \Big],\nonumber\\
    - HH' &=& \frac{1}{m^2_{p}}\rho^{\rm L}_{\rm K}\, ,\label{eqn:eom3LatProgramLin}
\end{eqnarray}
\endgroup
and the corresponding constraints,
\begingroup
\allowdisplaybreaks
\begin{eqnarray}
\sum_{i}\Delta_{i}^{-}E_{i} &=& 0 \; , \label{eq:GaussLatProgramLin}\vspace*{-10mm}\\
3H^2 &=& \frac{1}{m^2_{p}}(\rho^{\rm L}_{\rm K}+\rho^{\rm L}_{\rm V})\; ,\label{eq:HubbleLatProgramLin}
\end{eqnarray}
\endgroup
which mimic the continuum Eqs.~(\ref{eq:eoml1})-(\ref{eq:eoml3}) and the constraints~(\ref{eq:gauss})-(\ref{eq:ene}). We can run these linear EOM till the contribution of the terms that turn the dynamics non-linear starts becoming relevant. In that moment we switch to solving the full system of EOM~(\ref{eqn:eom1LatProgram})- (\ref{eqn:eom2LatProgram}) and (\ref{eqn:eom3LatProgram}), and their constraints (\ref{eq:GaussLatProgram}) and (\ref{eq:HubbleLatProgram}). We discuss this switch in Section~\ref{subsec:switch}.

A clear advantage of this method is that it allows to introduce the BD vacuum tail in the lattice, providing in this way an initial condition for all the different modes of the gauge field, while neglecting at the same time the gauge field backreaction into the inflationary dynamics, automatically preventing its vacuum tail to distort the physical evolution. Furthermore, since Eq.~(\ref{eqn:eom2LatProgramLin}) is a linear equation, we can compare its outcome against the solution obtained without the use of the lattice, from solving Eq.~(\ref{eq:linA}) in a 1D grid of discretized
comoving momenta $\{k_1, k_2, ..., k_n\}$, initializing each mode $k_i$ with the BD condition with fixed amplitudes, \ie with $A^\pm(k_i) = f(k_i)e^{-ik\tau}$ and $\dot A_i  =-i(k/a)f(k_i)e^{-ik\tau}$, where $f(k_i) \equiv 1/\sqrt{2k_i}$ is fixed (\ie not a random realization). A comparison against this 1D backreaction-less solution, which serves as a consistency check of our lattice runs, is presented in the next subsection.

One vital aspect that applies to any lattice simulation is the selection of the relevant momentum range or \textit{window}. Due to the exponential expansion nature of inflation, together with our inability to predict the duration of the non-linear dynamics during the last inflationary e-foldings, the momentum window required cannot be determined with full certainty in advance. Therefore, beforehand we even run a lattice simulation for a given coupling $\alpha_\Lambda$, we first study the aforementioned 1D backreaction-less solution for the same coupling (till the end of slow-roll inflation, as dictated by the homogeneous inflaton). From these solutions we determine:\vspace*{0.1cm}

$\bullet$ An  educated guess of the window $[k_{\rm IR},k_{\rm UV}]$ needed to capture at least the modes for which the gauge field power spectrum is significantly above the BD tail, assuming linear evolution of such modes until the end of slow-roll inflation.\vspace*{0.1cm}

$\bullet$ An estimation of the right moment to start our lattice simulations, say $\mathcal{N}_{\rm start}$ e-folds before the end of slow-roll inflation. As every gauge mode needs to be sufficiently sub-Hubble in the moment of the initialisation for the BD solution to be valid, we demand
\bea
k_{\rm IR}= w \times a(\mathcal{N}_{\rm start})H(\mathcal{N}_{\rm start})\;,\label{eqn:IRcriteria}
\eea
with $w \gg 1$ a {\it penetration factor} inside the Hubble radius. In practice we see that $w = 10$ suffices to capture well the initial linear dynamics.\vspace*{0.1cm}
    
$\bullet$ An estimation of the moment at which to switch from the linear EOM (\ref{eqn:eom1LatProgramLin})-(\ref{eqn:eom3LatProgramLin}) to those that capture the full non-linearities of the system, Eqs.~(\ref{eqn:eom1LatProgram})- (\ref{eqn:eom2LatProgram}) and (\ref{eqn:eom3LatProgram}), say $\mathcal{N}_{\rm switch}$ e-folds before the end of slow-roll inflation. The exact criteria to set this value is detailed in~Section~\ref{subsec:switch}.\vspace*{0.1cm}  

We note that, as in {\tt Paper I}, we measure $\mathcal{N}$ assuming that $\mathcal{N}=0$ corresponds to the end of slow-roll inflation, where we also set the scale factor to one.
\begin{figure}[t!]
\includegraphics[width=\columnwidth]{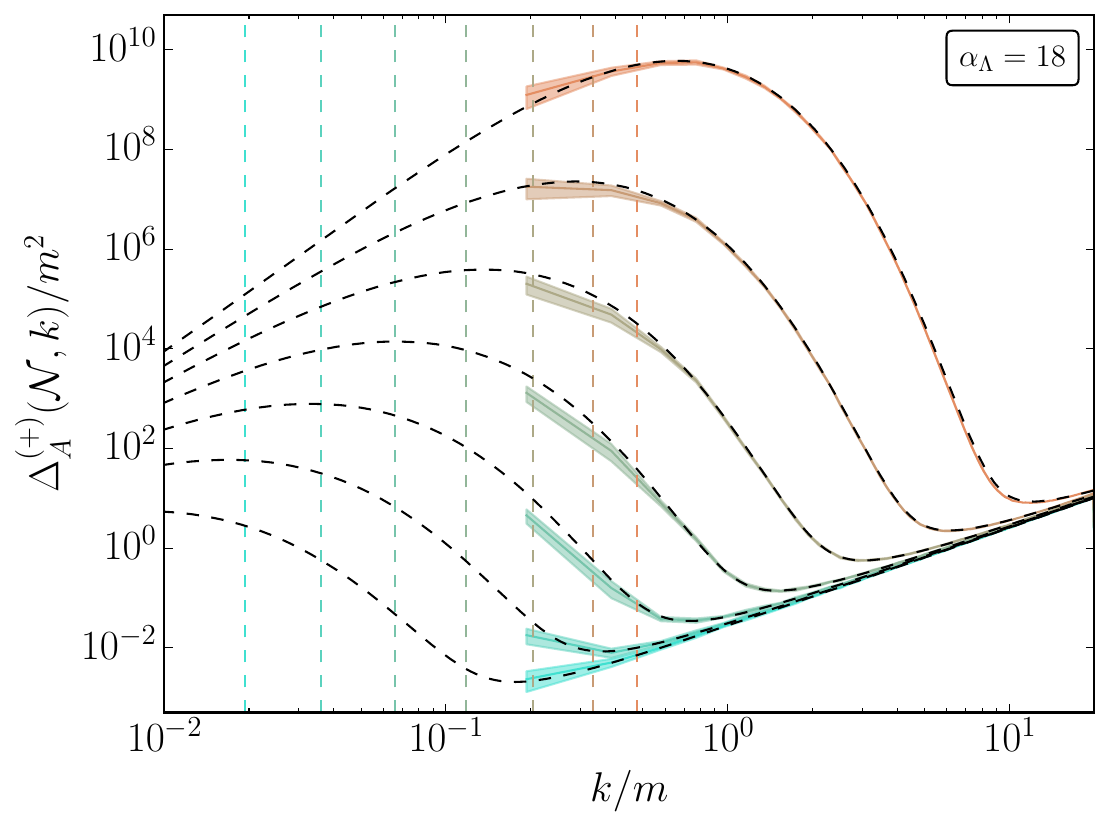}
\caption{Evolution of the tachyonic excitation of the gauge field for the coupling $\alpha_{\Lambda}=18$. Solid coloured lines correspond the mean value from averaging over 5 random realizations on the lattice, with shaded bands representing the $1\sigma$ deviations. The mode by mode numerical solution in the 1D method is shown in dashed. The colour gradient represents the evolution from $\mathcal{N}_{\text{start}}$ (blue) to $\mathcal{N}=-0.3$ (orange), with gaps of $0.7$ e-foldings.}
\label{fig:AllSpectraCompMathematicaVsLatticeIL18and20}
 \end{figure}

All the parameters, $k_{\rm IR}, k_{\rm UV}, \mathcal{N}_{\rm start}$ and $ \mathcal{N}_{\rm switch}$, are highly coupled and should be set altogether. From the backreaction-less solution we can securely set $k_{\rm IR}$, and hence also $\mathcal{N}_{\rm start}$. The lattice UV cutoff $k_{\rm UV}$, however, rather requires to simulate the subsequent non-linear dynamics in the lattice. As we highlighted in {\tt Paper I}, the full non-linear system exhibits a special sensitivity to the UV. As we increase $\alpha_\Lambda$, the number of e-folds in inflation increases, and smaller and smaller scales, which were expected to remain in vacuum in the linear picture, become excited during the non-linear dynamics. If the simulation does not include those scales a priori, no physically sensible description can be obtained. That is why only when simulating during the full non-linear (local) backreaction regime, we can correctly assess which $k_{\rm UV}$ will suffice to cover the relevant scales. We discuss the UV sensitivity of the system in Sec.~\ref{subsec:powUV}, where we address the local backreaction regime. Finally, to assess $\mathcal{N}_{\rm switch}$ we need to establish some criterium as to when the linear regime remains to be valid. We do this in Section~\ref{subsec:switch}, but only after we demonstrate in the next paragraphs the ability of our lattice formulation to capture the linear regime.\\

\hspace{0.7cm}{\small \bf Consistency checks of the linear dynamics}\vspace{0.2cm}

\begin{figure}[t]
\includegraphics[width=\columnwidth]{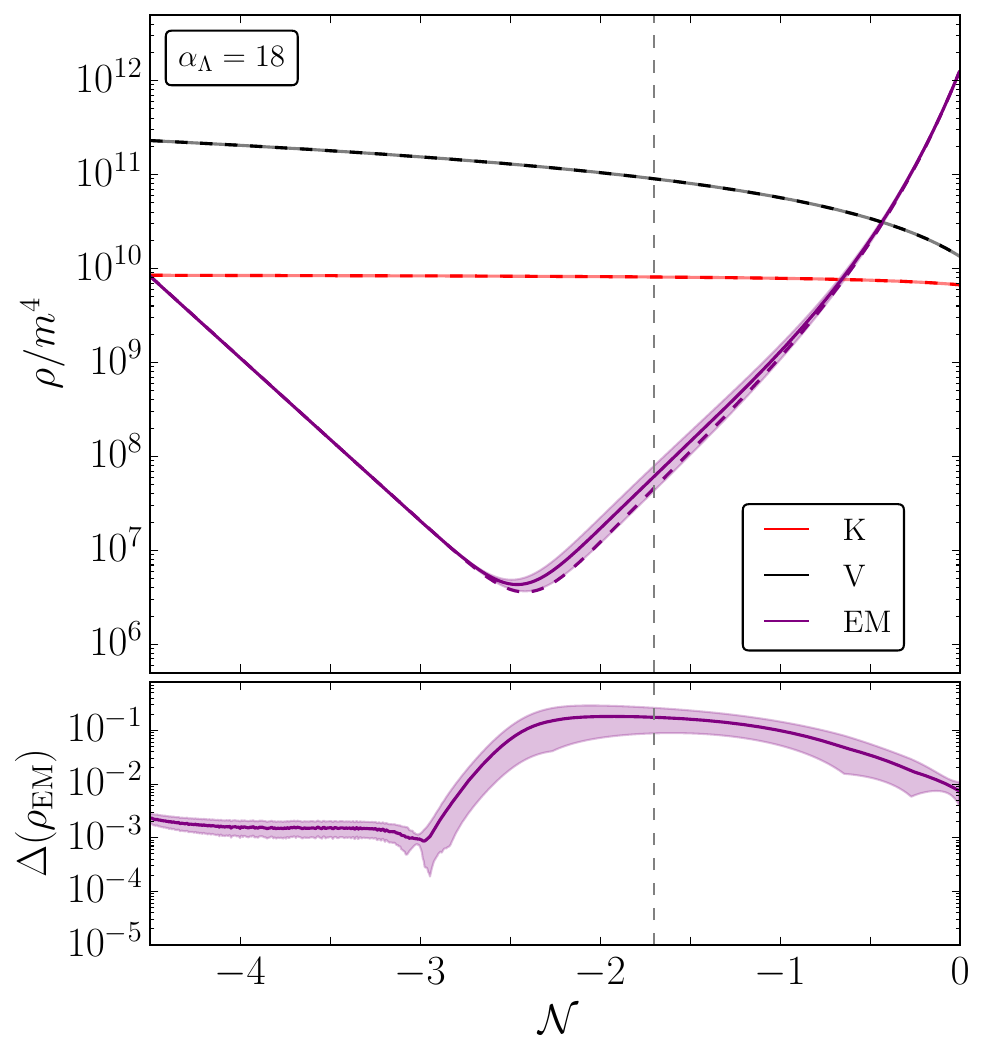}
\caption{Comparison for $\alpha_{\Lambda}=18$ in the linear regime between the different energy contributions: $\rho_{\rm K}$ in red, $\rho_{\rm V}$ in black, and $\rho_{\rm EM}$ in purple, between the lattice simulations (solid lines), and the numerical solutions of the mode-by-mode approach on a 1D grid (dashed lines). For the lattice simulations we include the mean over 5 random realizations and the corresponding $1\sigma$ deviations as a shaded band. We show the total energy density in the top panel, and the relative difference of the electromagnetic energy density between both methods in the bottom panel. The dashed vertical line corresponds to $\mathcal{N}_{\rm switch}$.
}
\label{fig:AllEnergCompMathematicaVsLatticeIL18and20}
\end{figure}
Our procedure to simulate the linear regime is based on setting up the initial condition according to Section~\ref{subsec:IC_Lattice} and running the simplified lattice EOM presented in Section~\ref{subsec:LinRegimeLatticeTH}. In Fig.~\ref{fig:AllSpectraCompMathematicaVsLatticeIL18and20} we compare the evolution of the power spectra for the representative coupling $\alpha_{\Lambda} = 18$, comparing the outcome from lattice simulations of the linear regime (coloured lines) and the 1D backreaction-less solution (dashed lines). The evolution corresponds to the period between $\mathcal{N}_{\text{start}} = -4.5$ and $\mathcal{N} = -0.3$, with steps of $0.7$ e-foldings. We use lattices with $N=640$ points per dimension and $k_{\text{IR}}/m = 0.1932$.

We show the power spectra for the positive helicity, which starts in the BD solution, using a fixed amplitude in the 1D grid method, and the average over random realizations in the lattice case. The figure shows the comparison between the mean value obtained over 5 random realizations (coloured solid lines), with $1\sigma$ standard deviation in shaded bands, and the 1D grid solution (dashed). We also include, with vertical lines and using the same colouring gradient, the scale of the comoving Hubble radius. 

We observe that the spectrum follows well the evolution obtained from the 1D-grid method. As expected, the spectral amplitudes at the largest comoving wavelengths are the first to start experiencing the tachyonic instability, due to the rolling of the inflaton. Namely, the amplitude $|A^+(\mathcal{N},k)|$ starts growing 
above the BD tail when, due to inflationary redshifting, the condition $k \lesssim 10aH$ is attained. As the wavelength approaches the Hubble radius, the spectrum reaches a maximum around Hubble crossing $k \simeq aH$. The maximum of the spectrum coincides roughly with the scale of the Hubble radius at every moment. With our choice of $k_{\rm IR}$ and $\mathcal{N}_{\rm switch}$ we ensure that we capture the maximum amplitude of the spectrum just before the non-linear terms are activated (see discussion in Section~\ref{subsec:switch}), knowing that the spectrum will continue drifting towards larger $k$'s, \ie shorter scales in the lattice. The apparent discrepancy in the most IR scales between the power spectra from the two methods, is essentially due to a {\it cosmic variance} effect: given the reduced number of modes inside the smallest-momentum bins in a lattice, the random realization of the spectrum's amplitude is expected to be more scattered around the actual theoretical value, since the statistical sampling is lower and lower, the more IR the scales are in a lattice. This is only evident in the lattice, as we only include fluctuations there. Even though, the most IR bin does not fully capture the growth in the first stages of the evolution, once $k_{\rm IR}<aH$ it gets the correct amplitude.

We compare the different energy contributions in the linear regime between the two methods, 1D and lattice, in Fig.~\ref{fig:AllEnergCompMathematicaVsLatticeIL18and20}. In the upper panel, we show $\rho_{\rm K}$ (red), $\rho_{\rm V}$ (black) and $\rho_{\rm EM}$ (purple). We obtain the latter for the 1D method from integrating the spectra shown in Fig.~\ref{fig:AllSpectraCompMathematicaVsLatticeIL18and20} (dashed), and in the lattice (solid), from the volume average of the contributions from $\vec{E}$ and $\vec{B}$, see $\rho_{\rm EM}^{\rm L}$ in Eq.~(\ref{eqn:energyDensityTermsl}). Here we also include the mean of all previous 5 random realizations, with the corresponding $1\sigma$ deviation. In the lower panel, we quantify the relative difference of $\rho_{\rm EM}$ between both methods, 
\begin{equation}
   \Delta (\rho_{\rm EM}) = \frac{| \rho_{\rm EM}^{\text{1D}}-\rho_{\rm EM}^{\rm{L}}|}{\rho_{\rm EM}^{\rm{L}}}\, ,
\end{equation}
where $\rho_{\rm EM}^{\rm{L}}$ represents the energy density obtained from our lattice simulations, and $\rho_{\rm EM}^{\text{1D}}$ is the energy density computed from the backreaction-less 1D  solution. 

It can be observed that initially there is a phase where the BD tail dominates over the tachyonic excitation in $\rho_{\rm EM}$. While $\rho_{\text{EM}}$ decreases with the expansion, IR modes also get excited, and eventually the excited IR range of the spectrum comes to dominate over the UV vacuum tail. In this example, this happens around $\mathcal{N} \sim -3$. Before that, the relative difference between methods is of the order of $\sim 0.1 \%$, which is caused by the randomness of the BD condition in the lattice. We have checked that increasing the accuracy of the integrator does not reduce this relative error. Once the excited part of the gauge spectrum dominates over the vacuum tail, the relative difference increases, reaching approximately $\sim 10\, \%$ depending on the random realization. This is due to the cosmic variance problem already noted for the difference between spectra in the IR modes in Fig.~\ref{fig:AllSpectraCompMathematicaVsLatticeIL18and20}. We highlight this issue by including the $1\sigma$ deviation in shaded, and we show that the variability of the fluctuation in the IR modes are causing this effect. Extending the evolution to $\mathcal{N}=0$ (while assuming inflaton slow-roll dominance), the mean relative error decreases again down to $\sim 1\,\%$, as the peak shifts to intermediate scales in the range $[k_{\text{IR}},k_{\text{UV}}]$, and the weight of the most discrepant IR bins becomes less significant. In summary, regardless of the seed for the BD fluctuations, we correctly capture the evolution of the electromagnetic energy density in the lattice. 

We note that while we can, of course, reduce the aforementioned $\sim 10\,\%$ discrepancy around $\mathcal{N}_{\rm switch}$ by pushing $k_{\rm IR}$ to smaller values, that would prevent us from having a sufficiently large $k_{\rm UV}$ to capture correctly the UV excitation of the spectrum during the subsequent non-linear regime, as we will explain in detail in Section~\ref{sec:localBR}. As the spectrum peak shifts towards intermediate lattice scales when the dynamics turn non-linear, we have checked that the accuracy of the outcome during the non-linear regime of simulations with different initial realizations is better than $\sim 1\,\%$ (see App.~\ref{App:NumUVstability}) given our choice of $k_{\rm IR}$ scales for all $\alpha_\Lambda$ coupling values, see Table~\ref{tab:AllSims}.

\subsection{Switching to non-linear dynamics}
\label{subsec:switch}

Now we concern ourselves with determining the right moment $\mathcal{N}_{\rm swtich}$ to switch from simulating Eqs.~(\ref{eqn:eom1LatProgramLin})-(\ref{eqn:eom3LatProgramLin}) to the full non-linear EOM (\ref{eqn:eom1LatProgram})-(\ref{eqn:eom2LatProgram}) and (\ref{eqn:eom3LatProgram}). We determine $\mathcal{N}_{\rm switch}$ by requiring \emph{simultaneously} to fulfil the following criteria at that moment:\vspace*{0.1cm}

$\bullet$ The integration over the excitation range of the gauge field spectrum must significantly dominate over the contribution from UV vacuum modes.\vspace*{0.1cm}

$\bullet$ The IR excitation captured on the lattice must represent the dominant support for the gauge field spectra, so that the contribution from more IR scales that are not captured on the lattice must be negligible.\vspace*{0.1cm}

$\bullet$ The electromagnetic energy density $\rho_{\rm EM}$ must remain sub-dominant with respect to the kinetic and potential energies of the inflaton, $\rho_{\rm K}$ and $\rho_{\rm V}$ respectively.\vspace*{0.1cm}

$\bullet$ The gauge source term $\propto F\tilde F$ in the inflaton's EOM must be sub-dominant compared to the other terms, \ie $\frac{\alpha_{\Lambda}}{a^3m_{p}}|\vec{E}\cdot\vec{B}|\,\ll\,$ $|\ddot{\phi}|$,  $3H|\pi_{\phi}|$, $m^2|\phi|$.\vspace*{0.1cm}

As our computer resources are finite, the level of dominance/sub-dominance of some terms versus others needs to be specifically considered for each coupling $\alpha_\Lambda$. The requisites ensure that the effect of the non-linear local terms will be gradual, initially remaining negligible, with the fields still evolving in the linear regime for a short period after $\mathcal{N}_{\rm switch}$, to later become the dynamics more and more non-linear.  

For the largest couplings considered, in order to prevent the unwanted contribution of the gauge field vacuum tail to affect the dynamics, an additional step is required. Due to the special sensitivity to the UV of the non-linear regime, the lattice maximum momentum $k_{\rm UV}$ (or the comoving spatial resolution) needs to be sufficient to cover all relevant modes that will be excited in such regime. If the BD vacuum solution was imposed to all those modes, the non-classical contribution to the electromagnetic backreaction terms would be too large and would become the evolution unphysical. We tackle this issue introducing an intermediate cutoff $k_{\text{IR}} < k_{\text{BD}} < k_{\text{UV}}$, so that only within the `IR range' $[k_{\rm IR},k_{\rm BD}]$ we set the BD initial condition for the modes that are expected to be excited during the linear regime. Within the `UV range' $[k_{\rm BD},k_{\rm UV}]$, instead, the gauge field and electric field modes are set to zero during the linear dynamics, as they are not expected to become excited from the tachyonic instability. The gauge field modes within $[k_{\rm BD},k_{\rm UV}]$ grow only stimulated out of the non-linear dynamics, when $\mathcal{N} > \mathcal{N}_{\rm switch}$. As a side effect of this procedure we can also tolerate to choose a larger time step $d\mathcal{N}$. We have performed numerical checks in order to ensure that applying such a cutoff does not alter the subsequent non-linear evolution, see App.~\ref{App:NumUVstability} for a detailed discussion.

\begin{figure*}[t]
\includegraphics[width=0.8\textwidth]{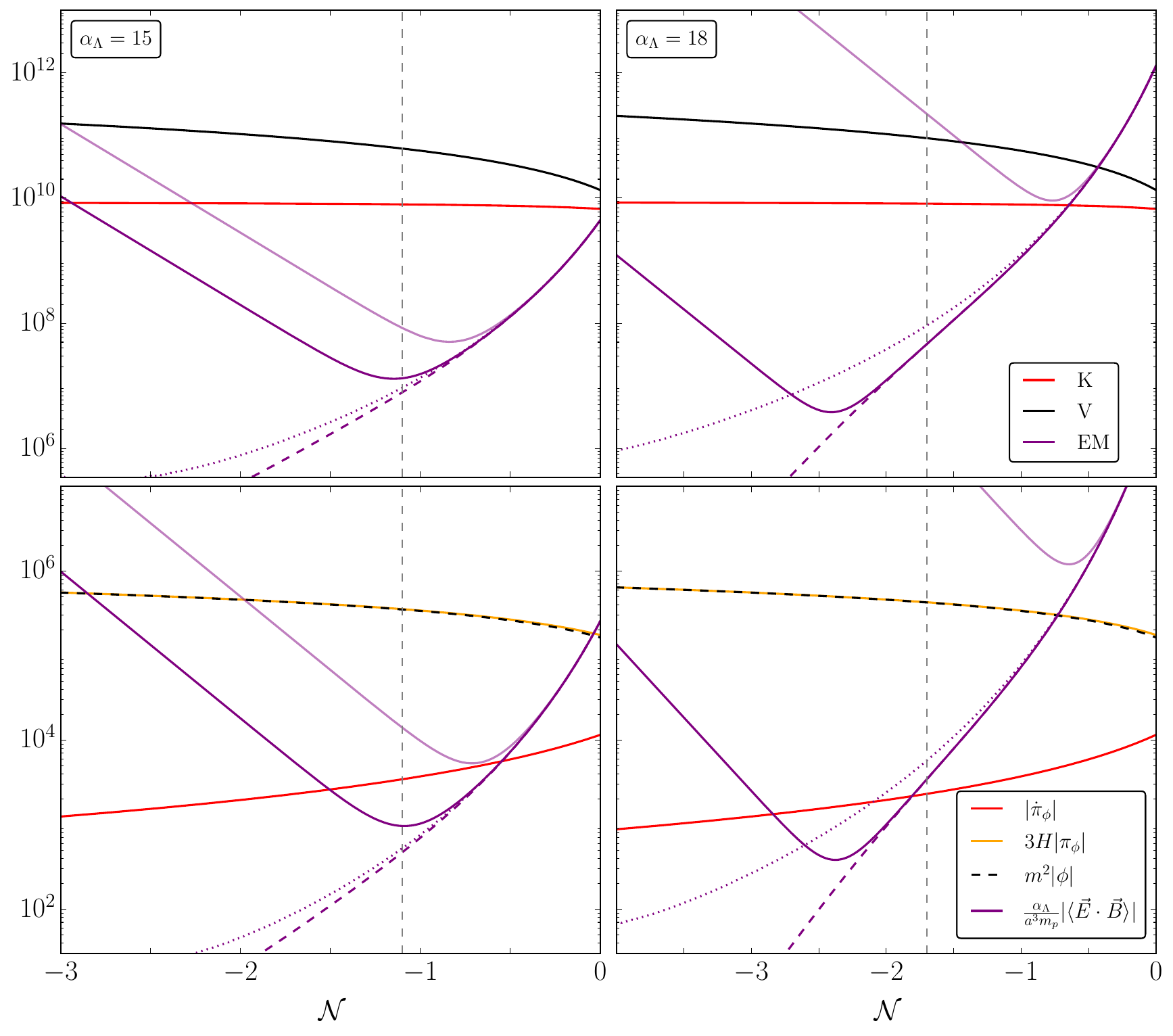}
\caption{Comparison for $\alpha_{\Lambda} = 15$ (left) and $18$ (right) of the terms related to the backreaction used to establish the time $\mathcal{N}_{\text{switch}}$. {\it Top:} the evolution of $\rho_{\rm K}$ (red), $\rho_{\rm V}$ (black) and $\rho_{\rm EM}$ (purple).
{\it Bottom:} the evolution of $|\dot{\pi}_{\phi}|$ (red), $3H|\pi_{\phi}|$ (orange), and $m^2|\phi|$ (black) of Eq.~(\ref{eqn:eom1}), together with $\frac{\alpha_{\Lambda}}{a^3 m_p} |\langle\vec{E} \cdot \vec{B}\rangle|$ (purple). The different purple lines represent different integration ranges: the dark solid purple lines are the contribution integrated for the range $[k_{\text{IR}}, k_{\text{BD}}]$, the light purple the integral for $[k_{\text{IR}}, k_{\text{UV}}]$, the dashed line is for only the excited IR contribution for $[k_{\rm IR}, k_{\rm cut}(\mathcal{N})]$, and the dotted purple line is the contribution of the excitation considering the whole cosmic history, \ie $[k_{\rm min}, k_{\rm cut}(\mathcal{N})]$ (see text). The vertical dashed gray lines correspond to the times $\mathcal{N}_{\text{switch}}$.}
\label{fig:mathematicaIntegratedOnset15vs18}
\end{figure*}

As illustrative examples of how to choose $\mathcal{N}_{\text{switch}}$ according to the above criteria, we discuss the specifics for $\alpha_{\Lambda} = 15$ and $18$, with $k_{\rm IR}$ and $k_{\rm BD}$ specified in Table~\ref{tab:AllSims}. In Fig.~\ref{fig:mathematicaIntegratedOnset15vs18}, we show the evolution of $\rho_{\rm K}$ (red), $\rho_{\rm V}$ (black) as well as $\rho_{\rm EM}$ (purple). In the bottom panel, we present the absolute value of the terms from the homogenous inflaton EOM: $|\dot{\pi}_{\phi}|$ (red), $3H|\pi_{\phi}|$ (orange), $m^2|\phi|$ (black), and $\frac{\alpha_{\Lambda}}{a^3 m_p} |\langle \vec{E} \cdot \vec{B}\rangle|$ (purple). We note that the electromagnetic energy density $\rho_{\rm EM}$, as well as the source term $\propto \langle \vec{E} \cdot \vec{B} \rangle$, are obtained from the 1D grid method from integrating the gauge field spectra. 

The different purple lines correspond to different integration ranges for the integrated electromagnetic contributions. The dark solid lines correspond to the contribution calculated in the range $[k_{\text{IR}}, k_{\text{BD}}]$ and the light ones in $[k_{\text{IR}}, k_{\text{UV}}]$. While the former is the contribution we measure in our simulations, the latter is what we would have if we set the BD solution for the whole dynamical range. The non-continuous lines correspond to the isolated contribution of the IR excitation, that is above the BD initial condition of the gauge spectra. We do this by integrating the spectra up to a time-increasing cutoff $k_{\rm cut}(\mathcal{N})$ that grows from $k_{\rm IR}$ up to $k_{\rm BD}$. The dashed lines correspond to the range $[k_{\rm IR},k_{\rm cut}(\mathcal{N})]$, whereas the dotted lines to $[k_{\rm min},k_{\rm cut}(\mathcal{N})]$, where $k_{\text{min}}$ corresponds to Hubble crossing of the CMB scales at $\mathcal{N} = -60$, and with which we capture using the 1D method the gauge excitation during the whole cosmic history.

The electromagnetic contributions captured by the lattice momentum window (solid purple lines) for $\alpha_{\Lambda} = 15$ (left panels) during the first e-folds are essentially determined by the high-frequency end contribution, around $k \lesssim k_{\rm BD}$, of the vacuum tail. As the inflationary expansion carries on they decrease in time, and eventually become subdominant with respect the inflaton's terms. We note that the EOM comparison (bottom panel) is the one that puts stricter constraints to fulfil our criteria. This is, of course, even worse if the intermediate cutoff scale $k_{\rm BD}$ is not set (light purple line), as the non-classical contribution is considerably higher. 

Additionally, we see that the IR excitation captured by our lattice momentum window (dashed lines) is initially very subdominant, but it eventually becomes comparable to the vacuum tail's contribution around $\mathcal{N} = -1.5$. Furthermore, it is also necessary that by the time we reach $\mathcal{N}_{\text{switch}}$, the IR excitation on the lattice dominates over that corresponding to modes below our IR threshold, \ie $k<k_{\text{IR}}$. We assess this by including the contribution of the excited part computed in the range $[k_{\rm min},k_{\rm cut}(\mathcal{N})]$ (dotted lines). The comparison with the dashed lines demonstrates that by the time the tachyonic excitation dominates over the UV tail, it also  acquires well the level of excitation that considers the full cosmic history.

For larger couplings, \eg for $\alpha_{\Lambda} = 18$ (right panels), although the evolution is qualitatively the same, establishing $\mathcal{N}_{\text{switch}}$ is more subtle. The contribution from the IR excitation above BD (dark solid purple lines) begins to dominate over the vacuum tail much earlier than for $\alpha_{\Lambda} = 15$, as the excitation of the gauge field are naturally stronger. We observe, however, that for the chosen $k_{\text{IR}}$, more e-folds of evolution are required for the excitation captured on the lattice (dashed lines) to reproduce the continuum (dotted lines) properly. While considering a smaller $k_{\text{IR}}$ scale would cure this problem, we cannot afford however such a simple solution, because as we will see in Sec.~\ref{subsec:powUV}, for these strong couplings we necessarily need to cover very UV scales, making unfeasible to have lattices with such a separation of scales.
\begin{figure}[t]
\includegraphics[width=\columnwidth]{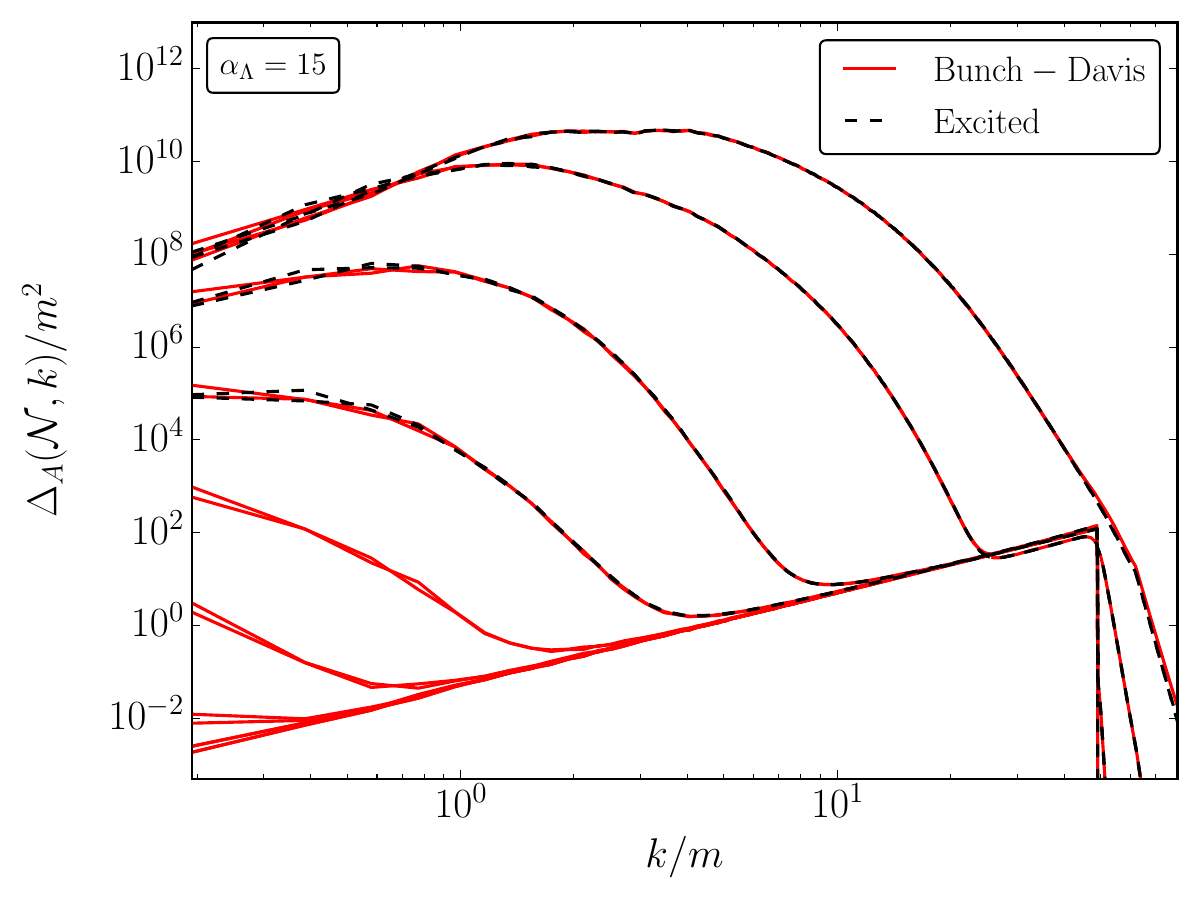}
\caption{Comparison of the evolution of the gauge power spectra for simulations with different initialisation techniques: BD vacuum solution (solid lines) and already excited (solid). We include two realizations for each initialization method with the same colour and line style. The spectra are extracted from $\mathcal{N}_{\rm start}$ until the end of non-linear inflation with $0.7$ e-folds between lines.}
\label{fig:SpectraIL15excitedVSbd}
 \end{figure}
Finally, we note that the effect of the action of the cutoff scale $k_{\rm BD}$ is more pronounced for the largest couplings. As in the previous example, we include the electromagnetic components computed considering that we set the BD vacuum solution all across the range $[k_{\text{IR}}, k_{\text{UV}}]$ using light solid purple lines. It clearly shows that without the cutoff at $k_{\rm BD}$ an incredible overestimation of the integrated contributions will occur, which would falsify completely the dynamics. We conclude that for the largest coupling the use of an intermediate cutoff is compulsory.

For these examples, we determine $\mathcal{N}_{\text{switch}} = -1.1$ ($\alpha_\Lambda = 15$) and $\mathcal{N}_{\text{switch}} = -1.7$ ($\alpha_\Lambda = 18$), as a compromise between the four criteria and the optimization of the dynamical range. For $\alpha_\Lambda = 15$, the exact switch moment is not so critical, since the conditions $\rho_{\rm EM} \ll \rho_{V}, \rho_{K}$ and $\frac{\alpha_{\Lambda}}{a^3 m_p} |\vec{E} \cdot \vec{B}| \ll 3H|\pi_{\phi}|, m^2|\phi|$, hold all the way down to $\mathcal{N} = 0$. In fact, for smaller couplings $\alpha_{\Lambda} < 15$, we have used the same choice $\mathcal{N}_{\text{switch}} = -1.1$, as the gauge field backreaction in such cases does not become significant until after the end of inflation. For $\alpha_\Lambda = 18$, however, this results in $\frac{\alpha_{\Lambda}}{a^3 m_p} |\vec{E} \cdot \vec{B}| \sim |\dot{\pi}_{\phi}|$ around the time $\mathcal{N}_{\rm switch}$. While this is not ideal, in practice these circumstances do not really pose a problem, as the acceleration term is still subdominant with respect to $3H|\pi_{\phi}|$ and $m^2|\phi|$, and $\propto |\vec{E} \cdot \vec{B}|$ does not even exceed a 1$\%$  contribution when the non-linearities are activated. Additionally, we ensure that the criterion $\rho_{\rm EM} \ll \rho_{\rm K}, \rho_{\rm V}$ continues to be met for these values of $\mathcal{N}_{\text{switch}}$.

We repeat the above analysis for the rest of the couplings considered in this work, for which we refer the reader to Table~\ref{tab:AllSims} for the exact values.
\\

\hspace{1.0cm}{\small \bf Initialization with $A_{\mu}$ already excited}\vspace{0.1cm}\label{subsec:excitedInit}

We have also developed an alternative procedure to start a simulation with a power spectrum for the gauge field (and the electric field) already excited above the vacuum tail within a range of IR scales. This is particularly useful for the largest couplings we consider, as it helps in two directions: it saves simulation time as we obtain the IR excited spectrum in the linear regime from the 1D grid solution, and allows us to tolerate slightly larger time steps, as we can start the simulations in a later stage of the inflationary period. 

In order to start a simulation with the gauge field already excited, there is a fundamental subtlety related to the relative phase between the gauge and the electric field. The BD vacuum forces the relative phase between the gauge field amplitude and the time derivative to be $\pi/2$, see Eqs.~\eqref{eqn:ABDcosmic} and~\eqref{eqn:EBDcosmic}. On the other hand, the chiral instability of $A^+$ leads to a vanishing relative phase between the gauge field and its time derivative. There is a smooth phase transition between the excited power spectrum in the IR scales and the vacuum tail in the UV scales. We have implemented, correspondingly, a methodology that initializes on the lattice a configuration of the gauge and electric fields, based on fluctuations in the reciprocal lattice drown from Gaussian random variable with variance equal to the following power spectra: on IR scales we take the form of the excited spectrum above BD (at $\mathcal{N}_{\rm switch}$) from the 1D grid solution, and at UV scales we imposed the BD vacuum form. We introduce a relative phase between the gauge field amplitude and its time derivative, so that $\varphi = 0$ for very excited modes in IR range, $\varphi = \pi/2$ in the deep UV vacuum tail, and we interpolate smoothly between those two values for the small range of modes in between the IR and UV regions. We include a detailed description of the procedure in Appendix~\ref{App:BD}.

As a prove of the validation of this technique, in Fig.~\ref{fig:SpectraIL15excitedVSbd} we include the comparison of the evolution of the gauge power spectra until the end of non-linear inflation, extracted from simulations initialised in the BD vacuum (red solid lines) and excited at $\mathcal{N}_{\rm switch}$ (black dashed lines) for $\alpha_\Lambda=15$. We have included two realizations for each methodology to stress that the difference between methods in the IR region is of the same order as the difference between realizations. We refer the reader to Appendix~\ref{App:NumUVstability} for more comparisons and validations tests in this regard.

\begin{figure*}[t!]
\includegraphics[width=0.8\textwidth]{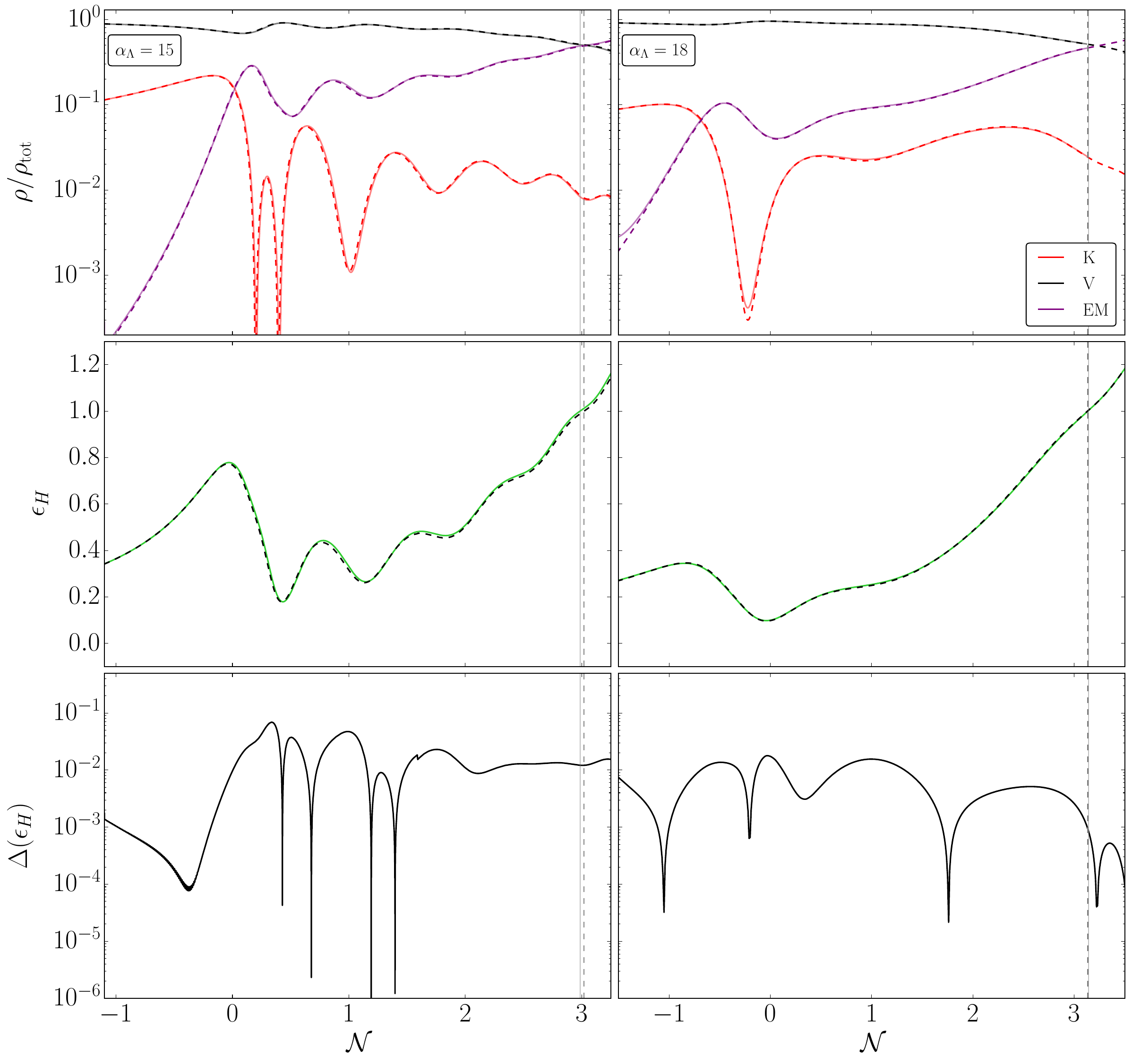}
\caption{Comparison between the predictions of the homogeneous backreaction scheme on the lattice (solid) and of the gradient expansions formalism~\cite{Gorbar:2021rlt} (dashed) for $\alpha_{\Lambda}=15$ (left) and $\alpha_{\Lambda}=18$ (right). {\it Top:} the evolution of the different energy components normalised with respect the total energy density. {\it Middle:} the evolution of the inflationary parameter $\epsilon_H$. {\it Bottom:} the relative difference in $\epsilon_H$ (\ref{eq:reldiff_epsilon}). The vertical lines represent the end of inflation for each method.}
\label{fig:homoComparisonPanel}
\end{figure*}

\subsection{Homogeneous Backreaction}
\label{subsec:resultsHomo}

Before we move on into the local backreaction regime of the system, which we discuss thoroughly in Section~\ref{sec:localBR}, it is convenient that we adapt our lattice procedure to reproduce the homogeneous backreaction regime. This will be useful, in the first place, as a consistency check on the robustness of our lattice formulation, as we will be able to compare our outcome to previous homogeneous backreaction results in the literature. Secondly, this will also help us to highlight the differences, in multiple aspects, with the truly local backreaction dynamics. 

\begin{figure*}[t]
\includegraphics[width=0.8\textwidth]{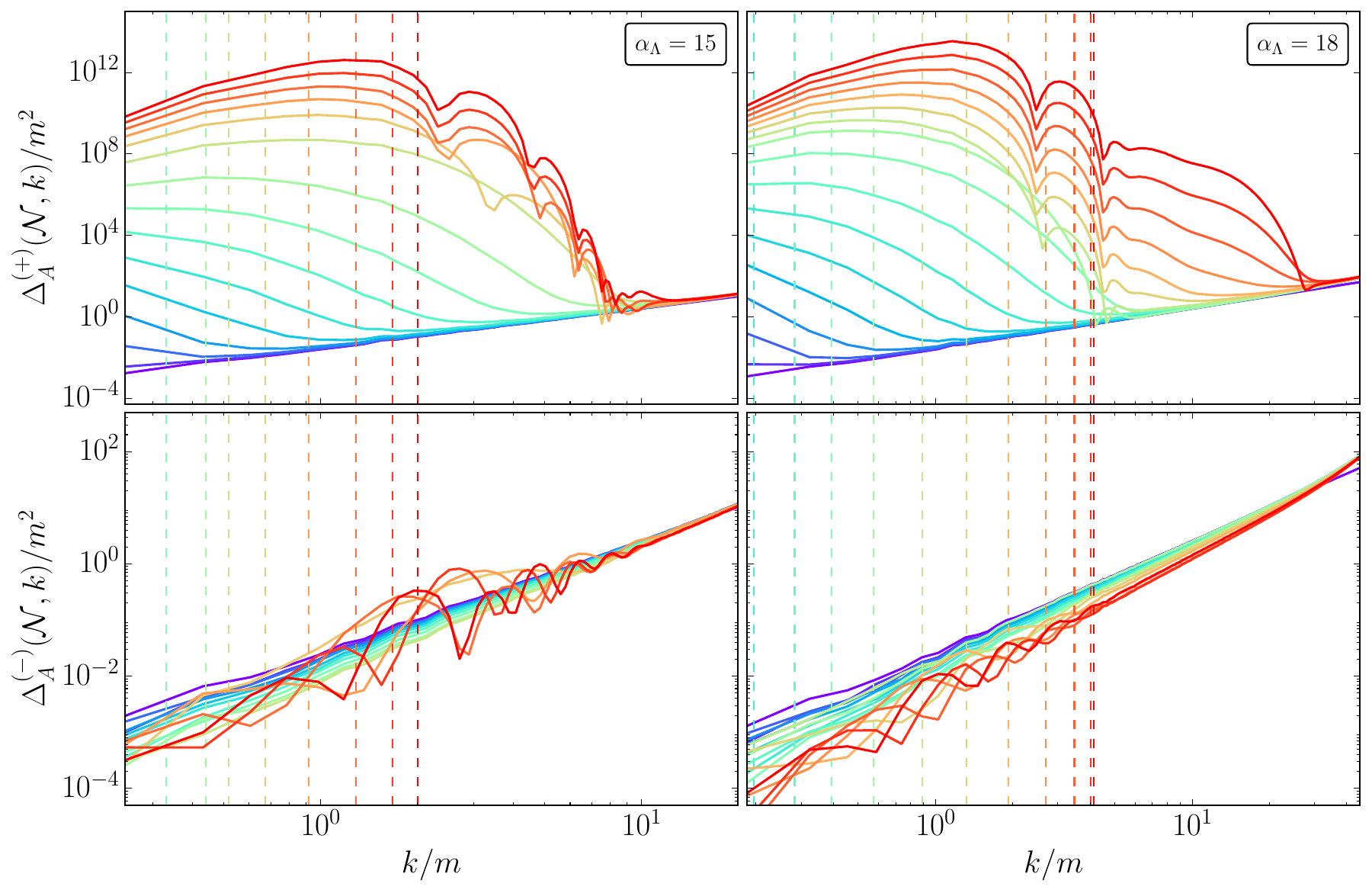}
\caption{Evolution of the power spectra of the positive (upper panels) and negative (lower panels) helicity components of the gauge field for  $\alpha_{\Lambda}=15$ (left) and $\alpha_{\Lambda}=18$ (right) for homogeneous backreaction scheme in the lattice. Color gradient goes from purple (early times) to red (late times) every $0.5$ efold from $\mathcal{N}_{\rm start}$ to the end of inflation. Vertical dashed lines correspond to the Hubble radius scale for each time respectively.
}
\label{fig:homoSpectra}
 \end{figure*}

The iterative technique \cite{Cheng:2015oqa,Notari:2016npn,DallAgata:2019yrr,Domcke:2020zez} and GEF~\cite{Sobol:2019xls,Gorbar:2021rlt} approaches to deal with homogeneous backreaction, share a common feature: the backreaction of the gauge field is modelled by considering expectation values of the source term,  $\langle \vec{E}\cdot\vec{B} \rangle$. We can easily implement this approximation on the lattice, by simply taking volume averages of the source term,
\be
\langle\vec{E}\cdot\vec{B}\rangle_{\rm V} 
\equiv \frac{1}{N^3}\sum_{\mathbf{n}}\sum_i E_i^{(2)}B_i^{(4)}\, .
\ee
We can therefore run the following system of equations
\begingroup
\allowdisplaybreaks
\begin{eqnarray}
\pi'_{\phi} &=& - 3\pi_{\phi} + \frac{1}{H}\Big(\frac{\alpha_\Lambda}{a^3 m_p}\langle\vec{E}\cdot\vec{B}\rangle_{\rm V} - m^2\phi\Big)\,, 
\label{eqn:eom1LatProgramHomBack}\\
E'_{i} &=& - E_{i} - {1\over H}\Big[\frac{1}{a^2}\sum_{j,k}\epsilon_{ijk}\Delta_{j}^{-}B_{k} \\
&& \hspace{1.45cm} +~ \frac{\alpha_{\Lambda}}{2am_{p}}\Big(\pi_{\phi}B_{i}^{(4)} + \pi_{\phi, +\hat{\imath}}B_{i,+\hat{\imath}}^{(4)}\Big) \Big],\nonumber \label{eqn:eom2LatProgramHomBack}\\
    - HH' &=& \frac{1}{m^2_{p}}\left(\rho^{L}_{\rm K}+{2\over3}\rho^{L}_{\rm EM}\right)\, ,\label{eqn:eom3LatProgramHomBack}
\end{eqnarray}
\endgroup
and check the veracity of the constraints
\begingroup
\allowdisplaybreaks
\begin{eqnarray}
\sum_{i}\Delta_{i}^{-}E_{i} &=& 0 \; , \label{eq:GaussLatProgramHomBack}\vspace*{-10mm}\\
3H^2 &=& \frac{1}{m^2_{p}}(\rho^{\rm L}_{\rm K}+\rho^{\rm L}_{\rm V} +\rho^{\rm L}_{\rm EM})\; . \label{eq:HubbleLatProgramHomBack}
\end{eqnarray}
\endgroup

In {\tt Paper~I}~\cite{Figueroa:2023oxc}, we demonstrated that the evolution of the parameter $\xi$ overlaps very well with the results from GEF~\cite{Gorbar:2021rlt}. Here we extend that comparison, also to the energy densities and the slow-roll parameter $\epsilon_{H}$. The upper panels of Fig~\ref{fig:homoComparisonPanel} shows the evolution of the different normalised energy density components: $\rho_{\rm K}$ (red), $\rho_{\rm V}$ (black) and $\rho_{\rm EM}$ (purple), for $\alpha_{\Lambda}=15$ and  $18$. 
The central panels include the evolution of $\epsilon_{H}=-\dot{H}/H^2$, which we identify as the most sensitive global variable, and we indicate the end of inflation with a vertical line at $\epsilon_{H}=1$.

As widely discussed in the literature, the backreaction of the excited gauge field onto the inflationary dynamics delays the end of inflation by several e-foldings. Our energy densities and $\epsilon_{H}$ also show the typical oscillatory features caused by resonant phenomena expected during homogenous backreaction, as described in~\cite{Domcke:2020zez}. This effect was already commented in {\tt Paper~I} for the instability parameter $\xi$. Here, we just note that it is clearly reproduced by our lattice technique for capturing the homogeneous backreaction regime on the lattice. As a quantitative measure of this, we compute the relative difference $\Delta(\epsilon_{H})$ between GEF and our simulations,
\begin{equation}
\Delta(\epsilon_H)= \frac{|\epsilon_H^{\rm GEF}-\epsilon_H^{\rm{L}}|}{\epsilon_H^{\rm{L}}}\, .
\label{eq:reldiff_epsilon}
\end{equation}
The lower panel of Fig.~\ref{fig:homoComparisonPanel} shows the relative difference for $\alpha_{\Lambda}=15$ and $18$. As it can be seen the differences between both approaches lie below $1\%$ by the end of inflation, which is of the order of the variation between different realisations among simulations. Thus, we can safely say that both techniques provide equivalent descriptions of the evolution with homogeneous backreaction.

Another interesting feature, which has been ignored in the previous literature, is the gauge field power spectrum in the homogenous backreaction case. Fig.~\ref{fig:homoSpectra} shows the power spectrum of both helicity modes $A^+$ (upper panel) and $A^-$ (lower panel) till the end of inflation (according to homogeneous backreaction). As a major feature, we observe that only one mode grows above the vacuum solution. Therefore, the gauge field excitation remains fully chiral in the regime of homogeneous backreaction. Also, we note that the spectra display oscillatory features for sub-Hubble modes. While in the linear regime and in the local non-linear regime, the maximum of the excitation follows roughly the Hubble scale, in the homogeneous backreaction regime the excitation freezes at a fixed comoving scale at the onset of non-linearities, and power spectrum peak remains super-Hubble at the end of inflation. We indicate the evolution of the comoving Hubble scale with dashed vertical lines, with the most right one representing the Hubble scale at the end of inflation.

\section{Lattice Simulations, Part II.\\ Local Backreaction}
\label{sec:localBR}

A precise description of the dynamics can only be obtained if we solve the set of lattice Eqs.~(\ref{eqn:eom1LatProgram}), (\ref{eqn:eom2LatProgram}) and
(\ref{eqn:eom3LatProgram}), which reproduce the continuum Eqs.~(\ref{eqn:eom1}), (\ref{eqn:eom2}) and (\ref{eqn:eom3}), accounting for a fully inhomogeneous description of both the inflaton and the gauge field. In this section we show the results of our campaign of simulations, all of which preserve the Gauss constraint~(\ref{eq:GaussLatProgram}) initially with machine precision, up to an accuracy better than $\mathcal{O}(10^{-6})$ during the evolution, and the Hubble constraint~(\ref{eq:HubbleLatProgram}) up to $\mathcal{O}(10^{-4})$. The corresponding relevant parameters used in our runs are indicated in Table~\ref{tab:AllSims}.
The coupling range considered, listed in the first column of the table, spans from `fairly low' values ($\alpha_{\Lambda}\lesssim12$), for which it has been shown that preheating is not effective~\cite{Adshead:2015pva,Cuissa:2018oiw}, to `strong' values ($\alpha_{\Lambda} \gtrsim 15$), for which we observe the strong backreaction regime emerging~\cite{Figueroa:2023oxc}. The main objectives of our simulations are twofold: first, to study the backreaction effect across a broad range of couplings, aiming to determine its characteristics and to identify the parameter space in which the {\it strong} backreaction regime occurs, as defined in {\tt Paper I}. Second, to perform a systematic analysis of the dynamics within the strong backreaction regime, with a detailed examination of the effects due to inflaton inhomogeneities. 

\begin{table}[t!]
\centering
{\renewcommand{\arraystretch}{1.5}%
\begin{tabular}{|c|c|c|c|c|c|c|c|c|}
\hline
$\alpha_{\Lambda}$ & $N\ (\geq)$ & $k_{\rm{IR}}/m$ & $k_{\rm{UV}}/m\ (\geq)$ &  $k_{\rm{BD}}/m$ &  $\mathcal{N}_{\rm{start}}$ & $\mathcal{N}_{\rm{switch}}$ \\
\hline \hline
10 & 320 & 0.1932 & 53.54& 30.91 & -4.5 & -1.1\\
\hline
11 & 320 & 0.1932 & 53.54& 30.91 & -4.5 & -1.1\\
\hline
12 & 320 & 0.1932 & 53.54& 30.91 & -4.5 & -1.1 \\
\hline
13 & 320 & 0.1932 & 53.54 & 30.91 & -4.5 & -1.1\\
\hline
14 & 480 & 0.1932 & 80.31 & 46.36 & -4.5 & -1.1\\
\hline
15 & 640 & 0.1932 & 107.08 & 46.36 & -4.5& -1.1\\
\hline
16 & 1152 & 0.1932 & 192.75 & 30. & -4.5 & -1.1 \\
\hline
17 & 2048 & 0.1932 & 342.66 & 20. & -4.5 & -1.4 \\
\hline
18 & 3072& 0.1932 & 514.00 & 10. & -4.5& -1.7 \\
\hline
19 & 3072 & 0.1544 & 410.81 & 10. & -4.75& -2.0\\
\hline
20 & 3072 & 0.1233 & 327.95 & 9. & -5 & -2.4\\
\hline
\end{tabular}
\caption{Summary of the relevant parameters  of our simulations, with $\alpha_{\Lambda}$ the dimensionless coupling constant, $N$ the number of lattice sites per dimension, $k_{\rm{IR}}$ the infrared momentum,  $k_{\rm{UV}}$ the ultraviolet momentum, $k_{\rm{BD}}$ the intermediate cutoff scale, $\mathcal{N}_{\rm{start}}$ the efold at which we start the simulation in the linear regime and $\mathcal{N}_{\rm{switch}}$ the efold in which we evolve the full equations of motion. For each coupling, only the cases with the largest separation of scales (or $N$) are included.}
    \label{tab:AllSims}
    }
\end{table}

\begin{figure*}[t]
\centering
\subfloat{\includegraphics[width=0.49\textwidth]{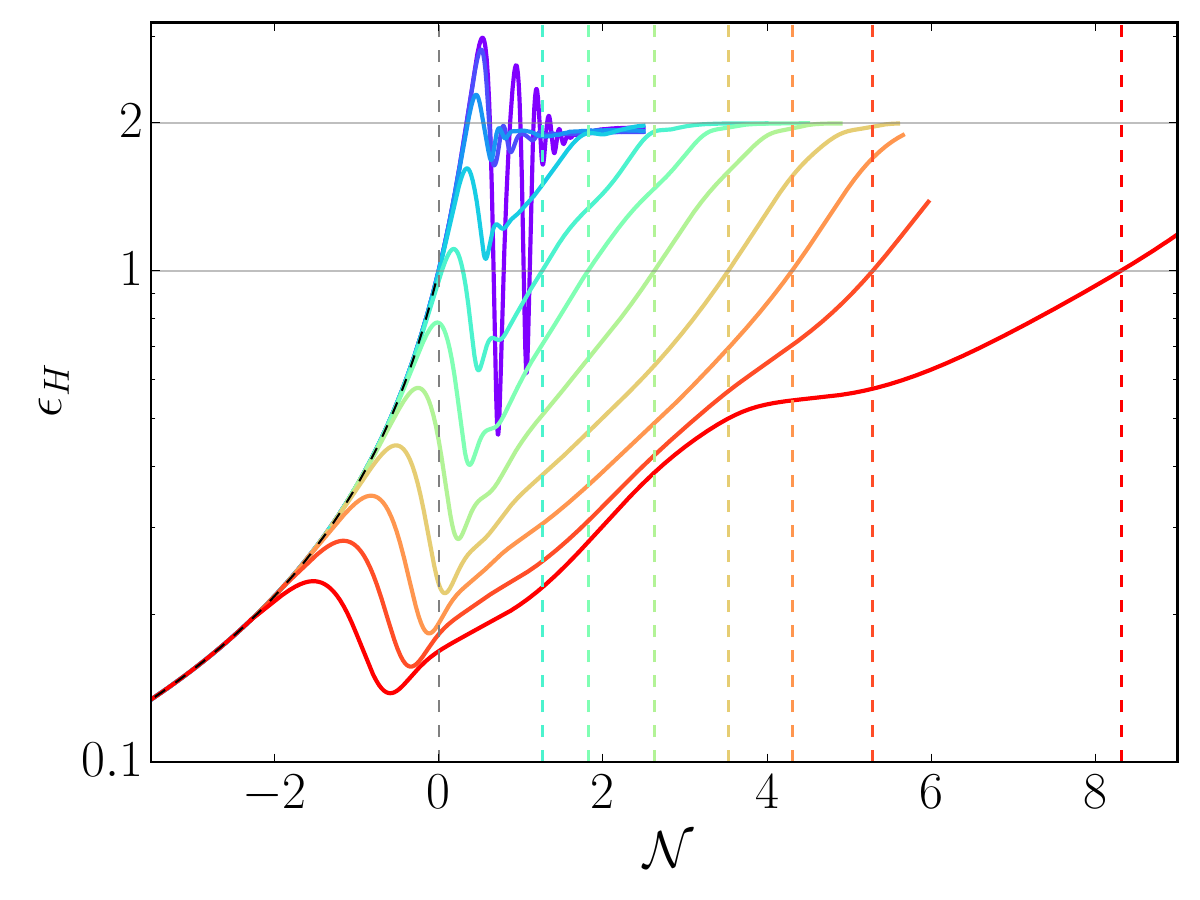}} \quad
\subfloat{\includegraphics[width=0.49\textwidth]{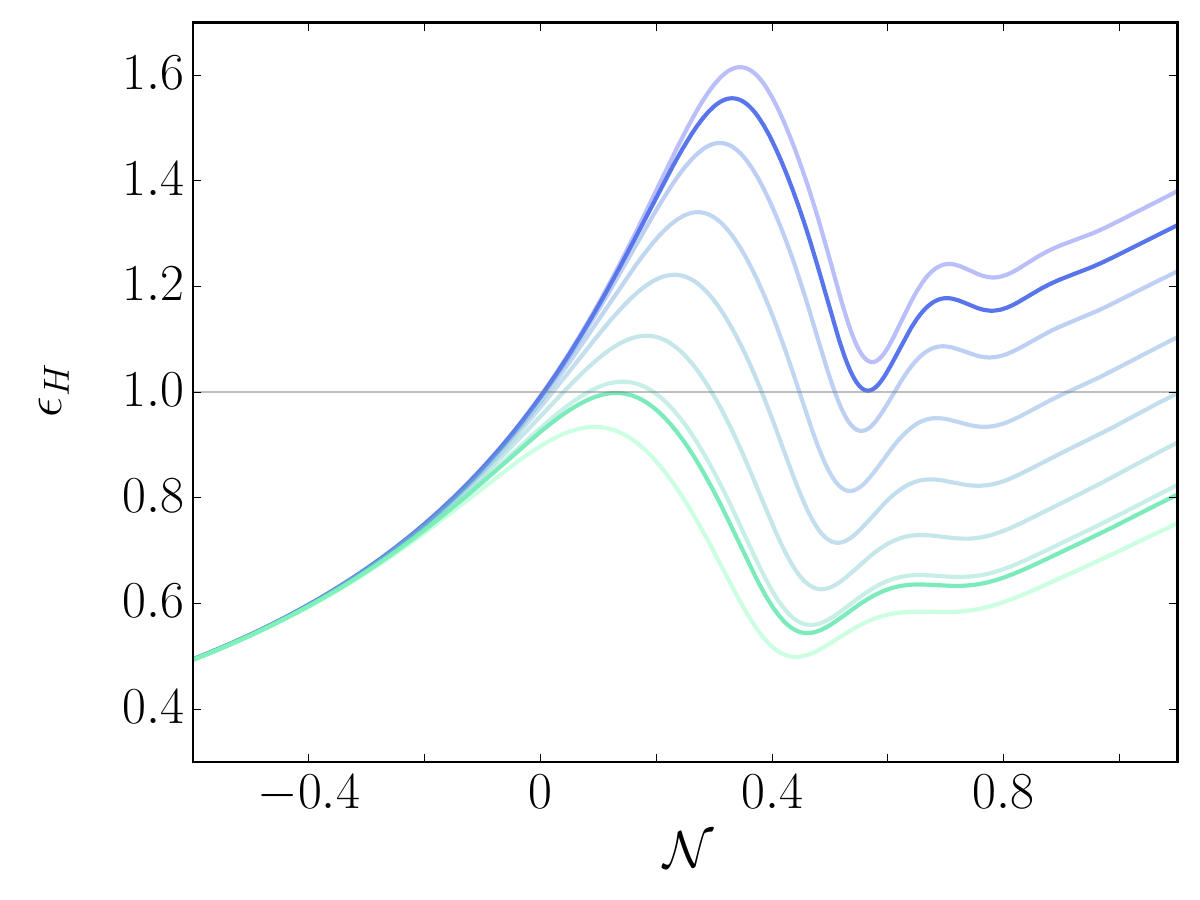}}
\caption{\textit{Left}: The evolution of $\epsilon_H=-\dot{H}/H^2$ for the simulations included in Table~\ref{tab:AllSims}, going from $\alpha_{\Lambda}=10$ (purple) to $\alpha_{\Lambda}=20$ (red) with steps of $\Delta\alpha_{\Lambda}=1$. The black dashed trajectory corresponds to the linear prediction. Coloured dashed lines correspond to the end of inflation for each case, while thin grey  solid horizontal lines indicate $\epsilon_H=1$ (end of inflation) and $\epsilon_H=2$ (radiation domination). \textit{Right}: Equivalent to left panel for couplings between $\alpha_\Lambda= 13$ and $14.5$ with a step of $\Delta\alpha_\Lambda=0.25$ represented with thin lines. Thick lines are for $\alpha_\Lambda=13.1$ (blue) and $\alpha_\Lambda=14.31$ (green), which correspond to the lower and upper ends of the mild backreaction regime, respectively.}
\label{fig:epsilonVScouplings}
\end{figure*} 

In order to illustrate the differences in the evolution during the backreaction regime, we plot the slow roll parameter $\epsilon_H$ for all couplings considered in the left panel of Fig.~\ref{fig:epsilonVScouplings}. The colour gradient goes from coldest to hottest as we increase the coupling, with purple corresponding to $\alpha_{\Lambda}=10$ and red to $\alpha_{\Lambda}=20$. The vertical dashed lines indicate the moment when the inflationary period ends for each coupling, \ie $\epsilon_H = 1$, using the same colour code. The figure shows the onset of the backreaction regime as compared to the linear prediction (black dashed line), as we vary the strength of the coupling. The deviation from the linear trajectory shifts to earlier times as we increase $\alpha_{\Lambda}$. Based on this, one can differentiate three different regimes, depending on the coupling strength:\vspace{0.26cm}

$\bullet$ {\it Weak backreaction regime} ($\alpha_{\Lambda}\lesssim 13.1$): The contribution of the gauge field to the inflationary dynamics is subdominant during inflation, with the system dynamics behaving much like in standard slow-roll regime. The deviation from the backreaction-less trajectory happens during reheating after inflation, which still ends at $\mathcal{N} = 0$. During the inflaton oscillations after inflation, there is some level of backreaction of the gauge field depending on the coupling. For lower couplings, say $\alpha_{\Lambda} \lesssim 8$, this post-inflationary backreaction becomes negligible \cite{Cuissa:2018oiw}. \vspace{0.1cm}

$\bullet$ {\it Mild backreaction regime} ($13.1 \lesssim \alpha_{\Lambda}\lesssim 14.31$): The effect of the backreaction on the expansion rate occurs monotonically closer to the end of slow-roll inflation the larger the coupling is, with $\epsilon_H$ exhibiting a bump at some moment $\mathcal{N} \sim 0.1-0.3$ after inflation, depending on the coupling. The end of inflation,  $\epsilon_H=1$, is still reached at around $\mathcal{N}\sim 0$, but there is a phase afterwards where the system re-enters back again into an inflationary regime. This re-entering coincides with similar behaviours observed in previous literature~\cite{Adshead:2019igv,Adshead:2019lbr}.\vspace{0.1cm}

$\bullet$ {\it Strong backreaction regime} ($\alpha_{\Lambda}\gtrsim 14.31$): More drastic changes are observed, as the backreaction occurs still inside the slow-roll regime, delaying the end of inflation by a number of e-folds $\Delta\mathcal{N}_{\rm br}$, which grows with the coupling. The vertical dashed lines shown in the left panel of Fig.~\ref{fig:epsilonVScouplings} indicate the location of the end of inflation for the couplings of this regime.\vspace{0.1cm}

Looking at the right panel of Fig.~\ref{fig:epsilonVScouplings} helps to identify the meaning of the coupling range that defines the mild backreaction regime. Namely, we identify the lower end value $\alpha_\Lambda \simeq 13.1$ (blue thick line) as the maximum coupling for which there is no further extension of inflation, so that the local minimum of the trajectory of $\epsilon_H$ after $\mathcal{N} = 0$ never becomes smaller than unity again. We identify the upper end value $\alpha_\Lambda \simeq 14.31$ (green thick line) as the minimum coupling for which the local maximum in the trajectory of $\epsilon_H$ does not go above unity close to $\mathcal{N} = 0$ (eventually it will become bigger than unity at the end of inflation in the new regime driven by the backreaction). 

\begin{figure*}[t]
\centering
\subfloat{\includegraphics[width=0.49\textwidth]{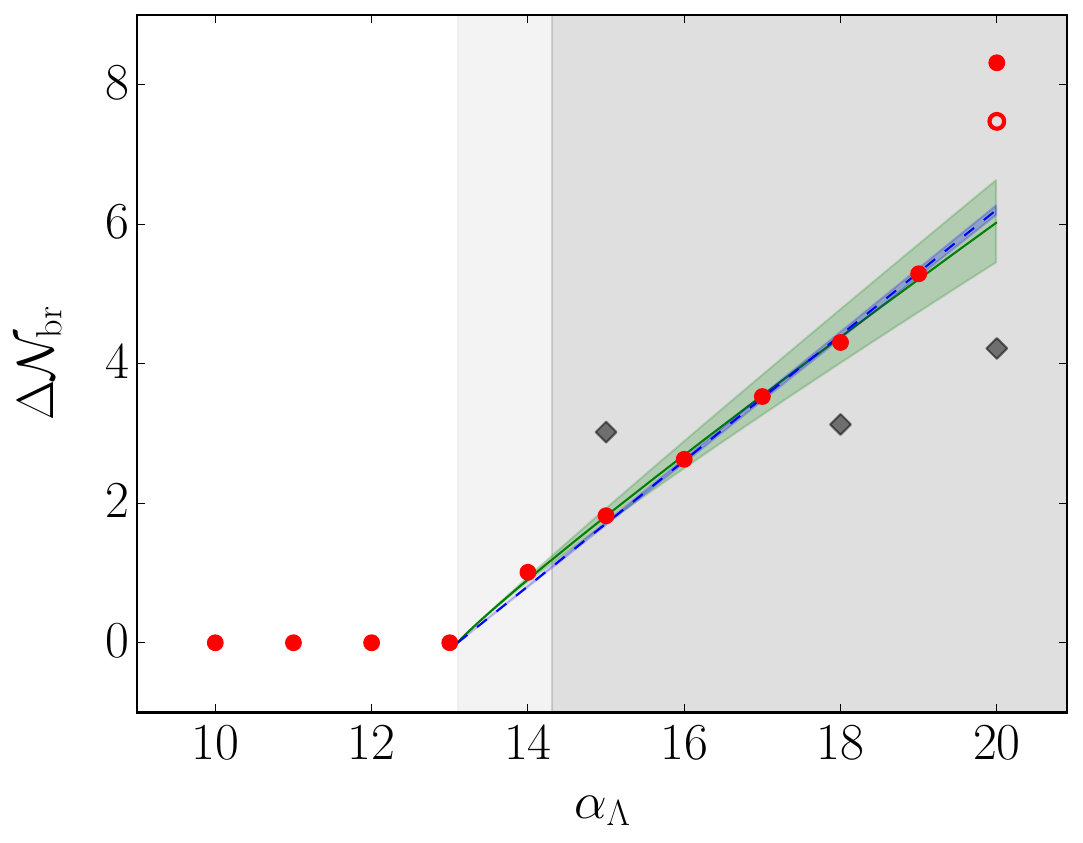}} \quad
\subfloat{\includegraphics[width=0.49\textwidth]{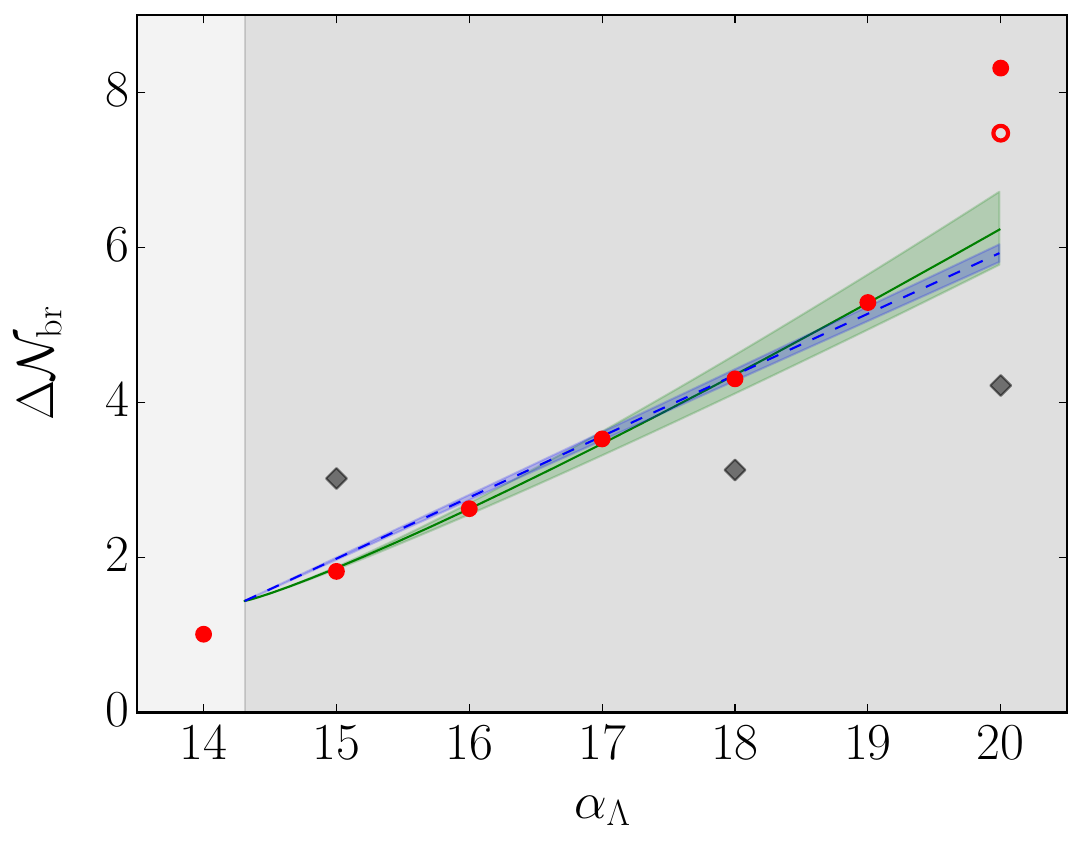}}
\caption{The number of extra e-folds in inflation with respect the linear regime $\Delta \mathcal{N}_{\rm br}$ versus $\alpha_{\Lambda}$. Red dots correspond to simulations with local backreaction, while grey diamonds with homogeneous backreaction. The empty circle corresponds to the simulation for $\alpha_\Lambda=20$ with the largest dynamical range that failed to reach $\epsilon_{H}=1$, see Sec.~\ref{subsec:powUV}. The dashed and solid lines correspond to the fits provided using a linear relation, Eqs.~(\ref{eq:fitDeltaN}), (\ref{eq:fitDeltaN_SBR}), and a power-law relation, Eqs.~(\ref{eq:fitDeltaN_PL}), (\ref{eq:fitDeltaN_PL_SBR}), fitting for $\alpha_\Lambda>13$ (left) and $\alpha_\Lambda>14.31$  (right) respectively. The coloured shaded bands represent the error from the fitting procedure.}
\label{fig:inflationends}
\end{figure*}

In the strong coupling regime, inflation is extended by a number of e-foldings $\Delta\mathcal{N}_{\rm br} \gtrsim 1$, beyond the end of slow-roll inflation expected at $\mathcal{N} = 0$. The dependence of $\Delta\mathcal{N}_{\rm br}$ with the coupling $\alpha_\Lambda$ can be observed in Fig.~\ref{fig:inflationends}. As shown in Fig.~\ref{fig:epsilonVScouplings}, the inflationary period does not extend beyond $\mathcal{N}=0$ for values $\alpha_{\Lambda} \leq 13.1$, so we fix $\Delta\mathcal{N}_{\rm{br}}=0$ for those couplings. For $\alpha_{\Lambda} > 13.1$, the value of $\Delta\mathcal{N}_{\rm{br}}$ increases monotonically with the coupling value. This indicates that the larger the axion-gauge coupling is, the stronger and earlier the backreaction of the gauge field takes place, and hence the larger the extension of inflation becomes. This behaviour differs notably from what happens in the homogeneous backreaction approach, also included in the figure with grey diamonds, where the lengthening of inflation is also observed, but the extra e-foldings are always of the order $\Delta\mathcal{N}_{\rm{br}}\sim 3-4$ for the couplings we consider.

For the mild and strong backreaction regimes, the growth of $\Delta\mathcal{N}_{\rm{br}}$ is roughly linear with the coupling, at least up to $\alpha_\Lambda=19$. We propose two possible fits:
\begin{eqnarray}
{\rm Linear:} && \Delta\mathcal{N}_{\rm{br}}=m_{1}(\alpha_{\Lambda}-13.1)\; ,
\label{eq:fitDeltaN}\\
{\rm Power-law:} && \Delta\mathcal{N}_{\rm{br}}=b_1(\alpha_{\Lambda}-13.1)^{a_1},
\label{eq:fitDeltaN_PL}
\end{eqnarray}
where we fix  $(\alpha_{\Lambda},\Delta\mathcal{N}_{\rm{br}}) = (13.1,0)$ as the lower end point for both, separating them from the weak backreaction regime for which $\Delta\mathcal{N}_{\rm br} = 0$. We obtain $m_{1}=0.90 \pm 0.01$, and $a_{1}=0.930 \pm 0.03$ and $b_{1}=1.00 \pm 0.04$, respectively, so that the power-law fit supports approximately the linear hypothesis, though preferring some mild curvature. These fits and their errors are included in the left panel of Fig.~\ref{fig:inflationends}, with the linear fit in blue and the power-law fit in green. We note that $\alpha_\Lambda=20$, which is completely off the trend from either linear or power-law behaviours, has been omitted from the fits on purpose. This is because, as we will explain later Sec.~\ref{subsec:powUV}, the dynamics for this coupling has not reached yet a certain quality criterium of convergence that we demand for every simulation. We also include, as a red empty circle, the value obtained by extrapolating the dynamics of the simulation for $\alpha_{\Lambda} = 20$, with the largest separation of scales that failed before achieving $\epsilon_H=1$. We will elaborate on this issue in Section~\ref{subsec:powUV}. If the value of $\Delta\mathcal{N}_{\rm br}$ for $\alpha_\Lambda = 20$ was to be obtained from either of the fits, we would expect it to lie somewhere within the range $\Delta\mathcal{N}_{\rm br} \approx 5.5-6.6$, whereas out best simulation of this large coupling gives $\Delta\mathcal{N}_{\rm br} \simeq 8.1$ (or $7.5$ for the extrapolated case).

Alternatively, as the strong backreaction is truly a distinctive regime, different from the mild backreaction, we also fit $\Delta\mathcal{N}_{\rm{br}}$ vs $\alpha_\Lambda$ using only the coupling range supporting the strong backreaction. Starting from the point 
$(\alpha_{\Lambda},\Delta\mathcal{N}_{\rm br}) =(14.31,1.44)$, which corresponds to the lower end of this regime, we impose
\begin{eqnarray}
\label{eq:fitDeltaN_SBR}
{\rm Linear}: && \hspace{-2mm}\Delta\mathcal{N}_{\rm{br}}=m_{2}(\alpha_{\Lambda}-14.31)+1.44\,,\\
\hspace{-5mm}{\rm Power-law}: && \hspace{-2mm}\Delta\mathcal{N}_{\rm{br}}=b_2(\alpha_{\Lambda}-14.31)^{a_2}+1.44\,,
\label{eq:fitDeltaN_PL_SBR}
\end{eqnarray}
and obtain $m_{2}=0.79 \pm 0.02$, and $a_{2}=1.15 \pm 0.03$ and $b_{2}=0.65 \pm 0.03$ (we note that we have excluded again the value for $\alpha_\Lambda = 20$). These fits and their errors are shown in the right panel of Fig.~\ref{fig:inflationends}, using the same colour code as in the left panel.

\begin{table}[t]
    \centering
    {\renewcommand{\arraystretch}{1.5}
    \begin{tabular}{|c|c|c|c|c|}
        \hline
        \multirow{ 2}{*}{$\alpha_{\Lambda}$} & \multicolumn{4}{c|}{$\Delta\mathcal{N}_{\rm{br}}$} \\
        \cline{2-5}
        & linear (\ref{eq:fitDeltaN}) & power-law (\ref{eq:fitDeltaN_PL}) & linear (\ref{eq:fitDeltaN_SBR}) & power-law (\ref{eq:fitDeltaN_PL_SBR}) \\
        \hline \hline
        20 & 6.21$\pm$0.07 & 6.03$_{-0.57}^{+0.61}$ & 5.9$\pm$0.1 & 5.85$_{-0.45}^{+0.45}$ \\
        22.5 & 8.46$\pm$0.09 & 8.04$_{-0.83}^{+0.90}$ & 7.9$\pm$0.2 & 8.88$_{-0.77}^{+0.81}$ \\
        25 & 10.7$\pm$0.1 & 10.0$_{-1.08}^{+1.21}$ & 9.9$\pm$0.2 & 12.06$_{-1.12}^{+1.20}$ \\
        30 & 15.2$\pm$0.2 & 13.9$_{-1.67}^{+1.80}$ & 13.8$\pm$0.3 & 18.75$_{-1.94}^{+2.11}$ \\
        35 & 19.7$\pm$0.2 & 17.6$_{-2.16}^{+2.53}$ & 17.8$\pm$0.4 & 25.76$_{-2.84}^{+3.15}$ \\
        \hline
    \end{tabular}
    \caption{Estimated amount of extra e-folds in inflation for selected couplings, obtained by extrapolating the fits of Eq.~(\ref{eq:fitDeltaN})-(\ref{eq:fitDeltaN_PL_SBR}).}
    \label{tab:extrap}
    }
\end{table}

 \begin{figure*}[t]
\includegraphics[width=\textwidth]{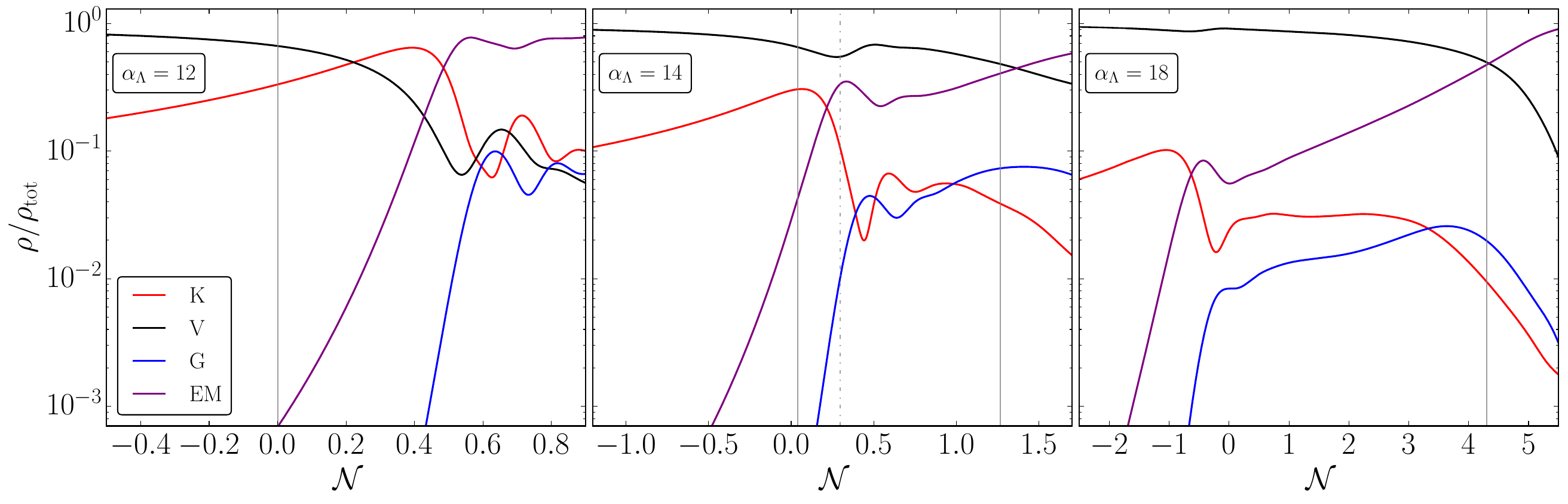}
\caption{The energy components normalised to the total energy density, for the couplings  $\alpha_{\Lambda}=12,\ 14$ and $18$, representative, respectively, of the weak, mild and strong backreaction regimes. Colours represent different components: black for potential, red for kinetic, blue for gradients, and purple for electromagnetic. The solid vertical lines corresponds to the end of inflation in each case. The dash-dotted line for $\alpha_{\Lambda}=14$ corresponds to the re-entering into the inflationary period.}
\label{fig:rhodensityRegimes}
 \end{figure*}

Thanks to the fits we can estimate the value of $\Delta\mathcal{N}_{\rm br}$ for larger couplings, extrapolating its growth with the proposed fits. We list such values with the corresponding errors, for $\alpha_\Lambda = 20-35$, in Table~\ref{tab:extrap}. These estimations must be taken, of course, with a grain of salt, as only a dedicated lattice study for such large couplings could give the correct number. Based on our current data, we suspect that the linear growth may slow down, meaning that our extrapolations in Table~\ref{tab:extrap} could be viewed as upper bounds on the inflation extension for the given couplings. Investigating this in detail requires however larger computational resources than our present capabilities. 

The separation between different regimes and, in particular, the return to inflation during mild-backreaction after $\mathcal{N} = 0$, can be qualitatively understood by analysing the inflationary parameter $\epsilon_H$ in terms of energy density components,
\begin{equation}
    \epsilon_H = -\frac{\dot{H}}{H^2} = 1+\frac{2\rho_{\rm{K}}-\rho_{\rm{V}}+\rho_{\rm{EM}}}{\rho_{\rm{tot}}}\;.
    \label{eqn:epsilonInEnergies}
\end{equation}
In Fig.~\ref{fig:rhodensityRegimes} we show the evolution of different energy contributions for couplings representative of each regime,  $\alpha_{\Lambda}=12,\, 14$ and $18$, which correspond (from left to right in the figure), to the weak, mild and strong coupling regimes, respectively. More specifically, we plot the evolution of different energy densities with respect the total one, where $\rho_{\rm{K}}/\rho_{\rm tot}$ is depicted in red, $\rho_{\rm{V}}/\rho_{\rm tot}$ in black, $\rho_{\rm{G}}/\rho_{\rm tot}$ in blue and $\rho_{\rm{EM}}/\rho_{\rm tot}$ in purple. The vertical grey solid lines in each panel indicate the point where $\epsilon_{H}=1$, signalling the end of inflation for each coupling.

\begin{figure*}[t]
\centering
\subfloat{\includegraphics[width=0.5\textwidth]{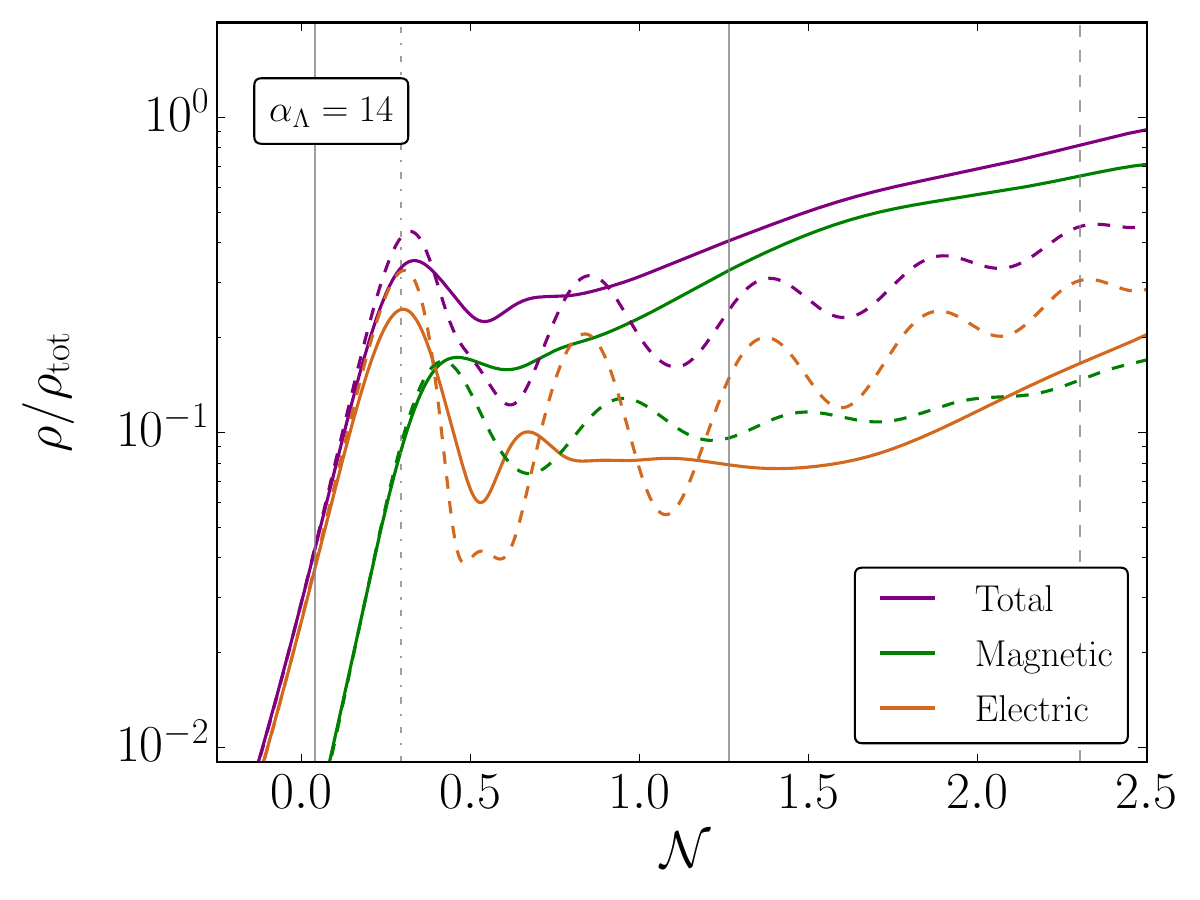}} 
\subfloat{\includegraphics[width=0.5\textwidth]{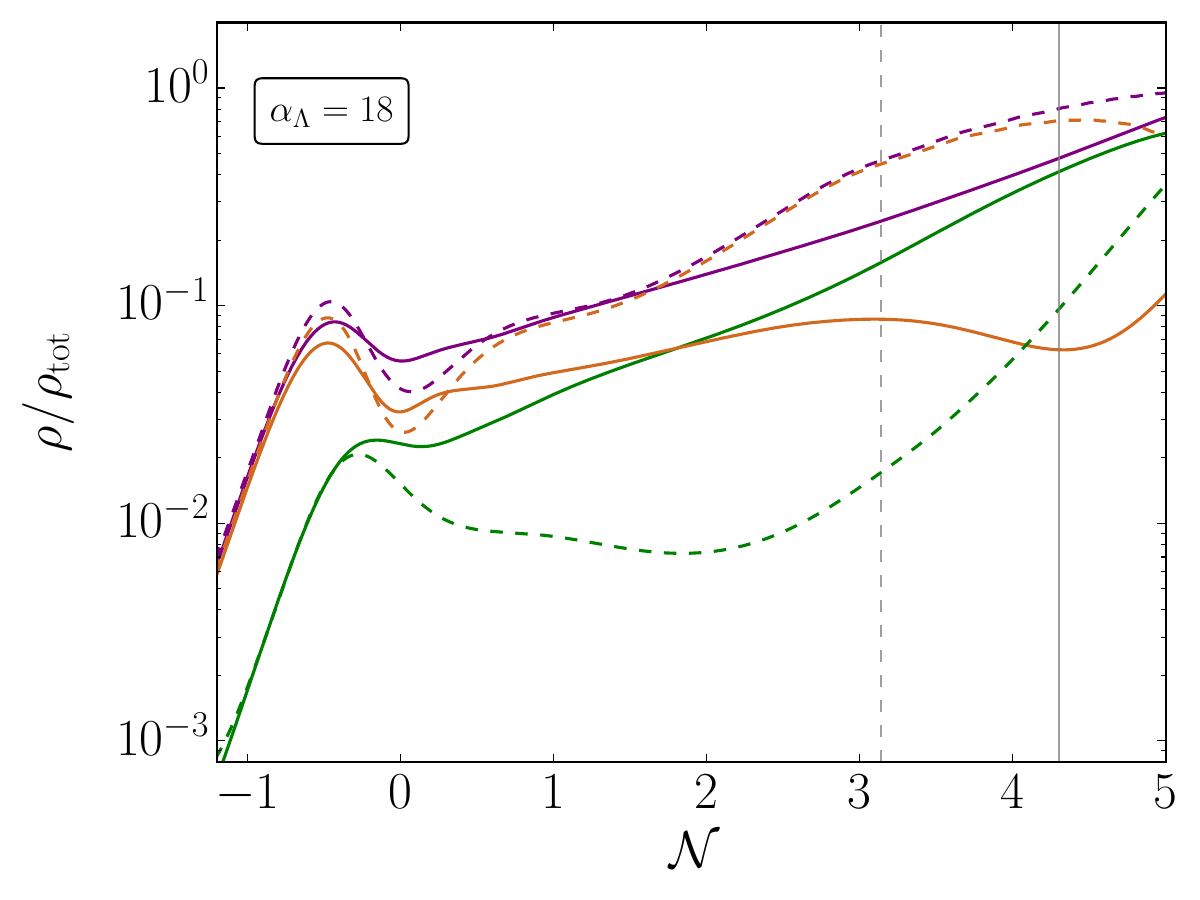}}
\caption{Comparison of the normalised electric energy density $\rho_{\rm{E}}/\rho_{\rm{tot}}$ (orange) and normalised magnetic energy density $\rho_{\rm{B}}/\rho_{\rm{tot}}$ (green) for $\alpha_{\Lambda}=14$ (left) and $\alpha_{\Lambda}=18$ (right). The total electromagnetic fraction is included in purple. The solid lines correspond to the full local backreaction simulations and dashed lines to the homogeneous approach. Vertical lines correspond to the end of inflation for each case, solid for local backreaction and dashed for homogeneous; the dash-dotted line represents the re-entering into inflation for $\alpha_\Lambda=14$.}
\label{fig:homovsfullEandB}
\end{figure*}

In the left panel of Fig.~\ref{fig:rhodensityRegimes}, we see that in the weak coupling regime the effect of the electromagnetic energy density is almost negligible during inflation; in the example, it only reaches a value $\sim 2$ orders of magnitude smaller than the kinetic at $\mathcal{N}=0$. In fact, if the electromagnetic contribution is neglected in Eq.~(\ref{eqn:epsilonInEnergies}), we observe that $2\rho_{\rm{K}}=\rho_{\rm{V}}$ corresponds to the end of inflation, which coincides with the observed behaviour in the figure and with the end of standard slow-roll. In this regime the inflaton's dynamics remain barely affected by the growth of the gauge field and follows the backreactionless trajectory during inflation. It is in the post-inflationary period, for $\mathcal{N}>0$, where the weight of the gauge field increases considerably and backreacts on the inflaton and background dynamics. Similarly, we observe that in the weak coupling regime, the inflaton gradients are not relevant during inflation, but as with the electromagnetic part, they become relevant afterwards. In fact, we see how both growths are completely correlated.

In the middle panel of Fig.~\ref{fig:rhodensityRegimes}, we observe that in the mild coupling regime the electromagnetic energy density weights in earlier in the dynamics than in the weak coupling regime. As indicated in Fig.~\ref{fig:epsilonVScouplings}, this regime exhibits an interesting feature where $\epsilon_{H}=1$ is reached at $\mathcal{N}\approx0$, but due to backreaction effects, there is afterwards another additional inflationary period that lasts approximately $\sim 1$ efold. Contrary to the weak regime, where inflation ends solely as a consequence of the growth of the axion kinetic energy, in this regime $\rho_{\rm{EM}}$ cannot longer be neglected and $\epsilon_H=1$ is obtained when $\rho_{\rm{V}}=2\rho_{\rm{K}}+\rho_{\rm{EM}}$ is satisfied. Subsequently, $\rho_{\rm{EM}}$ surpasses $\rho_{\rm{K}}$, which decreases considerably and becomes comparable to $\rho_{\rm{G}}$, both contributing around $5\%$ percent to the total energy density around $\mathcal{N}=0.5$. From then on the second most dominant contribution becomes $\rho_{\rm{EM}}$ (after $\rho_{V}$), and hence $\epsilon_{H}<1$ (dashed-dotted vertical line), re-inflating the universe once again for a short while, between $\mathcal{N}=0.3$ and $\mathcal{N}=1.3$. Inflation ends finally when $\rho_{\rm{EM}}\approx\rho_{\rm{V}}$, as indicated by the second vertical solid line (they are not exactly equal because the inflaton's kinetic and gradient energies still weight approximately $10\, \%$).

Finally, in the right panel of Fig.~\ref{fig:rhodensityRegimes}, we observe that in the strong backreaction, $\rho_{\rm{EM}}$ becomes large earlier during inflation, and in fact, it becomes comparable to the inflaton's kinetic energy, and even surpasses it, before $\mathcal{N}=0$. Similarly, the relative contribution of $\rho_{\rm{G}}$ grows gradually until it reaches even $\sim 50\%$ of $\rho_{\rm{K}}$ around $\mathcal{N} = 0$. Afterwards, the electromagnetic contribution gradually dominates more and more over the kinetic part, which settles down to a constant contribution of $\sim 2\%$ of the total energy budget, till almost the end of inflation. We coin this period (in the figure from $\mathcal{N} = 0$ to $ \mathcal{N}=\Delta\mathcal{N}_{\rm br} \simeq 4.1$) as the \textit{electromagetic slow-roll} regime; during this, the system is still dominated by the inflaton potential, but steadily transferring energy to the electromagnetic sector, actually in an exponential manner, though with a smaller rate that during the tachyonic growth in the linear regime. The end of inflation, $\epsilon_H=1$, is obtained when $\rho_{\rm{EM}}$ becomes comparable to $\rho_{\rm{V}}$, finishing inflation with the Universe almost reheated, as the photon's energy represents $\sim 50\%$ of the total by then, and becomes the dominant species reaching $\sim 90\%$ if the total energy, in less than an efold after the end of inflation.

As already noted in {\tt Paper I}, the stronger the coupling considered in the strong backreaction regime, the longer the inflationary expansion is prolonged. This feature can be qualitatively understood by looking at Eq.~(\ref{eqn:epsilonInEnergies}), which indicates that inflation ends when $2\rho_{\rm{K}}+\rho_{\rm{EM}} = \rho_{\rm{V}}$. In the context of the strong backreaction, the end of inflation occurs when the electromagnetic contribution dominates over the kinetic energy of the inflaton. Therefore, we can envisage that for larger couplings, as the strong backreaction is set earlier, larger extra inflationary expansion will emerge, as the rate of growth of the gauge field energy density is slower during the electromagnetic slow-roll regime, than during the linear regime.

\subsection{(Electro)Magnetic slow-roll}
\label{subsec:magneticSlowRoll}

\begin{figure*}[t]
\centering
\subfloat{\includegraphics[width=0.45\textwidth]{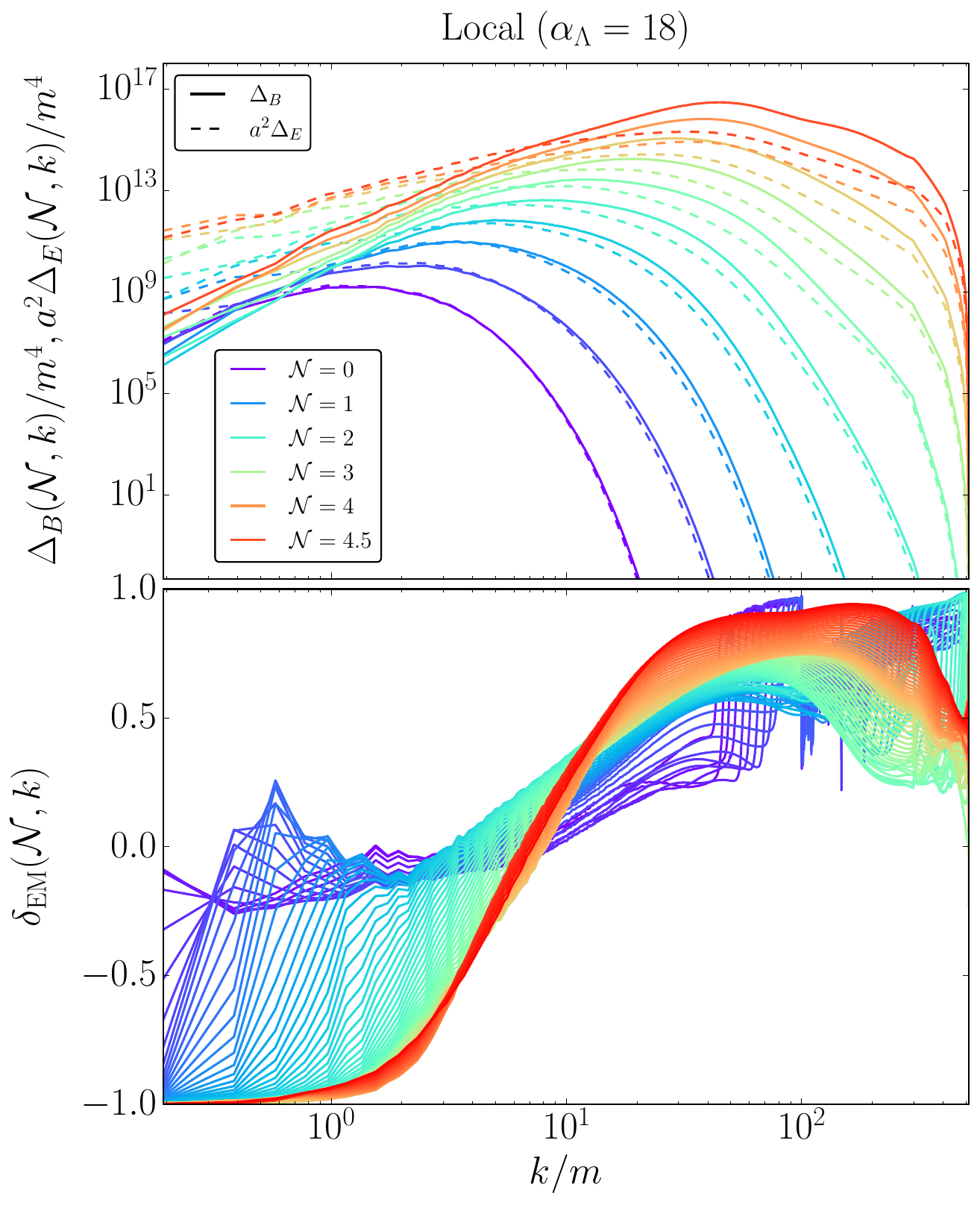}} \quad\quad\quad
\subfloat{\includegraphics[width=0.45\textwidth]{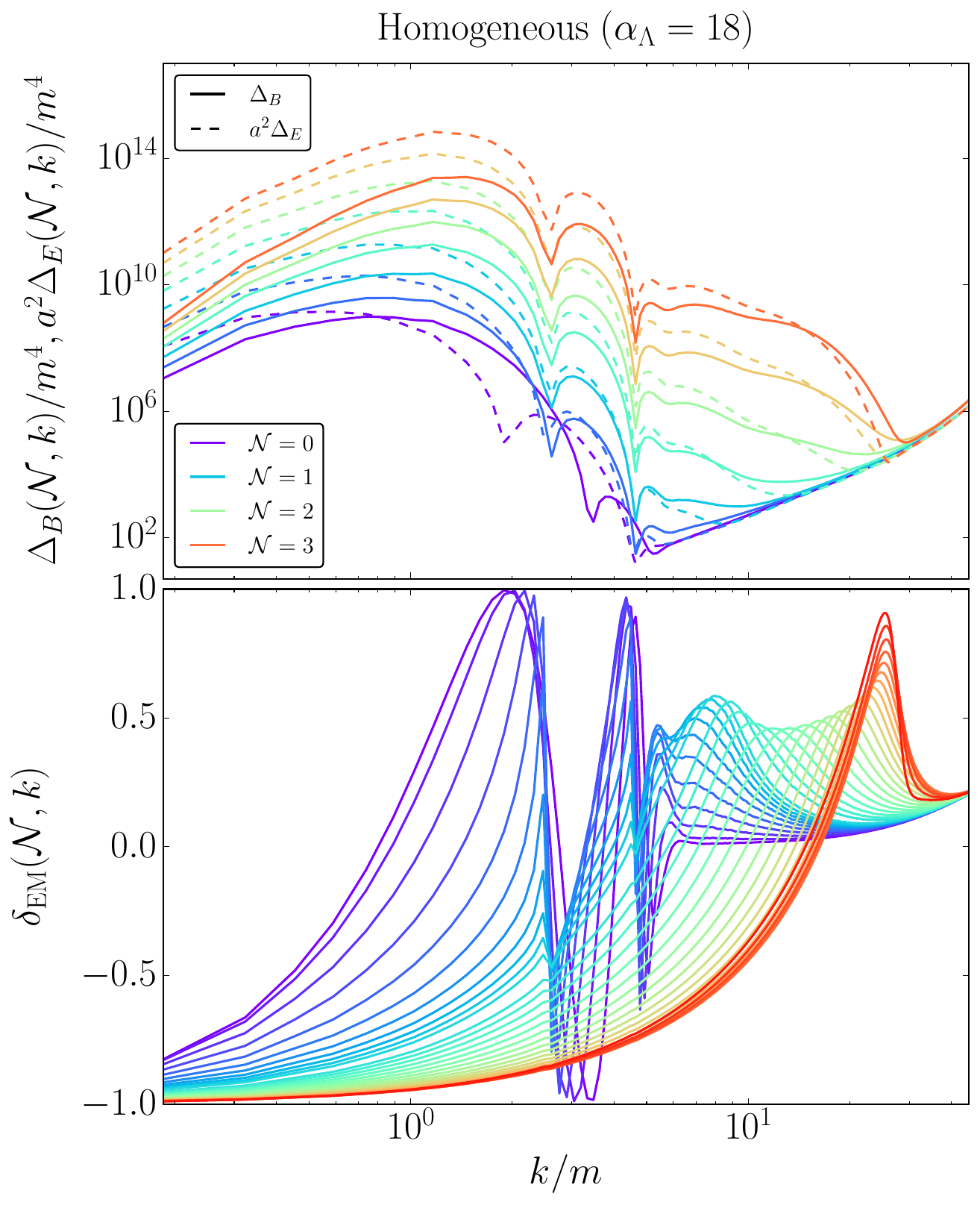}}
\caption{Comparison of the electric and magnetic power spectra during the $\Delta\mathcal{N}_{\rm br}$ extra e-foldings for $\alpha_{\Lambda}=18$. The evolution is given from the bluest to the reddest colours. In the upper panel we show the magnetic spectrum $\Delta_{B}$ in solid lines and the electric spectrum $a^2\Delta_{E}$ with a separation of $0.5$ e-folds between spectra. In the lower panel we present the normalized difference $\delta_{\rm EM}$ for gaps of $0.05$ e-folds.}
\label{fig:spectrumElectricMagneticStrong}
\end{figure*}

It is informative to study the decomposition of the gauge field energy density into the electric and the magnetic contributions. Fig.~\ref{fig:homovsfullEandB} shows this decomposition for $\alpha_{\Lambda}=14$ (left panel) and $\alpha_{\Lambda}=18$ (right panel), where the fractional electric energy density is plotted in orange, the magnetic in green and the total in purple. We also include, using the same colour scheme but in dashed, the same fractions for the homogeneous case. The end of inflation is also included for both cases by vertical lines, solid for the full case and dashed for the homogeneous approximation. 

During the tachyonic instability (linear regime) the photon's energy is dominated by the electric contribution, which is $\sim 10$ times larger than the magnetic counterpart for $\alpha_\Lambda = 18$. However, in the truly physical local regime once the strong backreaction regime becomes evident, the magnetic field acquires greater relevance and it eventually dominates over the electric part during the last extra e-folds of inflation. In fact, it is the magnetic contribution which grows in an exponential manner, whereas the electric contribution stops growing and even drops close to the end of inflation. It is noteworthy that for $\alpha_{\Lambda}=14$, which represents the mild backreaction regime, the same magnetic-vs-electric dominance occurs in a short period of time after re-entering back to inflation (at $\mathcal{N}\gtrsim0.5$).

We conclude that the noticeable lengthening of the inflationary period that we report for the local strong backreaction regime is strictly correlated with the growth (at exponential rate) of the magnetic energy density. This conclusion, however, changes completely if the inhomogeneities of the inflaton are not included in the system dynamics. In the homogeneous picture of backreaction, the electric field contribution (which was already dominating during the linear regime) largely continues dominating over the magnetic part, specially for larger couplings. 

In the upper pannels of Fig.~\ref{fig:spectrumElectricMagneticStrong} 
we show the power spectra of the electric and magnetic fields 
for $\alpha_{\Lambda}=18$, in the local backreaction (left) and in the homogeneous case (right). In particular, we plot the evolution from $\mathcal{N}=0$ to $\Delta \mathcal{N}_{\rm br}=4.5$ (left) and $3$ (right) of the magnetic comoving power spectrum $\Delta_B(\mathcal{N},k)$ 
with solid lines, and of the electric counterpart $a^{2}\Delta_E(\mathcal{N},k)$ with dashed lines, in intervals of $0.5$ e-foldings. In the truly local regime, at the earliest times (coldest colours), the electric modes dominate over magnetic ones in the most IR region of the spectra, while in the middle to  UV regions electric and magnetic modes are roughly of the same power. As the evolution progresses through the strong backreaction, the dominance of the electric modes in the most IR region becomes more pronounced, but at the same time the weight of such IR region becomes more subdominant against the power developed by either spectra at the mid-to-UV scales, as the spectra drifts to the right. During this drift to higher $k$'s, the magnetic power becomes more prominent around the peak scales. By $\mathcal{N}=2$ the peaks of the spectra of the magnetic field and the electric field are of the same order. This point coincides with the crossing of both energies in Fig.~\ref{fig:homovsfullEandB}. At the end of inflation, at $\mathcal{N}=4.5$, the peak of the spectra has shifted more than a decade into smaller scales, and although within the IR range $\rho_{\rm E}/\rho_{\rm B} \sim 10^3$, around the peak it holds that $\rho_{\rm B}/\rho_{\rm E} \sim 10$. As the peak of the spectra is $\sim 10^4$ higher than the IR tail, the contribution of the magnetic power significantly dominates over the electric power. The magnetic dominance during strong backreaction is therefore a consequence of underlying non-trivial scale dependent effects, which explain the results shown for the energy densities in Fig.~\ref{fig:homovsfullEandB}. 

Remarkably, this magnetic dominance around the peak scales does not occur in the homogeneous approximation, see the upper right panel of Fig.~\ref{fig:spectrumElectricMagneticStrong}. We observe that the peaks of the excitations for both components remain fixed around common mid comoving scales $k/m\sim 1-2$, as already noted for the gauge field in Fig.~\ref{fig:homoSpectra}. Additionally, it is the electric comoving contribution the one that dominates over the magnetic counterpart for all excited modes, explaining the evolution shown in Fig.~\ref{fig:homovsfullEandB}.

To quantify more clearly the scale-dependence difference between magnetic and electric power spectra, we define 
\begin{equation}
    \delta_{\rm EM}(\mathcal{N},k) \equiv \frac{\Delta_{B}(\mathcal{N},k)-a^2\Delta_{E}(\mathcal{N},k)}{\Delta_{B}(\mathcal{N},k)+a^2\Delta_{E}(\mathcal{N},k)}\, ,
\end{equation}
so that $\delta_{\rm EM} < 0$ indicates dominance of electric power, $\delta_{\rm EM} > 0$ dominance of magnetic power, and $\delta_{\rm EM} = 0$ indicates equal power. The values $\delta_{\rm EM} = -1$ and $\delta_{\rm EM} = +1$ correspond to maximal electric or magnetic dominance, respectively. We plot this quantity in the lower panels of Fig.~\ref{fig:spectrumElectricMagneticStrong}, in this case in intervals of $\Delta\mathcal{N}=0.05$ e-folds. Initially, in the local backreaction case, $\delta_{EM}$ is negative but close to zero at the dominant IR scales, while $\delta_{EM}$ approaches unity at UV modes, but the weight of this region of the spectrum at those moments is negligible. As the evolution progresses, the most IR modes show an evolution towards $\delta_{EM}=-1$, as the electric field dominates at that region, which however is becoming subdominant in terms of spectral power. As fields' spectra shift to smaller scales, an opposite trend occurs at mid-to-UV scales, as we see an evolution towards $\delta_{EM}=1$. At the end of inflation (reddest lines), the spectral shape of $\delta_{EM}$ resembles to a hyperbolic tangent, with the electric modes clearly dominating at IR region (with negligible weight in the power spectra), while the magnetic modes clearly dominate in the mid-to-UV scales (which sustain by then the dominant peak of the spectra). We note that the region $k/m > 3 \times 10^2$ corresponds to values greater than the Nyquist frequency of the given simulation, so the pattern just described is distorted as $\delta_{\rm EM}$ falls off towards the maximum $k$'s captured on the lattice, likely as a consequence of lack of UV resolution. 

The picture is completely different for the homogeneous case, see lower right panel Fig.~\ref{fig:spectrumElectricMagneticStrong}. Initially, $\delta_{\rm EM}$ reflects the oscillatory pattern of the spectra of the electric and magnetic field, and fluctuates around 0. By the end of inflation, however, a clear dominance of the electric field is observed for all excited scales. The magnetic field only seems to dominate at very high UV scales, where the excitation is barely above the BD vacuum tail. 

We conclude that the phase of (electro)magnetic slow-roll during the strong backreaction is driven by the significant growth of the magnetic field in the system. Interestingly, this growth is scale dependent and mostly sustained at the mid-to-UV scales captured on a lattice. This fact highlights the relevance of (and hence the need to capture) sufficiently small scales during the strong backreaction dynamics. Conversely, small scales play little to no role in the homogeneous approximation, as they are never excited during the extended inflationary period. For instance, Fig.~\ref{fig:homoSpectra} shows that most of the power of the gauge field in the homogeneous approach is located in the mid-IR region and, in fact, at super-Hubble modes at the end of inflation. This contrasts with the more UV dominated spectra in the truly physical local case, as the gauge field spectra there never ceases to drift to smaller scales during the extension of inflation, and at the end of it the electric and magnetic power spectra are actually peaked at sub-Hubble scales \cite{Figueroa:2023oxc}. 
\begin{figure}[t!]
\includegraphics[width=\columnwidth]{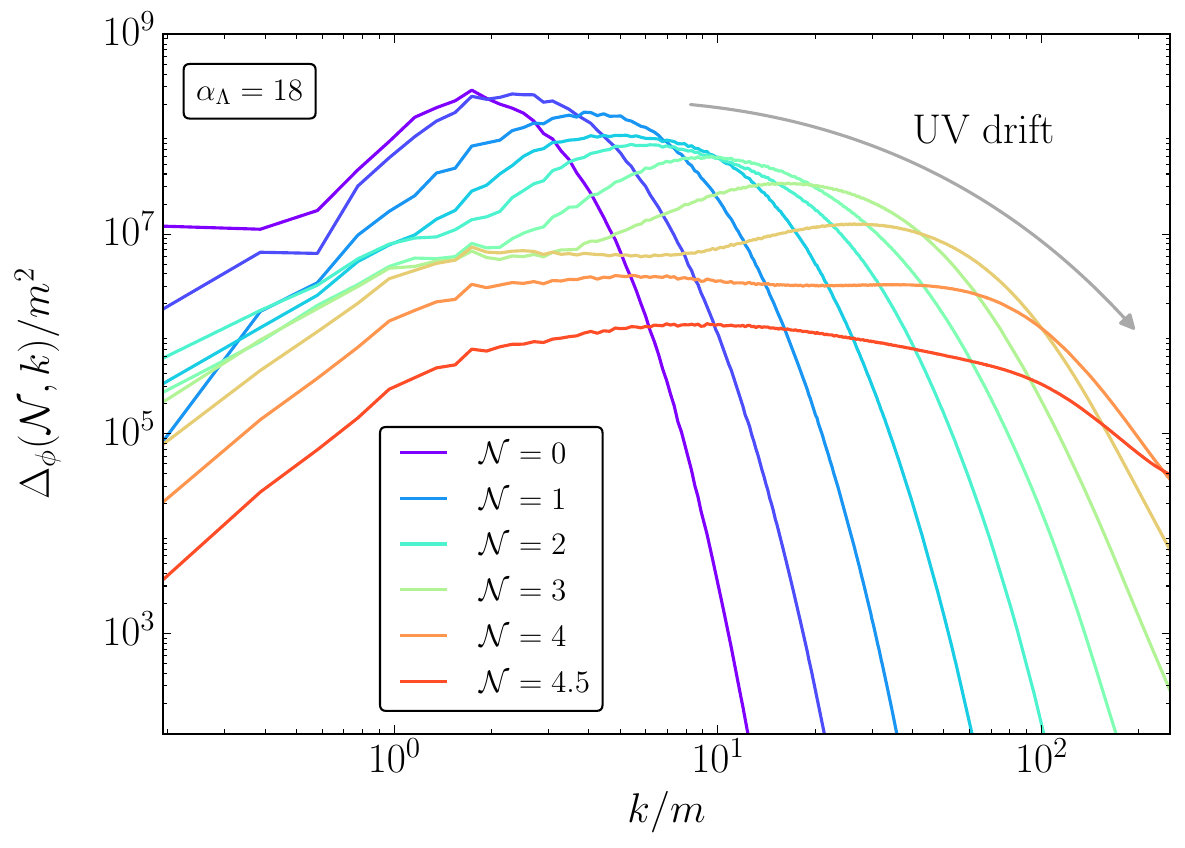}
\caption{Power spectrum of the inflaton for $\alpha_{\Lambda}=18$. The evolution shown covers the period $\mathcal{N} = 0$ to $\mathcal{N} = \Delta\mathcal{N}_{\rm br} = 4.5$ with gaps of $0.5$ e-folds, using a colour gradient from blue to red, exactly the same as in the top panel of Fig.~\ref{fig:spectrumElectricMagneticStrong} for the electric and magnetic fields.}
\label{fig:spectrumPhiStrong}
 \end{figure}

A major difference between the truly local dynamics and the homogeneous approximation, is that by including inhomogeneities in the inflaton sector, we allow successively smaller scales to be excited as time goes by. Fig.~\ref{fig:spectrumPhiStrong} shows the power spectrum of the inflaton fluctuations during the strong backreaction regime for $\alpha_{\Lambda}=18$. The evolution period and the gaps are the same as in the top-left panel of Fig.~\ref{fig:spectrumElectricMagneticStrong}, and we only plot after the linear regime is finished. The power spectra clearly spreads out towards UV scales, which then need to be present in the lattice in order to allow for this UV effect. Despite their modest contribution to the total energy density (see Fig.~\ref{fig:rhodensityRegimes}), the inflaton gradients directly affect the gauge field evolution through the term $\vec{\nabla}\phi\times\vec{E}$ in Eq.~(\ref{eqn:eom2}), which obviously is not present in the homogenous approach. Furthermore, the term $\pi_\phi\vec{B}$ in the same equation, cannot be factorized anymore as the product of a homogenous inflaton velocity times the magnetic field, as it captures instead a convolution of modes coupling different excited scales. This IR-UV feedback of scales lies at the heart of the non-linear nature of the dynamics during strong backreaction, and it is responsible for the spreading of power towards UV scales, which is not captured by the homogeneous approach.

\subsection{Inflating UV scales}
\label{subsec:powUV}

\begin{figure}[t]
\includegraphics[width=0.95\columnwidth]{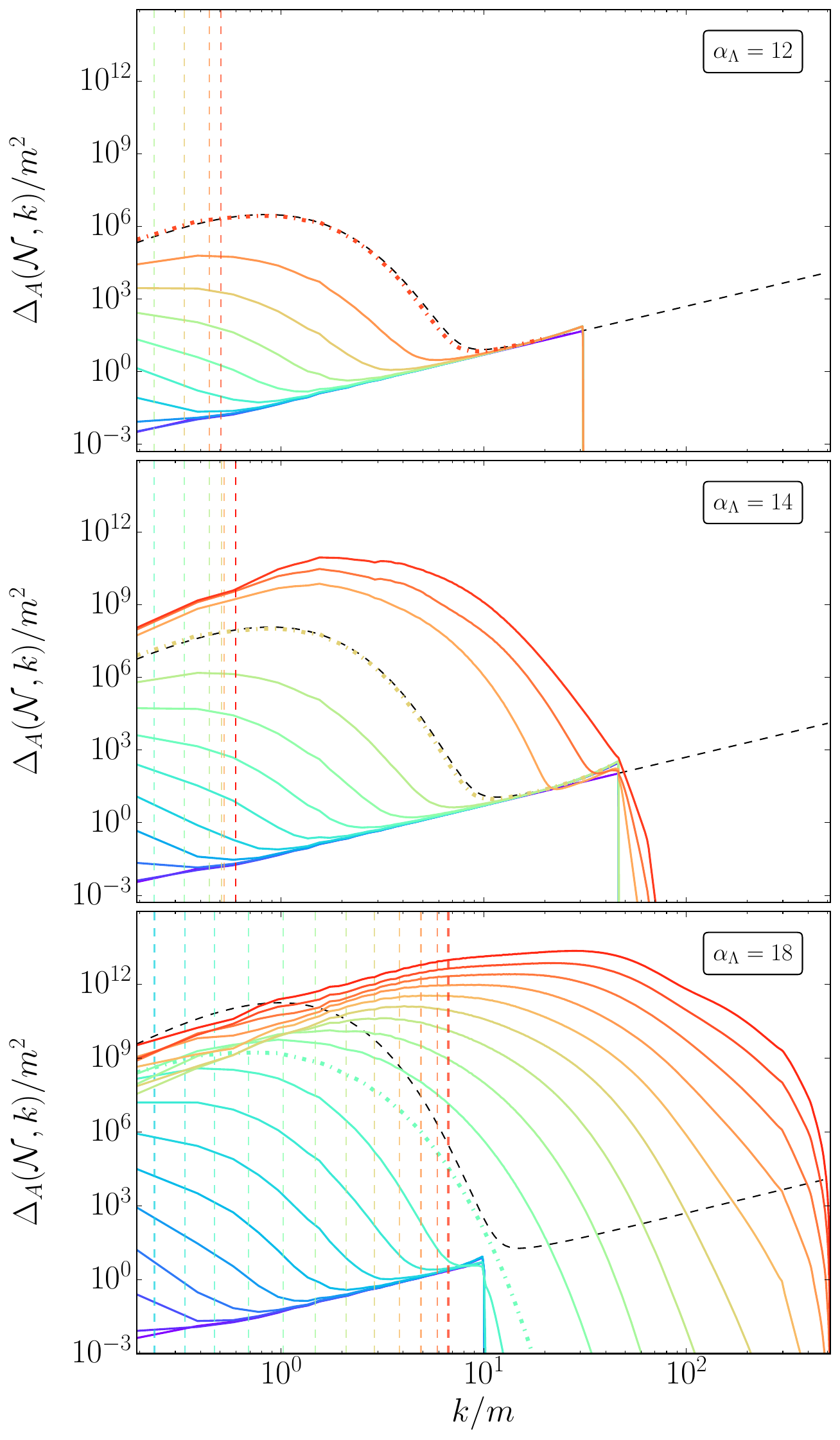}
\caption{Evolution of the gauge field power spectrum $\Delta_{A}(\mathcal{N},k)$ for the couplings $\alpha_{\Lambda}=12$ (top panel), $14$ (middle panel), and $18$ (bottom panel). The gap between spectra is of $0.5$ e-foldings from bluest to reddest colours, with the last spectrum corresponding to the end of inflation at $\Delta\mathcal{N}_{\rm br}=0$, $1.5$, and $4.5$, for $\alpha_{\Lambda}=12, 14,$ and $18$, respectively. For comparison purposes the range of modes in the plots are set to cover the UV region of $\alpha_{\Lambda}=18$. The evolution of the comoving Hubble scale is included with dashed vertical lines, using the same colour code of the spectra. The dashed black lines correspond to spectra, for each coupling, at the end of slow-roll inflation and computed in the linear regime. The coloured dotted-dashed lines correspond to spectra computed at the same times but obtained in the fully local and non-linear regime.}
\label{fig:AtotWEAKvsMILDvsSTRONG}
\end{figure}

One can obtain a deeper understanding of the amplification of UV scales by examining in closer detail the evolution of the gauge field power spectrum. In Fig.~\ref{fig:AtotWEAKvsMILDvsSTRONG}, we include the power spectrum $\Delta_{A}(\mathcal{N},k)$ of the gauge field for our representative set of couplings of each regime (weak, mild and strong backreaction). The separation between lines is of $0.5$ e-folds, and they are shown until the end of inflation for each coupling, starting from the initial Bunch-Davies configuration (coldest) till full excitation (warmest colours). It is important to note that inflation lasts differently for each coupling, so the same colours in different panels do not correspond to the same moment: the last spectrum, the reddest one, corresponds to $\Delta\mathcal{N}_{\rm br}=0$, $1.5$, and $4.5$ for $\alpha_{\Lambda}=12$, $14$, and $18$, respectively. Additionally, the comoving Hubble scale, $aH$, is included in each panel as dashed vertical lines, using the same colour gradient. As a reference, the spectrum corresponding to the end of slow-roll inflation ($\mathcal{N}=0$) is included in dashed black from a 1D grid computation of the linear regime, and with a dotted-dashed line (with the corresponding colour in that moment) from the fully local non-linear computation.

All panels of Fig.~\ref{fig:AtotWEAKvsMILDvsSTRONG} show that the position of the maximum amplification of the power spectra follows the drift of the comoving Hubble scale during inflation, including during the inflationary extension due to strong backreaction. The evolution for ``small" couplings, $\alpha_{\Lambda}=12$ (upper panel) and $14$ (middle panel), show little to no deviation from the linear prediction as $\mathcal{N}=0$ is reached, red line for $\alpha_{\Lambda}=12$ and yellow for $\alpha_{\Lambda}=14$. However, whereas in the former inflation ends precisely at $\mathcal{N}=0$, the latter is characterised by a re-entry into inflation, from $\mathcal{N}=0.3$ to $\mathcal{N}=1.3$. As appreciated in the middle panel, during the re-entry phase, with more inflation taking place in a quasi-electromagnetic slow-roll regime, the spectrum keeps growing and shifting towards more UV scales (lines in orange and red). Incidentally, the region above the cutoff scale $k_{\rm BD}$ is not excited in this case.

In the strong backreaction regime (lower panel of Fig.~\ref{fig:AtotWEAKvsMILDvsSTRONG}), we see that the evolution of the gauge field spectrum deviates from what is observed in the other two weaker couplings. In this case, and as explained before, the non-linear dynamics become relevant before $\mathcal{N}=0$, and therefore, we see how the physical spectrum at $\mathcal{N}=0$ (light green dotted-dashed line) deviates from the spectrum obtained from the linear analysis (dashed black line). From that moment on, the extra inflationary period beyond slow-roll inflation follows. This can be clearly observed in this figure, as the comoving Hubble scale moves towards UV scales at a roughly constant rate (over log scales), and so does the peak of the power spectrum. In this regime the modes higher than the artificial cutoff $k_{\rm BD}$ are considerably excited and become more and more dominant during the extra period of inflation. See App.~\ref{App:NumUVstability} for a discussion on the technical necessity for the artificial cutoff and its effect on the non-linear dynamics.

As well exemplified in Fig.~\ref{fig:AtotWEAKvsMILDvsSTRONG}, both the mild ($\alpha_\Lambda = 14$) and strong ($\alpha_\Lambda = 18$) backreaction cases, lead to extra e-foldings beyond slow-roll inflation. However, while the mild backreaction simulation only requires a slightly larger dynamical range above the initial cutoff $k_{\rm BD}$, in the strong backreaction simulation, for which the duration of inflation increases considerably, UV scales way above $k_{\rm BD}$ become excited and relevant in the dynamics. In general, it is crucial to cover the entire dynamical range to accurately capture the underlying physics, and in the strong coupling regime this becomes very demanding, as nonlinear effects extend the range of modes needed towards the UV. We highlight this by plotting all three panels of Fig.~\ref{fig:AtotWEAKvsMILDvsSTRONG} with a common momenta range $k/m\in [0.193, 391.497]$, which is the minimal required range for the largest coupling of the three panels, \ie $\alpha_{\Lambda}=18$. 

In terms of computational costs, the strong backreaction dynamics are the most expensive by far, as a simultaneous coverage of the relevant IR and UV scales of the problem needs to be considered from the beginning of a simulation, requiring successively larger lattices for larger couplings. In fact, one cannot predict a priori the final UV scale where the spectrum's peak will settle down at the end of inflation. We thus discuss now in detail this particular UV sensitivity of the strong backreaction regime and discuss the additional measures required on the lattice to accurately capture the dynamics.\\

\hspace{1.2cm}{\small \bf UV sensitivity and resolution tests}\vspace{0.2cm}

During the local backreaction regime, the peak of the gauge field spectra (\eg the power spectrum of the electric and magnetic fields), remains always inside the comoving Hubble radius. This is opposite to the homogeneous case, where the peak remains at super-Hubble scales till the end of inflation, recall Fig.~\ref{fig:homoSpectra}. A fundamental requirement, therefore, for accurately describing local backreaction is ensuring sufficient coverage of sub-Hubble modes within the lattice. The lower panel of Fig.~\ref{fig:AtotWEAKvsMILDvsSTRONG} shows however that, for a representative example of the strong backreaction, with $\alpha_{\Lambda}=18$, the comoving Hubble scale $aH$ grows almost an order of magnitude, thus leaving less room for capturing a wide enough range of modes between the Hubble scale and the natural lattice UV cutoff.

\begin{figure}[t]
\includegraphics[width=0.95\columnwidth]{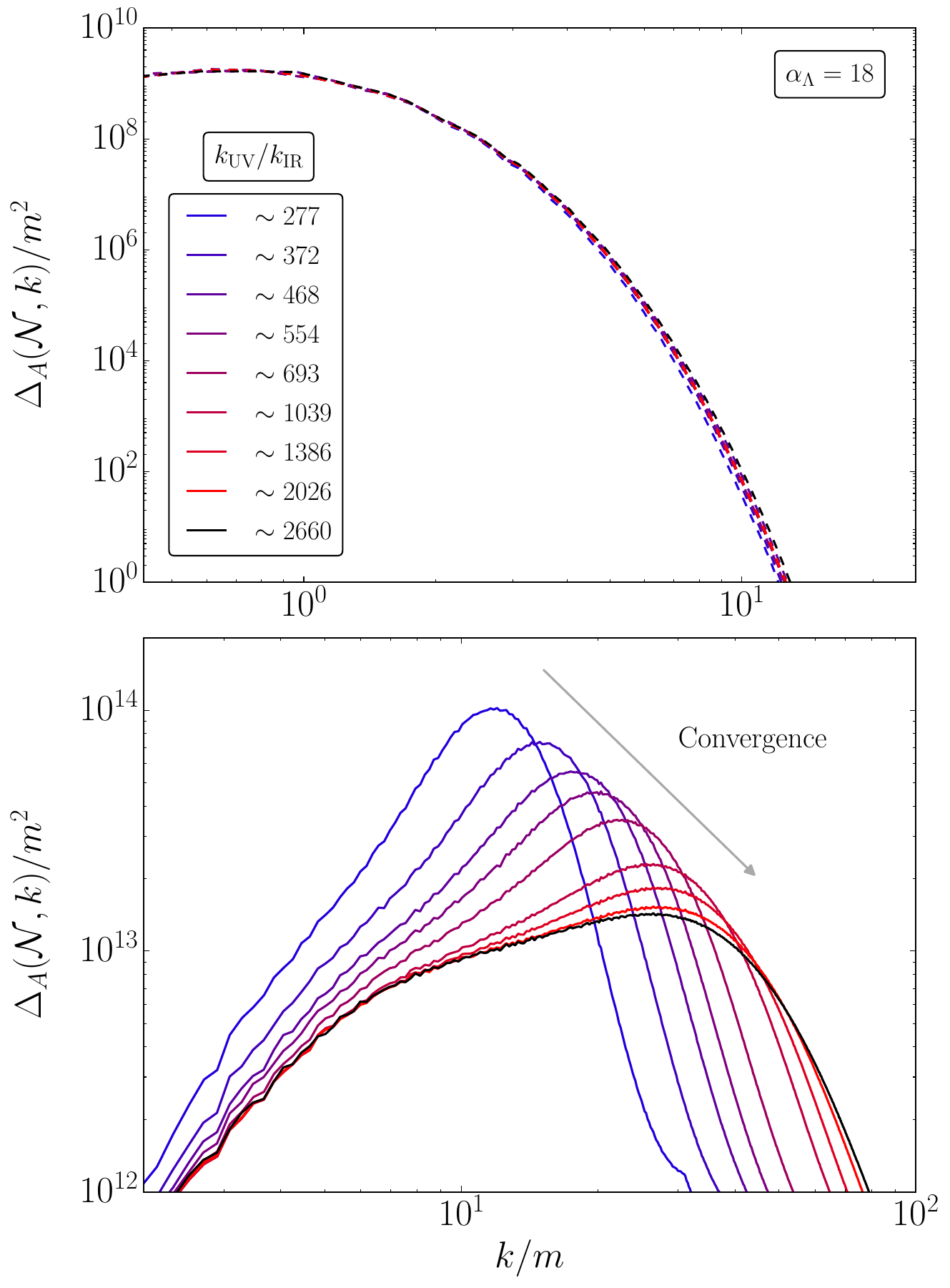}
\caption{Convergence study for $\alpha_\Lambda=18$. Gauge field power spectra against comoving momenta at $\mathcal{N}=0$ (upper panel) and at the end of inflation that corresponds to the largest simulation (lower panel). Different colours represent simulations with different separation of scales, as indicated in the label.}
\label{fig:SpectraInflationEndsIL18lattices}
 \end{figure}

Fig.~\ref{fig:SpectraInflationEndsIL18lattices} shows the comparison of the power spectra for $\alpha_\Lambda = 18$, for different lattice sizes with a common comoving volume but increasing UV resolution. They correspond to simulations with $k_{\rm IR}/m = 0.1932$ and $N= 320$, 430, 540, 640, 800, 1200, 1600, 2340 and 3072, corresponding to $k_{\rm UV}/m =$ 53.54, 71.95, 90.35, 107.98, 133.85, 200.78, 267.71, 391.51, and 514.00, respectively. In the upper panel, spectra are evaluated at $\mathcal{N}=0$, whereas in the lower panel, they are evaluated at the end of inflation according to the simulation with the finest resolution ($N=3072$). Each colour represents different $k_{\rm UV}/k_{\rm IR}$ ratios. 

The comparison between panels shows that at $\mathcal{N} = 0$, \ie when the system has barely become non-linear and the spectra are all well captured independently of the UV resolution, all simulations coincide. However, as the system progresses deep inside the strong backreaction regime, the spectra separate from each other, depending on whether the UV coverage is sufficient or not. Indeed, a ``UV barrier" effect is evident in the smallest simulations, as if the lack of support for higher modes would ``squeeze" and push the spectra towards smaller $k$'s. This UV artifact distorts both the amplitude and location of the spectrum's peak, thus faking the physics; the more severe the lack of UV support (blue-ish curves), the greater the distortion. As we increase the resolution (purple to reddish curves), the range of UV modes that a simulation can capture increases, and we start seeing a convergence of spectra at the end of inflation, as the reddest curve almost overlaps with the black one. 

\begin{figure*}[t]
\centering
\subfloat{\includegraphics[width=0.45\textwidth]{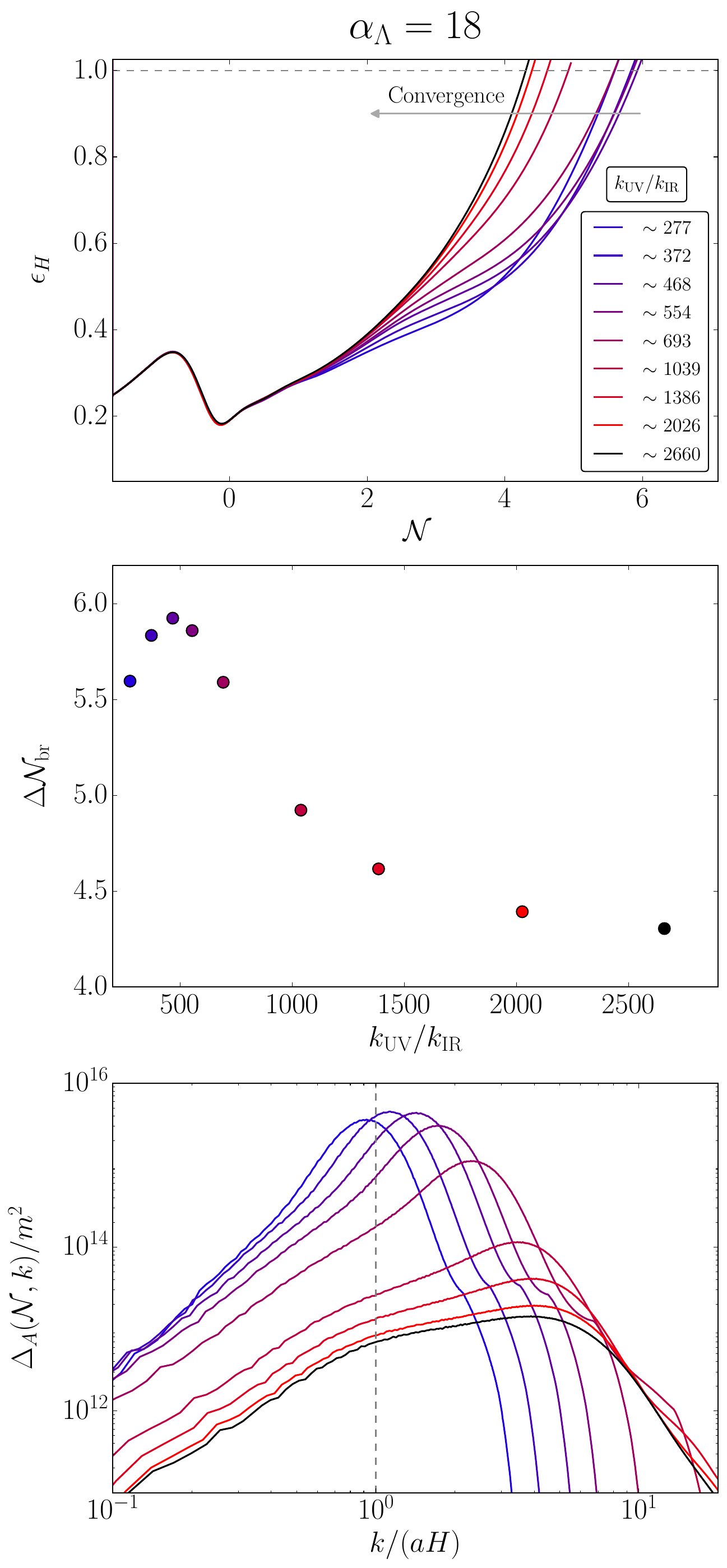}} \quad\quad\quad
\subfloat{\includegraphics[width=0.45\textwidth]{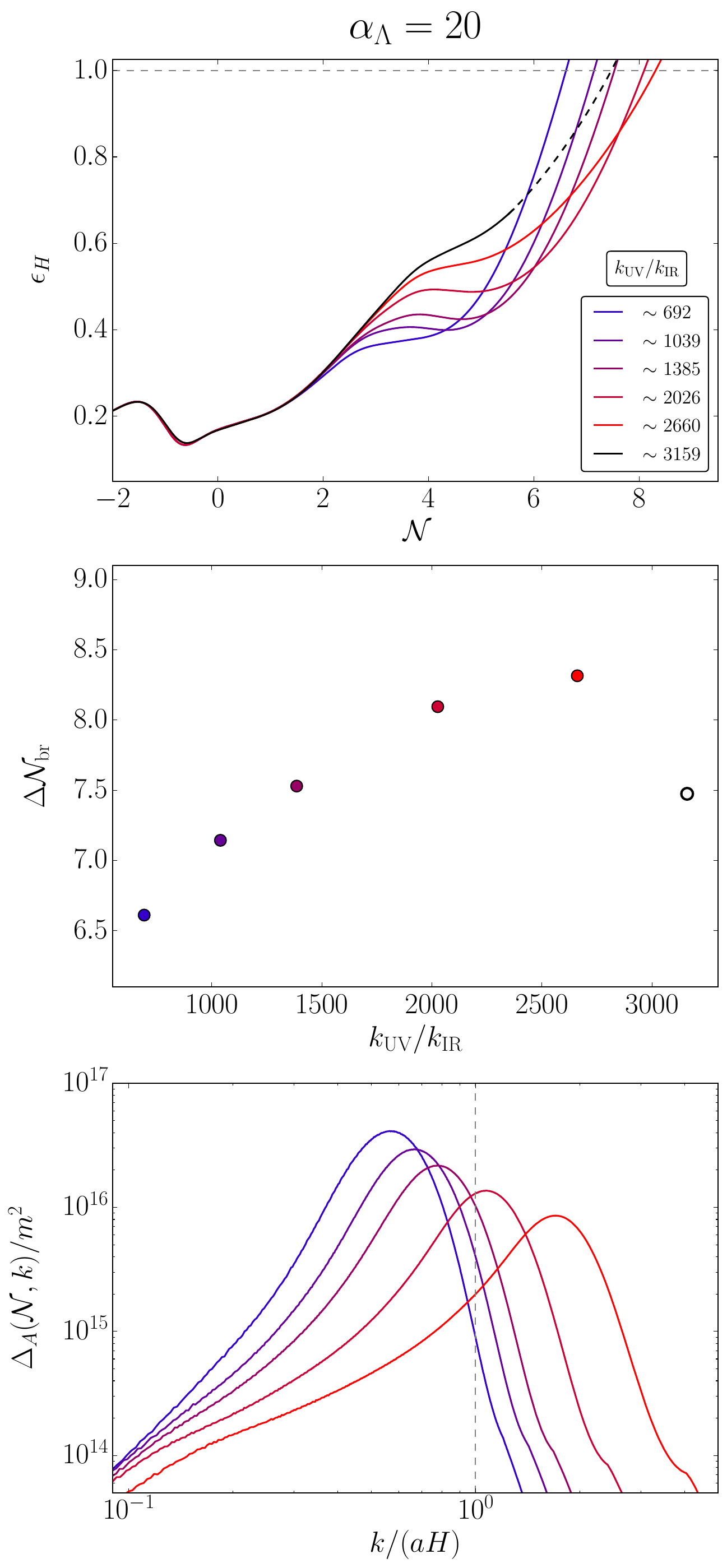}}
\caption{\textit{Upper panels}: the evolution of $\epsilon_H$ for simulations with increasing IR-to-UV ratio (see label for colouring prescription). \textit{Middle panels}: the extra number of e-folds for each resolution. \textit{Lower panels}: comparison of power spectra of the gauge field versus $k/(aH)$ extracted at the end of inflation for each case. In the left we show $\alpha_\Lambda=18$ and in the right $\alpha_\Lambda=20$}
\label{fig:InflationEndsIL18lattices}
\end{figure*}

The relevance of a correct UV coverage does not apply solely to the gauge field spectral shape. As our approach for the expansion of the universe is self-consistent, every change in the evolution of the gauge field will inevitably impact on the background evolution, specially in the (electro)magnetic slow-roll regime. This is precisely what Fig.~\ref{fig:InflationEndsIL18lattices} shows, where we plot the evolution of $\epsilon_H$ (upper panels) and the number of extra e-folds of inflation (mid panels) against the UV-to-IR ratio for simulations with $\alpha_\Lambda = 18$ (left panels) and  $\alpha_\Lambda = 20$ (right panels). We use the same colour scheme as in the previous figure. 

The example for $\alpha_\Lambda = 18$ shows that the lengthening of inflation depends non-monotonically in the separation of scales, showing a growing trend until $k_{\rm UV}/k_{\rm IR}\sim 500$, and decreasing behaviour towards an asymptotic value (representing convergence) for larger resolutions. In conclusion, if the UV coverage of the simulation of a given coupling is not enough, the overall dynamics are \textit{faked} and non-physical results are obtained. 

In the lower left panel of Fig.~\ref{fig:InflationEndsIL18lattices} we show this effect in the spectra in an alternative manner, by choosing the moment at which inflation ends in each case, which is different for every simulation. Additionally we also plot the spectra as a function of $k/(aH)$, in order to highlight the hierarchy of the position of the maximum of the spectra with respect to the comoving Hubble scale. The panel clearly shows that for the simulations with the worst dynamical separation of scales, the main excitation of the gauge field cannot remain inside the comoving Hubble radius. In fact, for the worst resolution case for $\alpha_\Lambda=18$, it is slightly out. As we increase the dynamical range, there is more space in the UV and the peak progressively accommodates to its real scale. A fairly good convergence is obtained for separation of scales $k_{\rm UV}/k_{\rm IR} \gtrsim 2000$, which corresponds to lattices with $N\gtrsim2300$ points per dimension. Though the overlap of spectra among the highest resolution cases is not perfect, their relative discrepancies only amount to $\mathcal{O}(1)\%$ changes in the end of inflation time scale.

A major problem of the  UV sensitivity of strong backreaction is that, a priori, we cannot know how much we need to push the UV scale of the lattice so that the output is a trustworthy represention of the true physics. In traditional simulations of post-inflationary dynamics (\eg of the Standard Model Higgs~\cite{Enqvist:2015sua,Figueroa:2015rqa,Figueroa:2017slm}, topological defects~\cite{Correia:2024cpk,Baeza-Ballesteros:2024otj,Baeza-Ballesteros:2023say,Hindmarsh:2021mnl,Blanco-Pillado:2023sap,Hindmarsh:2021vih,Gorghetto:2020qws,Figueroa:2020lvo,Hindmarsh:2019csc,Hindmarsh:2017qff,Gorghetto:2018myk,Figueroa:2012kw}, or (p)reheating~\cite{Figueroa:2024yja,Figueroa:2024asq,Figueroa:2021iwm,vandeVis:2020qcp,Nguyen:2019kbm,Antusch:2020iyq,Antusch:2021aiw,Antusch:2022mqv,Dufaux:2006ee,Cosme:2022htl}), the rate of expansion is not exponential, as it is typically a power-law in time, when not even vanishing, if the simulations are in Minkowski. The importance of a good resolution in those cases, therefore, amounts to how well one captures in a given simulation the UV tail of the excited spectra; a tail that is assumed to be suppressed as compared to the peak of the excitation. However, in our present simulations of inflation via (electro)magnetic slow-roll during strong backreaction, because of the continuous flow from IR to UV scales due to the exponential expansion, the difference between a good or a bad UV coverage simply amounts to capturing or not truthfully the physics. For the strong couplings we have explored in this work, a poor resolution can misplace the peak position of the spectra by a factor $\mathcal{O}(0.1)$, while enhancing artificially its amplitude by a factor $\lesssim \mathcal{O}(10)$, see \eg bottom panel of Fig.~\ref{fig:SpectraInflationEndsIL18lattices}.

As an illustrative example of a case where no convergence is yet reached, we plot the same relevant quantities for $\alpha_\Lambda=20$ in the right panels of Fig.~\ref{fig:InflationEndsIL18lattices}. We include simulations with 6 different resolutions, ranging from $k_{\rm UV}/k_{\rm IR}\sim 692$ (blue) to $k_{\rm UV}/k_{\rm IR}\sim$ 3150 (black), which correspond to box sizes from $N=800$ to $N=3648$ respectively. We note that the simulation with the highest resolution failed to reach $\epsilon_H=1$ (stopping at approximately $\epsilon_H\sim0.6$). Nonetheless, we include it here for informative purposes. To extrapolate its evolution (dashed lines), we assume that beyond the stopping point the trajectory follows the same trend as the simulation with the next highest resolution (red line). We include the value for $\Delta\mathcal{N}_{\rm br}$ obtained from such extrapolation as an empty black dot in the middle panel.

We observe that although there is a convergence in $\epsilon_H$ in the initial phase of backreaction, $-2\lesssim\mathcal{N}\lesssim2$, trajectories start diverging during the final stages towards the end of inflation. Simulations with the worst resolution (smaller $k_{\rm UV}/k_{\rm IR}$ ratio) deviate earlier, whereas the ones with better resolution sustain convergence up to $\mathcal{N} \sim 3.5$, and then diverge afterwards. 

Phenomenologically, we have identified as a characteristic feature of simulations with insufficient UV coverage, the emergence of a plateau in the $\epsilon_H$ trajectory, which deviates from the true (steeper) trajectory. As the resolution increases, the plateau gradually fades out. Incidentally, when the plateau first appears in a simulation with a very poor resolution, we see that as we improve the UV coverage, $\Delta\mathcal{N}_{\rm br}$ displays an initial growing trend till it reaches a ``turn-over" point for a critical resolution. When increasing further the resolution, the trend is reverted, and $\Delta\mathcal{N}_{\rm br}$  starts decreasing, till it settles down to a constant value, corresponding to convergence (see this effect in Fig.~\ref{fig:InflationEndsIL18lattices} for $\alpha_\Lambda = 18$). Due to our limited computational resources we could not launch simulations with a separation of IR-UV scales as large as the dynamics really require for $\alpha_\Lambda = 20$. We think that it is likely, based on the pattern of slightly smaller couplings, that the real evolution for $\alpha_\Lambda = 20$ will be described by a smooth growth similar to that depicted for $\alpha_\Lambda=18$, but our current data only describe the curves/points shown in Fig.~\ref{fig:InflationEndsIL18lattices}, which suggest that we barely reach the turn-over point.

Finally the lower right panel of Fig.~\ref{fig:InflationEndsIL18lattices} shows the spectral distortions caused due to the lack of UV resolution for $\alpha_\Lambda = 20$. Using the same colouring scheme, we plot $\Delta_A(\mathcal{N},k)$ for all resolutions at the moment corresponding to the simulation with the worst resolution, which then leads to the smallest extension of inflation, $\Delta\mathcal{N}_{\rm br} \simeq 6.6$. The smallest simulations (cold colours) do not have enough space in the UV tail of the spectrum which is forced to remain below the corresponding lattice UV cutoff $k_{\rm UV}$. As we increase the resolution, higher modes become available in the UV region and the excitation accommodates power naturally in those scales. Full convergence and definitive results can only be obtained if the excitation is well inside the comoving Hubble radius, as it is the case for $\alpha_\Lambda=18$, but not for $\alpha_\Lambda=20$. Hence, we estimate that for $\alpha_\Lambda = 20$ a much larger separation of scales is still required to achieve convergence of the output.

\subsection{Partial restoration of chirality balance and loss of transversality}

\begin{figure*}[t!]
\includegraphics[width=0.9\textwidth]{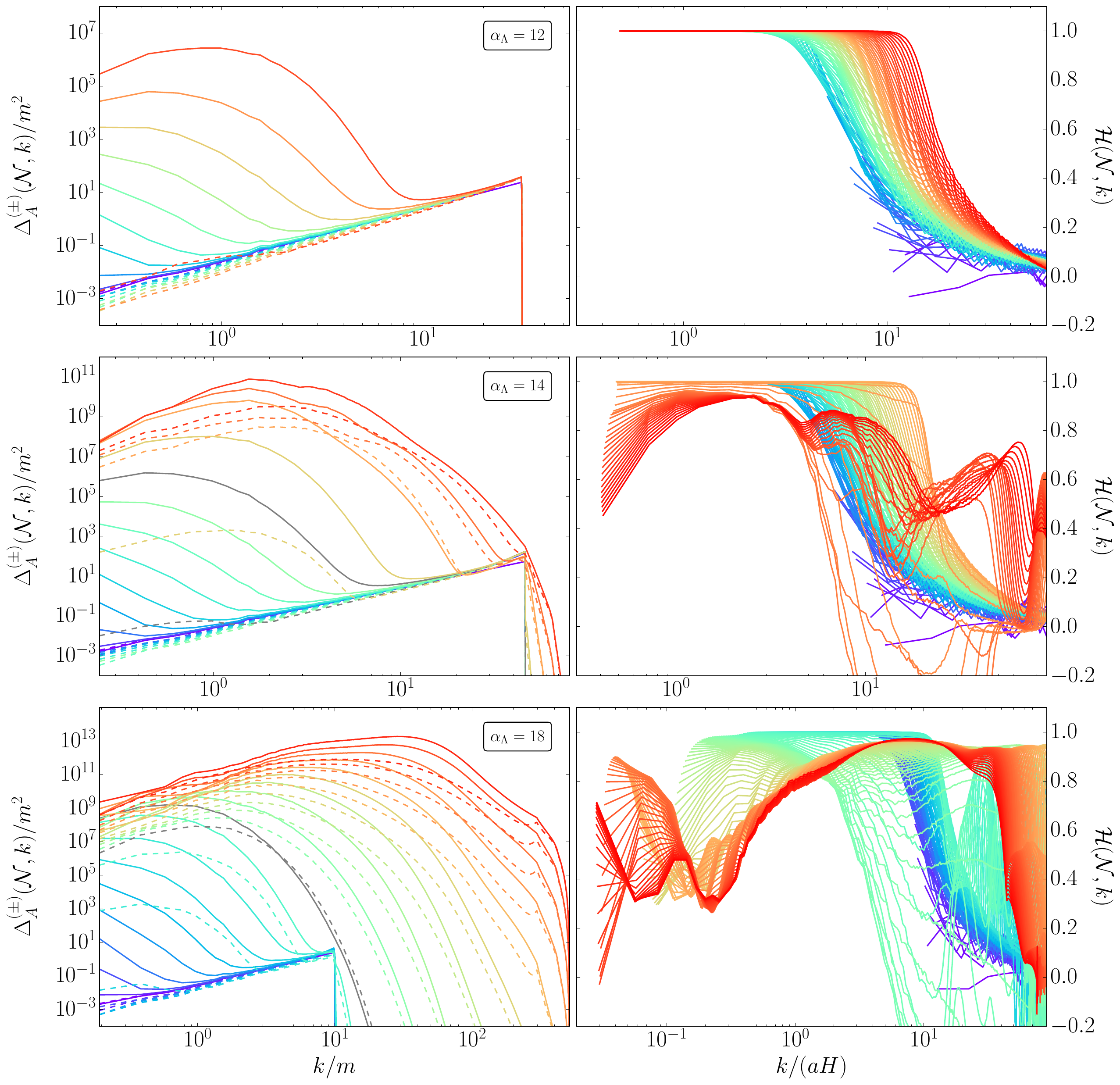}
\caption{Comparison between the positive and negative helicity modes of the gauge fields for the reference couplings $\alpha_{\Lambda}=12$, $14$ and $18$. \textit{Left}: gauge power spectrum evolution for $A^{+}$ (solid) and $A^{-}$ (dashed). We plot starting from the Bunch Davis vacuum (the coldest colour) with gaps of $0.5$ e-folds, until the end of inflation (the warmest colour). Grey dashed and solid lines indicate the spectra at $\mathcal{N}=0$. \textit{Right}: normalized spectral helicity evolution for the same period of the evolution as in the right panel but with gaps of $0.05$ e-folds.}
\label{fig:PlusMinusSpectrumAndHelicity}
 \end{figure*}

The chiral nature of axion inflation has been regarded as one of its key features. Even though chiral restoration effects at sub-Hubble scales are expected in (p)reheating \cite{Adshead:2015pva,Adshead:2016iae}, traditionally it has been always expected that the gauge field excitation is maximally chiral during inflation. In our previous work, {\tt Paper I}~\cite{Figueroa:2023oxc}, we revised this aspect during strong backreaction, and discovered a partial restoration of the balance between the two chiralities, namely a \textit{scale dependent chirality (im)balance}. We also discovered that the gauge field does not remain transverse during strong backreaction, as a longitudinal component is developed, also with a scale-dependent spectrum. In this section we address these issues in full detail and provide a physical explanation of their origin.

Figure \ref{fig:PlusMinusSpectrumAndHelicity} shows, in the left panel, the evolution of the power spectra of both helicities,  $\Delta^{(+)}_A(\mathcal{N},k)$ in solid and $\Delta^{(-)}_A(\mathcal{N},k)$ in dashed, plotted against $k/m$. In the right panel, we show the relative spectral helicity difference, defined as
\begin{equation}
\mathcal{H}(\mathcal{N},k)=\frac{\Delta^{(+)}_{A}(\mathcal{N},k)-\Delta^{(-)}_{A}(\mathcal{N},k)}{\Delta^{(+)}_{A}(\mathcal{N},k)+\Delta^{(-)}_{A}(\mathcal{N},k)}\, ,
\end{equation}
and plotted against $k/(aH)$. We compare the evolution for the couplings $\alpha_{\Lambda}=12$ (upper panels), 14 (middle panels) and 18 (bottom panels), representative of the weak, mild, and strong backreaction regimes, respectively. In the weak backreaction case we see the tachyonic component $A^{+}(\mathcal{N},k)$ growing exponentially, while the opposite helicity $A^{-}(\mathcal{N},k)$ remains in vacuum. This is reflected in the corresponding right panel, which shows that by the end of inflation $\mathcal{H}(\mathcal{N},k) \simeq 1$ regardless of whether the modes are sub- or super-Hubble. In the very UV part of the spectra, $\mathcal{H}(\mathcal{N},k)\to 0$ because there are modes in the lattice where both helicities are still in vacuum ($k>10m$ in the left panel). This behaviour is in agreement with previous comparisons, specially with the evolution of the energy component shown in Fig.~\ref{fig:rhodensityRegimes}, where little to no deviation from the linear regime prediction is observed during inflation.

In the mild backreaction case (middle panels), a significant change is however appreciated. Initially, as expected, only the positive helicity grows while the negative helicity remains unexcited. However, at the onset of non-linearities, even before $\mathcal{N}=0$, $\Delta^{(-)}_{A}(\mathcal{N},k)$ starts growing. This growths is initially very mild, and by the time $\mathcal{N}=0$ and $\epsilon_H=1$ for the first time (dashed grey line), the excitation of $A^-(\mathcal{N},k)$ is till very subdominant compared to the excitation of $A^+(\mathcal{N},k)$. Nevertheless, upon re-entering inflation, during the additional $1.5$ extra e-foldings (the last three lines in the figure), $\Delta^{(-)}_{A}(\mathcal{N},k)$ then grows considerably and, even though it remains below $\Delta^{(+)}_{A}(\mathcal{N},k)$, it becomes comparable in magnitude at certain scales. This effect is clearly appreciated in the right panel, where from $\mathcal{N}=0$ onwards (orange lines) there is a clear evolution towards lower values of $\mathcal{H}$ at all scales, and some sub-Hubble modes even reach $\mathcal{H}<0$.  By the end of inflation, indicated by the reddest lines, all relevant scales involved in the gauge field excitation have departed from $\mathcal{H}=1$, and a partial chiral restoration trend is visible in the most IR modes.

Finally, in the strong backreaction case (lower panels), we note that the evolution is initially similar to that in the mild backreaction case when it re-enters in inflation. The growth of the negative helicity happens across all scales from the onset of non-linearities and till the end of inflation, with the spectrum $\Delta^{(-)}_{A}(\mathcal{N},k)$ following a similar evolution as $\Delta^{(+)}_{A}(\mathcal{N},k)$, and developing a similar spread. In this case, a broader range of super-Hubble modes develop values as $\mathcal{H} \approx 0.5$. Around the region of the spectras's peak, the chiral imbalance is however reduced to $\mathcal{H} \approx 0.9$.

In summary, we observe that for larger couplings the excitation of $A^{-}$ occurs earlier, as the non-linearities develop sooner. The strength of the excitation of $A^{-}$ is also stronger the larger the coupling. The chirality imbalance is reduced across all scales of interest, but most notably, as the modes exit the Hubble radius, they settle to values oscillating around $\mathcal{H} \approx 0.3-0.7$, indicating that between one-third to one-half of the contribution comes from the opposite helicity. Furthermore, besides the negative helicity excitation, we also observe in our simulations the excitation of longitudinal modes $A^{\parallel}(\mathcal{N},k)$, which develop a wiggly scale-dependent spectrum similarly as that of $A^-(\mathcal{N},k)$. This implies that the gauge field $\vec A (\mathcal{N},k)$ ceases to be transverse, with $\partial_i A_i = \partial_i A_i^{\parallel} \neq 0$. 

The excitation of both $A^{-}(\mathcal{N},k)$ and $A^{\parallel}(\mathcal{N},k)$ during the non-linear dynamics, can be understood as a power transfer from the positive helicity $A^{+}(\mathcal{N},k)$, which by the onset of non-linearities is already exponentially excited (way above the vacuum tail), within a finite range of momenta. This transfer of energy is actually driven by the inflaton inhomogeneities, and hence it is only possible to capture it in the truly local backreaction regime, and not in the homogeneous approximation. In order to illustrate this, we write both inflaton-gauge interaction terms from Eq.~(\ref{eqn:eom2}) in Fourier space,
\begin{eqnarray}
\label{eq:convolution_2}
&&\big(\pi_\phi \vec B\,\big)_{\vec k} 
= -i\sum_{\lambda=\pm}\lambda \int \hspace{-0.5mm}d^3 q\,\,\dot\phi^*_{(\vec k\shortminus \vec q)}\,q \,A^{\lambda}_{\vec q}\,\vec\varepsilon^{\,\lambda}_{\vec q}\,,\\
\label{eq:convolution_3}
&&\hspace{-0.5cm}\big(\vec\nabla \phi \times \vec E\,\big)_{\vec k} = i\sum_{\sigma=\pm\, , \parallel}\int \hspace{-0.5mm}d^3 q\,\,\phi^*_{(\vec k\shortminus \vec q)}\dot A^{\sigma}_{\vec q}\Big((\vec k-\vec q\,)\times\vec\varepsilon^{\,\,\sigma}_{\vec q}\Big),\nonumber\\
\end{eqnarray}
where we explicitly show that $\vec{E}$ can hold projections in all three components of the chiral basis ($\sigma = -, +, \parallel$), whereas $\vec{B}$ is transverse ($\lambda = +,-$), as $\vec\nabla \cdot\vec B = 0$ always holds. Since the integrals in the convolutions span over all momenta $\vec q$, we observe that for $\vec q\not\propto \vec k$, the following projections ${\vec k}\cdot \vec{\varepsilon}^{\,\lambda}_{\vec q}\neq 0$ and $\vec{\varepsilon}^{\,-}_{\vec k}\cdot\vec{\varepsilon}^{\,\lambda}_{\vec q}\neq 0$ are non-vanishing, as they are a function of the (cosine of the) angle between the external momentum $\vec k$ and the internal modes $\vec q$, $\cos\theta \equiv (\vec k\cdot\vec q\,)/(|\vec k\,||\vec q\,|)$, which sweeps all possible values between -1 to +1. This implies that there will always be non-vanishing projections of $(\pi_\phi \vec B)_{\vec k}$ and $(\vec\nabla \phi \times \vec E)_{\vec k}$ onto both chiralities $\pm$ and onto the longitudinal mode. For example, in the EOM for the gauge field, the negative helicity $A^-_{\vec k}$ will now be sourced by the component $\big(\pi_\phi \vec B\,\big)_{\vec k}^- \,\equiv\, \vec \varepsilon^{\,\,+}_{\vec k}\cdot\big(\pi_\phi \vec B\,\big)_{\vec k}$, which by construction is given by
\begin{eqnarray}\label{eq:piBproject}
&& \hspace{-0.5cm} \big(\pi_\phi \vec B\,\big)_{\vec k}^- \equiv -\,i\int \hspace{-0.5mm}d^3 \vec q\,\,\dot\phi^*_{(\vec k\shortminus \vec q)}\,q\,\Big(A^{+}_{\vec q}(\vec\varepsilon^{\,+}_{\vec k}\hspace{-1mm}\cdot\hspace{-0.5mm} \vec\varepsilon^{\,+}_{\vec q}) \shortminus A^{-}_{\vec q}(\vec\varepsilon^{\,+}_{\vec k}\hspace{-1mm}\cdot\hspace{-0.5mm} \vec\varepsilon^{\,-}_{\vec q})\Big)\nonumber\\
&& \hspace{0.88cm} \simeq -\,{i\over2}\int \hspace{-0.3mm} d^2\hat q\,dq \,q^3 \,\,\dot\phi^*_{(\vec k\shortminus \vec q)}\,(1-\cos\theta)\,A^{+}_{\vec q}\,,
\end{eqnarray}
where we have used $\vec\varepsilon^{\,\,+}_{\vec k}\hspace{-0.5mm}\cdot \vec\varepsilon^{\,\lambda}_{\vec q} = (1-\lambda\cos\theta)/2$, and in the second line we have exploited the fact that $|A_{\vec q}^+| \gg |A_{\vec q}^-|$ holds during the initial stages of non-linear dynamics, over the region of excitation. Similarly, we can obtain analogous expressions for $\big(\pi_\phi \vec B\,\big)_{\vec k}^{\parallel} \,\equiv\, \hat k\cdot\big(\pi_\phi \vec B\,\big)_{\vec k}$, which will source $A_{\vec k}^{\parallel}$, and also for $\big(\vec\nabla \phi \times \vec E\,\big)_{\vec k}^- \,\equiv\, \vec \varepsilon^{\,\,+}_{\vec k}\cdot\big(\vec\nabla \phi \times \vec E\,\big)_{\vec k}$ and $\big(\vec\nabla \phi \times \vec E\,\big)_{\vec k}^{\parallel} \,\equiv\, \hat k\cdot\big(\vec\nabla \phi \times \vec E\,\big)_{\vec k}$, which will further source $A_{\vec k}^{-}$ and $A_{\vec k}^{\parallel}$, respectively.  Remarkably, this means that the negative and longitudinal components of the gauge field will be sourced to high large amplitudes by the exponentially excited tachyonic modes $A_{\vec q}^+$, as \eg shown explicitly by the expression in the second line of Eq.~(\ref{eq:piBproject}).  
\begin{figure}[t]
\includegraphics[width=\columnwidth]{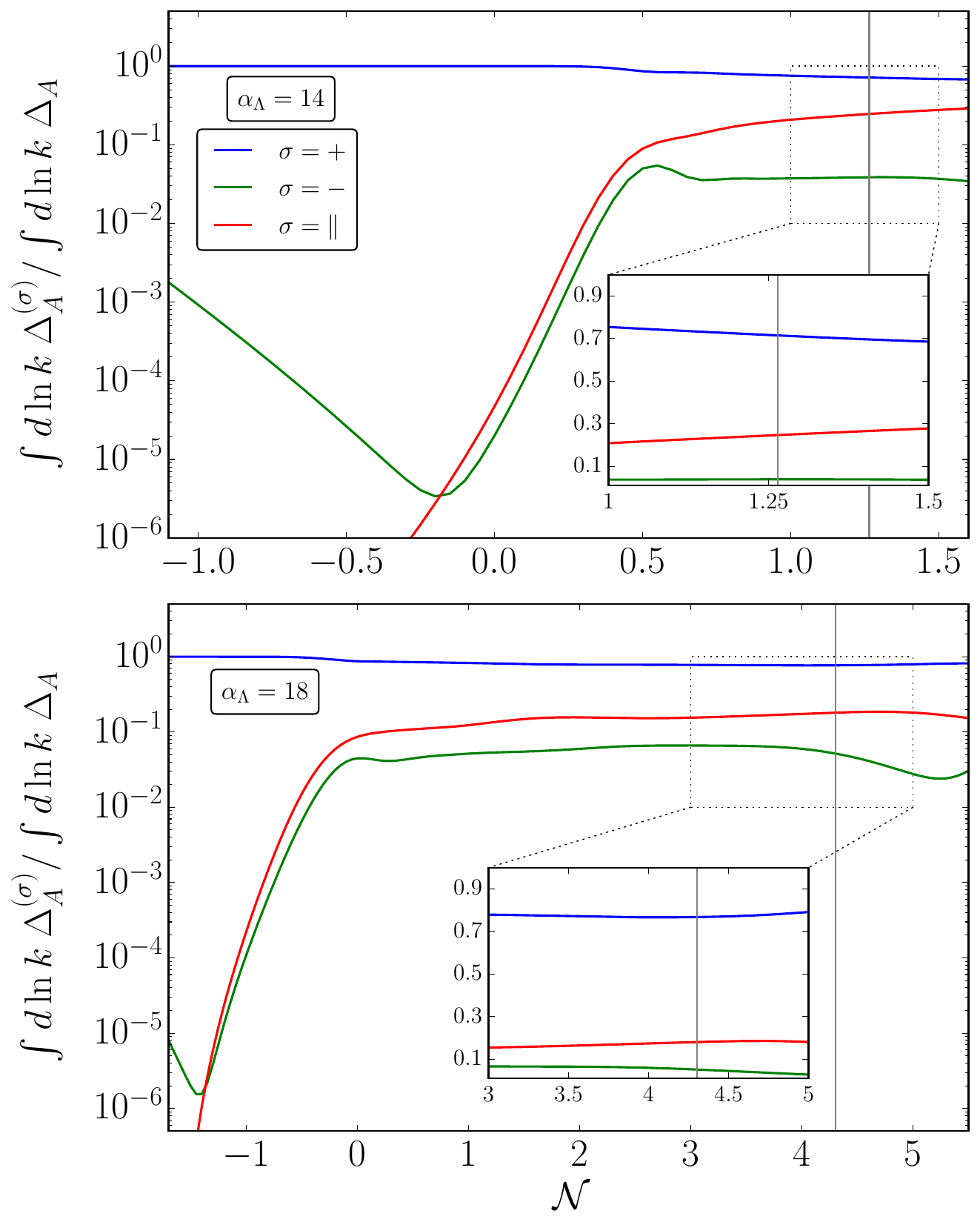}
\caption{Evolution of the integrated fractional contribution of the different components of the gauge fields with respect to the total for $\alpha_\Lambda=14$ (top) and $\alpha_\Lambda=18$ (bottom). Blue lines correspond to $A^{+}$, green line to $A^{-}$ and red line to $A^{\parallel}$.}
\label{fig:integratedPlusMinusLongWeakMildStrong}
 \end{figure}
Conversely, for a homogeneous inflaton, $\vec q$ collapses into $\vec k$, and there is no power transfer between different components. In Eq.~(\ref{eq:piBproject}) this can be understood manisfestly as $\cos\theta \rightarrow 1$ in the second line, whereas more generically, all components $\big(\pi_\phi \vec B\,\big)_{\vec k}^{-}$, $\big(\pi_\phi \vec B\,\big)_{\vec k}^{\parallel}$,  $\big(\vec\nabla \phi \times \vec E\,\big)_{\vec k}^{-}$, and $\big(\vec\nabla \phi \times \vec E\,\big)_{\vec k}^{\parallel}$ collapse to zero if the inflaton is homogeneous, because it holds that $\vec\varepsilon^{\,\,+}_{\vec k}\hspace{-0.5mm}\cdot \vec\varepsilon^{\,+}_{\vec k} = 0$ and $\hat k\hspace{-0.5mm}\cdot \vec\varepsilon^{\,\pm}_{\vec k} = 0$. In other words, only if the inflaton presents inhomogeneities, will the inflaton-gauge interactions terms act as sources for the longitudinal and negative helicity components. 

In conclusion, we highlight that the inhomogeneities generated during the local backreaction dynamics are responsible for the excitation of the longitudinal as well as the negative helicity components, and therefore this is a genuine effect that occurs during the truly inhomogeneous non-linear dynamics of axion inflation. In the homogeneous approximation, the longitudinal component remains at zero, and the negative helicity in vacuum.

We provide a quantitative description of the growth of the non-tachyonic modes in Fig.~\ref{fig:integratedPlusMinusLongWeakMildStrong}. There we plot the evolution of the integrated fractional contribution of the power spectrum corresponding to $A^{+}(\mathcal{N},k)$, $A^{-}(\mathcal{N},k)$ and $A^{\parallel}(\mathcal{N},k)$, for $\alpha_\Lambda=14$ (top) and $\alpha_\Lambda=18$ (bottom). The total contribution is largely dominated by the $(+)$ component, but there is an interesting interplay between the $(-)$ and the longitudinal component. The former remains near the vacuum solution until the inflaton inhomegeneities begin to affect its evolution, at around $\mathcal{N}\sim -0.5$ (top) and $\mathcal{N}\sim -1.5$ (bottom), and then starts growing exponentially, reaching rapidly amplitudes way above the vacuum solution. At those moments the longitudinal mode, which was vanishing during the linear regime, 
grows also exponentially, reaches the level of the negative helicity, and keeps growing at the same rate. Both $A^-(\mathcal{N},k)$ and $A^{\parallel}(\mathcal{N},k)$ are driven by the tachyonic growth of $A^{+}(\mathcal{N},k)$, once the inflaton inhomogeneities become noticeable, soon after the onset of non-linearities in the dynamics. The growth of the relative contributions stabilize at around $15-20 \%$ (longitudinal component) and $2-3 \%$ (negative helicity) during the (electro)magnetic slow-roll regime, and persists like that till the end of inflation (vertical line), for both couplings.

\section{Discussion}

Axion inflation provides a natural framework for a realistic particle physics realization of inflation. In particular, thanks to its shift symmetry, inflation is protected from quantum corrections. It is also phenomenologically very rich, offering a variety of unique observational signatures. In this respect, the accurate description of the dynamics is a necessary condition to correctly assess the observational predictions that can be used to probe this scenario. Previous approaches to study the dynamics of axion inflation, based on increasing levels of complexity, somehow fail to capture the true behaviour of the system, in particular of the strong backreaction dynamics. In order to describe accurately the latter regime, it is necessary to resort to a fully local treatment of the dynamics, which only real time field theory lattice simulations can provide. 

In this paper we present the details of a proper lattice formulation of axion inflation that makes the symmetries of the theory explicit, and satisfies continuum constraints to a very high degree at the lattice level. Besides, our lattice formulation agrees very well with the linear regime, which is exactly solvable, and with the homogeneous backreaction description, which can be solved by non-lattice methods. Our lattice description is validated in this way by these tests and sets the standard for performing lattice simulations for axion inflation. A series of simulations have been performed for a range of couplings, identifying three regimes: weak-, mild- and strong-backreaction, depending on the extension of inflation. For each coupling, except for the biggest coupling considered, we achieve convergence of simulations with increasingly better UV resolution. Our results show the difficulty in performing simulations for large couplings ($\alpha_\Lambda \geq 20$), as these require an increasingly larger separation of scales, which translates into increasingly expensive computational capabilities. In particular, our quantitative results, summarized \eg in our fits (\ref{eq:fitDeltaN})-(\ref{eq:fitDeltaN_PL}), are the only accurate parametrisation of the delay of the end of inflation for large couplings that have reached convergence. 

By studying the truly local nature of strong backreaction, we have understood how the electromagnetic energy density eventually drives inflation towards an end, in a new regime that we coin ``(electro)magnetic slow-roll". We observe that the magnetic field actually contributes the most to the gauge field energy, way above the electric counterpart (especially around the peak of the spectra). Our lattice study show the need to resolve well the peak of excitation, which is very sensitive to the lattice UV cutoff. In this respect, we show the dependence of the end of inflation on the UV lattice resolution, as we obtain only convergence among simulations (for a given fixed coupling) if the UV scales simulated are sufficiently separated away from the initial IR scales required for the linear regime. We demonstrate in full detail how not resolving properly the necessary UV coverage leads to physically wrong results. The non-trivial evolution of the comoving Hubble scale is another feature that shows how the strong backreaction dynamics require a very large separation of IR-to-UV scales, in order to resolve properly the peak of excitation of the gauge field. We also find that, in contrast to any inflaton-homogeneous approach (\eg during linear dynamics or homogenous backreaction), the system does not remain completely chiral, as the non-tachyonic chirality and the longitudinal mode are generated during the non-linear dynamics of inflation. We show that this is effect is driven uniquely by the fact that the inflaton becomes inhomogeneous, as this leads to mode-couplings between the chiral and longitudinal components of the gauge field, that would not be present if the inflaton remained homogeneous. We observe that in the strong backreaction cases the system reached radiation domination soon after the end of inflation, with the gauge field dominating the total energy budget, presenting a non-trivial coupling- and scale-dependent partial balance between chiralities.   

In the light of our results, we conclude that a proper description of the dynamics of strong backreaction must be done considering a fully inhomogeneous treatment of the dynamics, including the local fluctuations of $F\tilde F$, and the presence of gradients of the inflaton. This is necessary in order to make truly reliable predictions of the phenomenological signatures. For example, the amplitude and frequency today of the gravitational wave background sourced during inflation by the excited gauge field, strongly depends on how long inflation is extended. Furthermore, the production of scalar metric perturbations and the density of primordial black holes these could lead to, are strongly sensitive to the truly local nature of the non-linear dynamics. A detailed study of both phenomenological consequences using our lattice formulation, will be presented elsewhere.   \\

{\bf Acknowledgements: }We thank R. von Eckardstein, K. Schmitz, O. Sobol, V. Domcke, Y. Ema and S. Sandner for discussion and for kindly providing output data from the gradient expansion formalism. DGF (ORCID 0000-0002-4005-8915) and NL (ORCID 0000-0003-4285-4701) are supported by Generalitat Valenciana grant PROMETEO/2021/083, and by Spanish Ministerio de Ciencia e Innovaci\'on grant PID2023-148162NB-C21 and PID2023-148162NB-C22. NL (ORCID 0000-0003-4285-4701) is supported by ``Sabor y Origen de la Materia (Flavour and Origin of Matter) CPI-24-295" project. JL (ORCID 0000-0002-1198-3191), AU (ORCID 0000-0002-0238-8390) and JU (ORCID 0000- 0002-4221-2859) acknowledge support from Eusko Jaurlaritza (IT1628-22) and by the PID2021-123703NB-C21 grant funded by MCIN/AEI/10.13039/501100011033/ and by ERDF; ``A way of making Europe”. In particular, AU gratefully acknowledges the support from the University of the Basque Country grant (PIF20/151). This work has been possible thanks to the computing infrastructure of the ARINA and Solaris clusters at the University of the Basque Country, UPV/EHU, the Hyperion cluster from the DIPC Supercomputing Center, the FinisTerrae III cluster at Centro de Supercomputación de Galicia (CESGA), the Lluis Vives cluster at the University of Valencia and the \texttt{Graviton} cluster at Instituto de Física Corpuscular (IFIC), with funding contributions from EUROPA EXCELENCIA-2022 Grant No.~EUR2022-134028 and the European Union NextGenerationEU (PRTR-C17.I01) and Generalitat Valenciana Grant No.~ASFAE/2022/020.

\appendix
\section{Discrete action and equations of motion}
\label{App:LatticeFirstPrinciples}

We follow the spatial discretisation proposed in \cite{Figueroa:2017qmv} for the axial coupling between a shift-symmetric axion and a U(1) gauge field, \ie $\phi F\tilde{F}$. This method was latter adapted in \cite{Cuissa:2018oiw} to expanding backgrounds, where the expansion of the universe was evolved in a self-consistent manner. In a nutshell, this prescription ensures: i) discrete gauge invariance, ii) reproduction of the continuum at $\mathcal{O}(\delta x^2)$, iii) fulfillment the Bianchi identities and iv) exact discrete shift symmetry.

We embed the spatially discretised action in a hybrid discrete action
\begin{equation}
\label{eq:latticeAction}
\begin{split}
	S_{\rm m}^L = \int dt \sum_{\vec{n}}\delta x^3 \left[ \frac{1}{2}a^3\pi_{\phi}^2 - \sum_i a\frac{1}{2}(\Delta^+_i\phi)^2  -V(\phi) \right. \\
	 \left. + \sum_i \frac{1}{2}a\left(E_i^2 - a^{-2}B_i^2\right)  + \frac{\alpha_\Lambda\phi}{m_{\rm p}} \sum_i E_i^{(2)}B_i^{(4)} \right] \; ,
\end{split}
\end{equation}
where $\pi_{\phi}=\dot{\phi}$, $E_i=\dot{A}_i$, $B_i=\epsilon_{ijk}\Delta^+_jA_k$, and $E_i^{(2)}$ and $B_i^{(4)}$ are defined as in Eq.~(\ref{eq:E2B4definition}). See main text in Sec.~\ref{sec:lattice_approach} for the definitions of the lattice operators.

The time is maintained as a continuum variable, and the variation of this hybrid action yields to the discrete EOM for the matter sector that we already employed in {\tt Paper I} and described in Sec.~\ref{sec:lattice_approach}:
\begin{widetext}
\begin{eqnarray}
	\dot{\pi}_\phi &=& -3H\pi_\phi + \frac{1}{a^2} \sum_i \Delta_i^-\Delta_i^+ \phi - \frac{dV(\phi)}{d\phi} + \frac{\alpha_\Lambda}{a^3m_{\rm p}} \sum_i E_i^{(2)}B_i^{(4)} \; ,\\
	\dot{E}_i &=& -HE_i - \frac{1}{a^2} \sum_{j,k} \epsilon_{ijk} \nabla_j^- B_k - \frac{\alpha_\Lambda}{2am_{\rm p}} \left(\pi_\phi B_i^{(4)} + \pi_{\phi,+i}B^{(4)}_{i,+i} \right) \nonumber \nonumber\\
&& \hspace{5cm} 
	 + \frac{\alpha_\Lambda}{4am_{\rm p}} \sum_\pm \sum_{j,k} \epsilon_{ijk}  \left\{ \left[ (\Delta_j^\pm \phi) E_{k,\pm j}^{(2)} \right]_{+i} +  \left[ (\Delta_j^\pm \phi) E_{k,\pm j}^{(2)}  \right]   \right\} \; , 
\end{eqnarray}
\end{widetext}
with the discrete Gauss constraint 
\begin{equation}
    \sum_i \Delta_i^- E_i = - \frac{\alpha_\Lambda}{2am_{\rm p}} \sum_\pm \sum_i (\Delta_i^\pm \phi) B_{i,\pm i}^{(4)} \; .
\end{equation}

With this prescription at hand, as time is not discretised {\it a priori}, one is not restricted to use only symplectic time integrators such as leapfrog or velocity-verlet schemes (see \cite{Figueroa:2020rrl} for a review). This issue is particularly relevant in axion inflation because the gauge conjugate momentum appear in its EOM, in which case symplectic integrators can only provide an implicit solving scheme \cite{Cuissa:2018oiw}. Therefore, the prescription we present here is naturally suitable for non-sympletic integrators, which allow an explicit scheme.

In this work we choose a 2nd order Runge-Kutta (RK2) integrator, with which the continuous equations are satisfied up to $\mathcal{O}(\delta t^2, \delta x^2)$, or $\mathcal{O}(\delta \mathcal{N}^2, \delta x^2)$ if e-folds are used as the time variable. In the Appendix~\ref{App:NumUVstability} we show numerical stability tests and a comparison with a higher order integrator, RK3.

\section{Chiral Basis and Projection}
\label{app:chiralBaseProj}
In this appendix we show the procedure followed to construct the helicity basis in the reciprocal lattice.

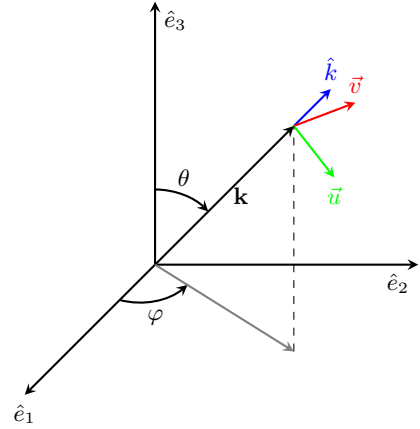
\begin{figure}[t]
\begin{tikzpicture}
    \tikzset{axis/.style={thick,->,>=stealth}}
    \tikzset{vector/.style={->,thick,>=stealth,black}}
    \tikzset{vectork/.style={->,thick,>=stealth,black}}
    \tikzset{vectorkhat/.style={->,thick,>=stealth,blue}}
    \tikzset{vectoru/.style={->,thick,>=stealth,green}}
    \tikzset{vectorv/.style={->,thick,>=stealth,red}}
    \tikzset{angle/.style={thick,->,>=stealth,black}}
    
    \draw[axis] (0,0,0) -- (3.5,0,0) node[anchor=north east] {$\hat{e}_{2}$};
    \draw[axis] (0,0,0) -- (0,3.5,0) node[anchor=north west] {$\hat{e}_{3}$};
    \draw[axis] (0,0,0) -- (0,0,4.5) node[anchor=north] {$\hat{e}_{1}$};
    
    \draw[vectork] (0,0,0) -- (3,3,3) node[midway, right] {$\mathbf{k}$};

    \draw[vectorkhat] (3,3,3) -- (3.8,3.8,3.8)  node[anchor=south]{$\hat{k}$};
    
    \draw[vectorv] (3,3,3) -- (3.507107,3,2.19289) node[anchor=south]{$\vec{v}$};
    
    \draw[vectoru] (3,3,3) -- (3.30825,2.0835,2.40175) node[anchor=north]{$\vec{u}$};
    
    \draw[angle] (0,1,0) arc[start angle=90,end angle=45,radius=1] node[midway, above] {$\theta$};
    
     \draw[angle] (0,0,1.22) arc[start angle=255,end angle=310,radius=1] node[midway, below] {$\varphi$};
    
    \draw[vector, gray] (0,0,0) -- (3,0,3);

    \draw[dashed] (3,0,3) -- (3,3,3);

\end{tikzpicture}

\caption{Definition of the orthonormal basis around the direction of the vector $\hat{\mathbf{k}}$ in the reciprocal space used to define the polarization vectors $\vec{\varepsilon}^{\, \pm}$.}
\label{fig:chiralbasis}
\end{figure}

The initial conditions for the gauge vectors require to write the orthonormal basis spanning the reciprocal space around the position vector ${\bf k} = (k_1,k_2,k_3)$ with $k = \sqrt{k_1^2 + k_2^2 + k_3^2}$. We can write the position vector in terms of the spherical coordinates ($\theta, \varphi$),
\begin{eqnarray}
{\bf k} = k(\sin \theta \cos \varphi, \sin \theta \sin \varphi, \cos \theta)\;.
\end{eqnarray}
We choose $\hat{e}_3$ as the reference direction to define the orthonormal basis around $\hat{\bf k}={\bf k}/k$. We define a plane orthogonal to $\hat{\bf k}$ with a couple of unitary vectors $(\vec{u},\vec{v})$ (see Fig.~\ref{fig:chiralbasis}):
\begin{eqnarray}
    \label{eq:v_k}
    \hspace*{-0.5cm}\vec{v} (\hat{\bf k})&=&\frac{\hat{e}_3 \times \hat{\bf k}}{|\hat{e}_3 \times \hat{\bf k}|} = (-\sin \varphi, \cos \varphi, 0)\;, \\
    \label{eq:u_k}\hspace*{-0.5cm}\vec{u}(\hat{\bf k}) &=& \vec{v} \times \hat{\bf k} =(\cos \theta \cos \varphi, \cos \theta \sin \varphi, - \sin \theta) \; .
\end{eqnarray}
The chiral vectors are then constructed as
\begin{equation}
    \vec{\varepsilon}^{\,\pm} = \frac{\vec{u} \pm i \vec{v}}{\sqrt{2}} \; ,
    \label{eq:epsilon_uv}
\end{equation}
which together with $\hat{\mathbf{k}}$ form the chiral basis $\{\vec{\varepsilon}^{\, +},\ \vec{\varepsilon}^{\, -},\ \hat{\mathbf{k}} \}$, and satisfy all properties of Eq.~\eqref{eqn:polarisationvectors}. Any vector can be written in the following manner
\begin{eqnarray}
A_i = \varepsilon_i^{\, +}A^+ + \varepsilon_i^{\, -}A^-  + \hat{k}_iA^{\parallel} \; ,
\end{eqnarray}
where $A^{\pm}$ stand for the positive and negative chiral projections of $A_i$ and the $A^{\parallel}$ is its longitudinal component. 

We explicitly construct the chiral basis on the lattice as follows. On the lattice, our choice of forward/backward lattice derivative interpreted as living at mid points gives a lattice momentum
\begin{eqnarray}\label{eqn:app:klattice}
    k_{{\rm L},i} = 2\frac{\sin (\pi \tilde{n}_i / N)}{ \delta x} \, .
\end{eqnarray}

In order to set the initial conditions, we have to build the chiral orthonormal basis using the lattice momenta $\hat{\mathbf{k}}_{\rm L}$ where the angles defining Eqs.~(\ref{eq:v_k})-(\ref{eq:u_k}) in the orthonormal basis $\{\vec u_{\rm L}, \vec v_{\rm L}, \hat{\bf k}_{\rm L}\}$ are given by the following expressions:
\begin{eqnarray}\label{eqn:app:shpericallatticecoords}
    \sin \theta_{\rm L} &=& \frac{\sqrt{k_{{\rm L},1}^2 + k_{{\rm L},2}^2}}{\sqrt{k_{{\rm L},1}^2 + k_{{\rm L},2}^2 +k_{{\rm L},3}^2 }}\, , \label{eqn:sinThetaLat}\\ 
    \sin \varphi_{\rm L} &=& \frac{k_{{\rm L},2}}{\sqrt{k_{{\rm L},1}^2 + k_{{\rm L},2}^2}}\, , \label{eqn:sinPhiLat}\\
    \cos \theta_{\rm L} &=& \frac{k_{{\rm L},3}}{\sqrt{k_{{\rm L},1}^2 + k_{{\rm L},2}^2 +k_{{\rm L},3}^2 }}\, ,  \label{eqn:cosThetaLat}\\ 
    \cos \varphi_{\rm L} &=& \frac{k_{{\rm L},1}}{\sqrt{k_{{\rm L},1}^2 + k_{{\rm L},2}^2}}\, .\label{eqn:cosPhiLat}
\end{eqnarray}
Finally, the lattice version of the chiral vectors (\ref{eq:epsilon_uv}):
\begin{equation}
\label{eqn:app:chiralvec}
    \vec{\varepsilon}_{\rm L}^{\,\pm} = \frac{\vec{u}_{\rm L} \pm i \vec{v}_{\rm L}}{\sqrt{2}} \; .
\end{equation}

We note that for the vector $\hat{\mathbf{k}}_{\rm L}$ pointing in the $\hat{e}_3$ direction, $\sin \varphi_{\rm L}$ and $\cos\varphi_{\rm L}$ are ill-defined. Nevertheless, this is just a consequence of choosing $\hat{e}_3$ as our reference direction to define the chiral basis, therefore we can just choose any value for $\varphi_{\rm L}$ as long as $\{\vec u_{\rm L}, \vec v_{\rm L},\hat{\bf k}_{\rm L}\}$ are orthogonal among them. We can just replace all the continuos definitions of the chiral vectors and the projector by its lattice counterpart to obtain the desired properties.   

Similarly, as described in Sec.~\ref{sec:lattice_approach}, we can construct the chiral projector on the lattice using the appropriate lattice momentum, Eq.~(\ref{eq:projector_lattice}), for which similar properties to those of the chiral vectors are exactly satisfied on the lattice:

\bea
&& k_{{\rm L}, i}\Pi^{{\rm L}, \lambda}_{ij}(\hat{\mathbf{k}}_{\rm L})=0\; , \\
&& \epsilon_{ijl}\hat{k}_{{\rm L},j}\Pi^{{\rm L}, \lambda}_{lm}(\hat{\mathbf{k}}_{\rm L}) = -i\lambda \Pi^{{\rm L}, \lambda}_{ij}(\hat{\mathbf{k}}_{\rm L})\; ,
\label{eqn:HelProp}\\
&&\Pi^{{\rm L}, \lambda}_{ij}(\hat{\mathbf{k}}_{\rm L})\Pi^{{\rm L}, \lambda'}_{ij}(\hat{\mathbf{k}}_{\rm L})=\delta_{\lambda\lambda'}\; , \\
&&\Pi^{{\rm L}, \lambda}_{ij}(\hat{\mathbf{k}}_{\rm L})=\Pi^{{\rm L}, \lambda}_{ij}(-\hat{\mathbf{k}}_{\rm L})=\Pi^{{\rm L}, \lambda}_{ij}(\hat{\mathbf{k}}_{\rm L})^*=\Pi^{{\rm L}, \lambda}_{ji}(\hat{\mathbf{k}}_{\rm L})\; . \nonumber \\  
\eea

In addition, we make sure that the Fourier transforms/anti-transforms are consistent with our lattice prescription. In our formulation, we use backward/forward spatial derivatives that live at the intermediate points relative to the position of the field they act upon. This implies that if the scalar field $\phi$ lives at ${\bf n}$, $\Delta_i^{\pm}\phi$ lives at ${\bf n}\pm\hat{\imath}/2$, so does the $i$th component of the gauge field in order to preserve gauge invariance. We make this explicit in the transformation of the gauge fields

\be
 A_i(\textbf{n}+\hat{\imath}/2) \equiv \frac{1}{N^3}\sum_{\tilde{\textbf{n}}}e^{-i\frac{2\pi}{N}\tilde{\textbf{n}}(\textbf{n}+\hat{\imath}/2)}A_i(\tilde{\textbf{n}})\; ,
 \ee
 \be
 A_i(\tilde{\textbf{n}}) \equiv \sum_{\textbf{n}}e^{+i\frac{2\pi}{N}(\textbf{n}+\hat{\imath}/2)\tilde{\textbf{n}}}A_i(\textbf{n}+\hat{\imath}/2)\; ,
\ee 
and similarly for the electric field. With this redefinitions we can naturally use (\ref{eqn:app:klattice}) as the lattice momentum to construct the chiral vectors (\ref{eqn:app:chiralvec}) trough the angles ~(\ref{eqn:sinThetaLat})-(\ref{eqn:cosPhiLat}), and the lattice helicity projector (\ref{eq:projector_lattice}). Furthermore, we ensure that Gauss's law is satisfied to machine precision during the initialization, as shown in Sec.~\ref{App:NumUVstability}. A similar procedure was used in \cite{Morgante:2021bks} to make explicit the locations of gauge fields for Fourier transforms.

Finally, we note that this procedure is equivalent to that shown in \cite{Figueroa:2020rrl} for pure gauge theories, where a complex lattice momentum is used for gauge initialisation. However, the complex nature of the chiral vectors and the chiral projector prevent the use of complex lattice momenta.

\section{Bunch-Davies phase for the initial conditions}
\label{App:BD}

The equations (\ref{eqn:ABDcosmic}) and (\ref{eqn:EBDcosmic}) of Sec.~\ref{subsec:IC_Lattice} dictate the initial conditions of the gauge field and its conjugate momentum. Aside from the amplitude of the BD solution, these conditions set the phase difference between $A^{\pm}$ and $E^{\pm}$ to be exactly $\pi/2$. In this Appendix we show the relevance of setting exactly those configurations.

\begin{figure}[t]
\includegraphics[width=8.8cm]{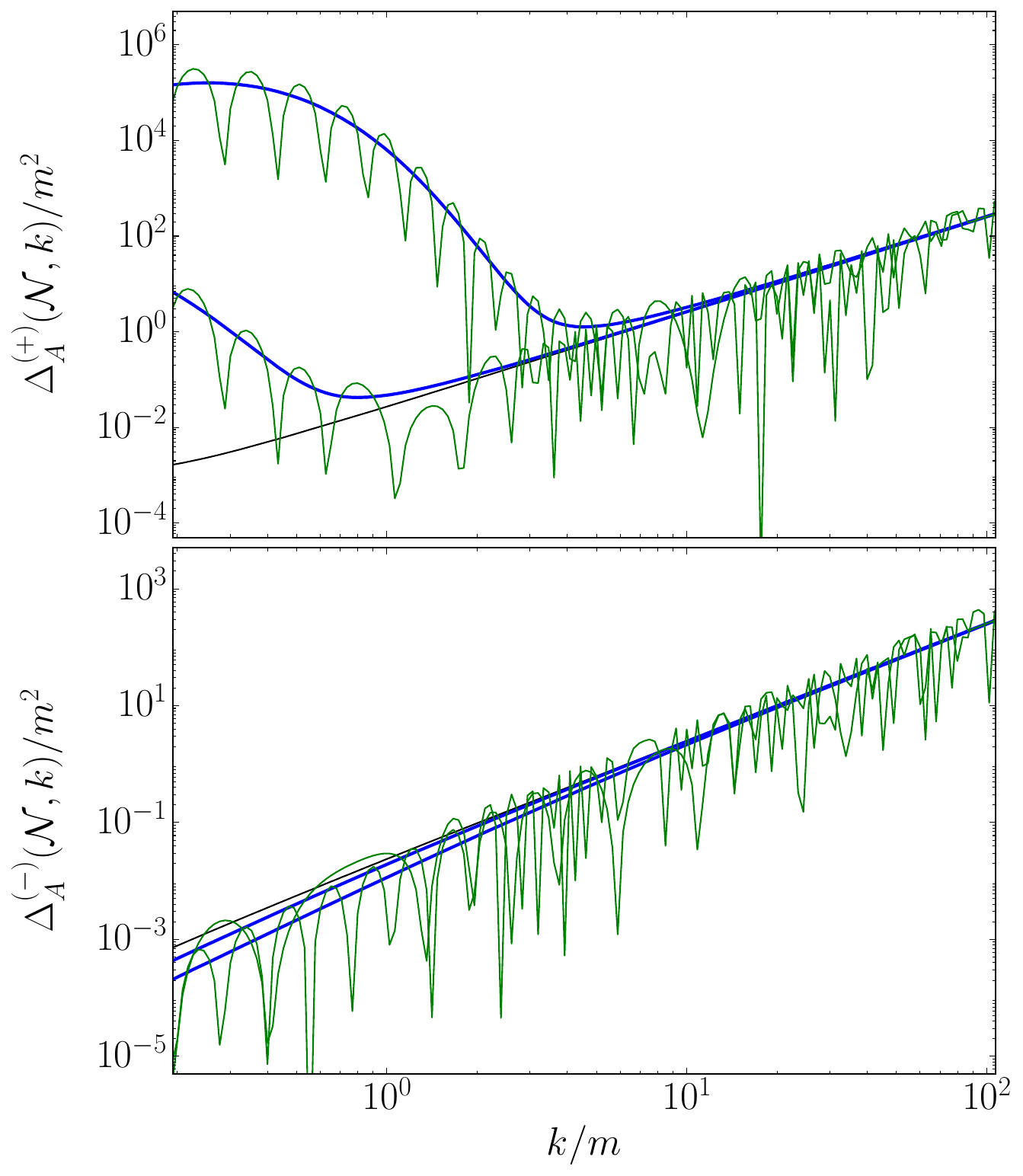}
\caption{Evolution of the power spectrum of the gauge field for the positive helicity (top) and the negative helicity (bottom), showing a temporal gap of $\Delta\mathcal{N}=1.7$. Different colours correspond to different initial phase differences between the gauge and electric fields: $\pi/2$ in blue and $0$ in green.
.}
\label{fig:linearInitialFaseProblem}
 \end{figure}

We solve the linear regime for the gauge field on a 1D grid for modes well inside the Hubble radius, $k \gg aH$ for different initial conditions. In the top panel of Fig.~\ref{fig:linearInitialFaseProblem}, we show the evolution of $\Delta_A^{(+)}$ at intervals of $\Delta \mathcal{N} = 1.7$, and in the bottom panel, the evolution of $\Delta_A^{(-)}$. The solution using the initial conditions of Eqs.~(\ref{eqn:ABDcosmic})-(\ref{eqn:EBDcosmic}) is depicted in blue, whereas we use green for distorted initial conditions, where we modify the relative complex phase between the gauge field and the electric field to be $0$. The distorted initial conditions show oscillatory artifacts in the spectra. 

The second initialisation procedure presented in Sec.~\ref{subsec:excitedInit} requires some additional refinements. In this case, we start the simulations at $\mathcal{N}_{\rm switch}$ and the infrared momentum is set by $k_{\rm IR} = 10 a(\mathcal{N}_{\rm start})H(\mathcal{N}_{\rm start})$. In this setup the positive helicity is excited for a range of momenta $k \lesssim aH$ up to a transition scale $k_{\rm tr}$, above which the fields remain in vacuum. 

\begin{figure}[t]
\includegraphics[width=8.8cm]{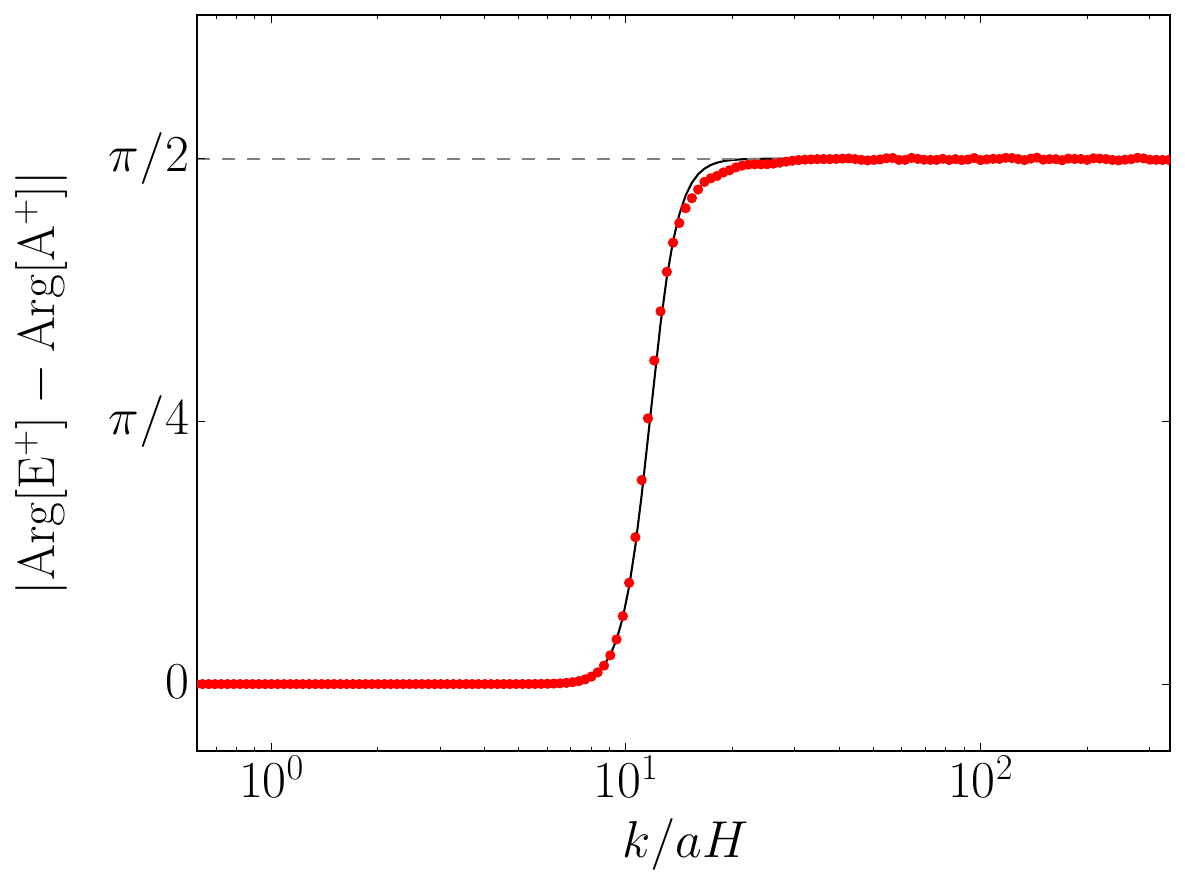}
\caption{Phase difference, shown with red dots, between the modes of the gauge field and the electric field for the positive helicity at $\mathcal{N}_{\rm switch}=-1.1$, corresponding to the coupling $\alpha_{\Lambda}=15$. The black line in the top panel represents the fit (\ref{eqn:interpolationFunc}) with parameters $k_{\rm tr}=3.65$ and $\epsilon=12.75$, while the horizontal line corresponds to $\pi/2$.}
\label{fig:excitedFaseDiff}
 \end{figure}

In Fig.~\ref{fig:excitedFaseDiff} we show the measured relative phase difference between $A^+$ and $E^+$ for $\alpha_{\Lambda} = 15$ obtained by solving the equations in a 1D grid (red dots). It can be observed that the relative phase of the positive helicity component is different in the excited IR region (super-Hubble modes) as compared to the non-excited region (sub-Hubble modes), where, as expected, it continues to be $\pi/2$. In fact, it appears that the exponential excitation removes any phase difference. For the negative mode, in turn, a global phase difference of around $\pi/2$ is still present.

We include this effect by using initial conditions
\bea
A^{+}(\mathcal{N},\tilde{\bf n}) &\simeq& A^{+}_{\rm{PS}(k)} e^{ik/(aH)}\,, 
\label{eqn:AplusExcitedcosmic}\\ 
A^{-}(\mathcal{N},\tilde{\bf n}) &\simeq& A^{-}_{\rm{PS}(k)} e^{ik/(aH)}\;,
\label{eqn:AminusExcitedcosmic}\\
E^{+}(\mathcal{N},\tilde{\bf n}) &\simeq& E^{+}_{\rm{PS}(k)} e^{i[k/(aH)-\Theta(k,k_{\rm tr},\epsilon)\pi/2]}\,, 
\label{eqn:EplusExcitedcosmic}\\ 
E^{-}(\mathcal{N},\tilde{\bf n}) &\simeq& E^{-}_{\rm{PS}(k)} e^{i[k/(aH)-\pi/2]}\,,
\label{eqn:EminusExcitedcosmic}
\eea
where $k = |\vk| = k_{\rm IR}|\tilde{\bf n}|$, and $A^{\pm}_{\rm{PS}(k)}$ and $E^{\pm}_{\rm{PS}(k)}$ correspond to realizations of the spectra extracted from the mode by mode linear solution at $\mathcal{N}_{\rm switch}$. The function $\Theta(k,k_{\rm tr},\epsilon)$ accounts for the smooth transition of the relative phase in the excited mode, which we model by
\begin{eqnarray}
    \Theta(k,k_{\rm tr},\epsilon)=\frac{1}{2}\left[\tanh{\left(\epsilon \log{\frac{k}{k_{\rm tr}}}\right)}+1\right]\; ,\label{eqn:interpolationFunc}
\end{eqnarray}
with $k_{\rm tr}$ and $\epsilon$ free parameters. We show using a solid black line the result obtained by the modelling function with $(k_{\rm tr}, \epsilon) = (3.65,12.75)$. We refer to Sec.~\ref{subsec:switch} and App.~\ref{App:NumUVstability} for consistency checks that validate this initialization method. 

We note that a similar initialisation procedure, departing from an excited configuration, was employed in \cite{Cuissa:2018oiw}. The power spectra of the electric and magnetic fields of that work exhibit some oscillatory features similar to those shown in Fig.~\ref{fig:linearInitialFaseProblem}, see \eg Fig.~9. We attribute the appearance of such oscillations to not considering the transition in the relative phase. 

\section{Numerical stability and consistency checks}
\label{App:NumUVstability}

We present several consistency checks and stability analyses to demonstrate the validity of our numerical framework. For all tests we choose a common baseline simulation, namely $\alpha_\Lambda=15$, $\mathcal{N}_{\rm start} = -4.5$, $N=640$ and $k_{\rm IR}/m = 0.1932$.

\subsection{Time step, $\delta\mathcal{N}$}
\begin{figure}[t]
\includegraphics[width=8.8cm]{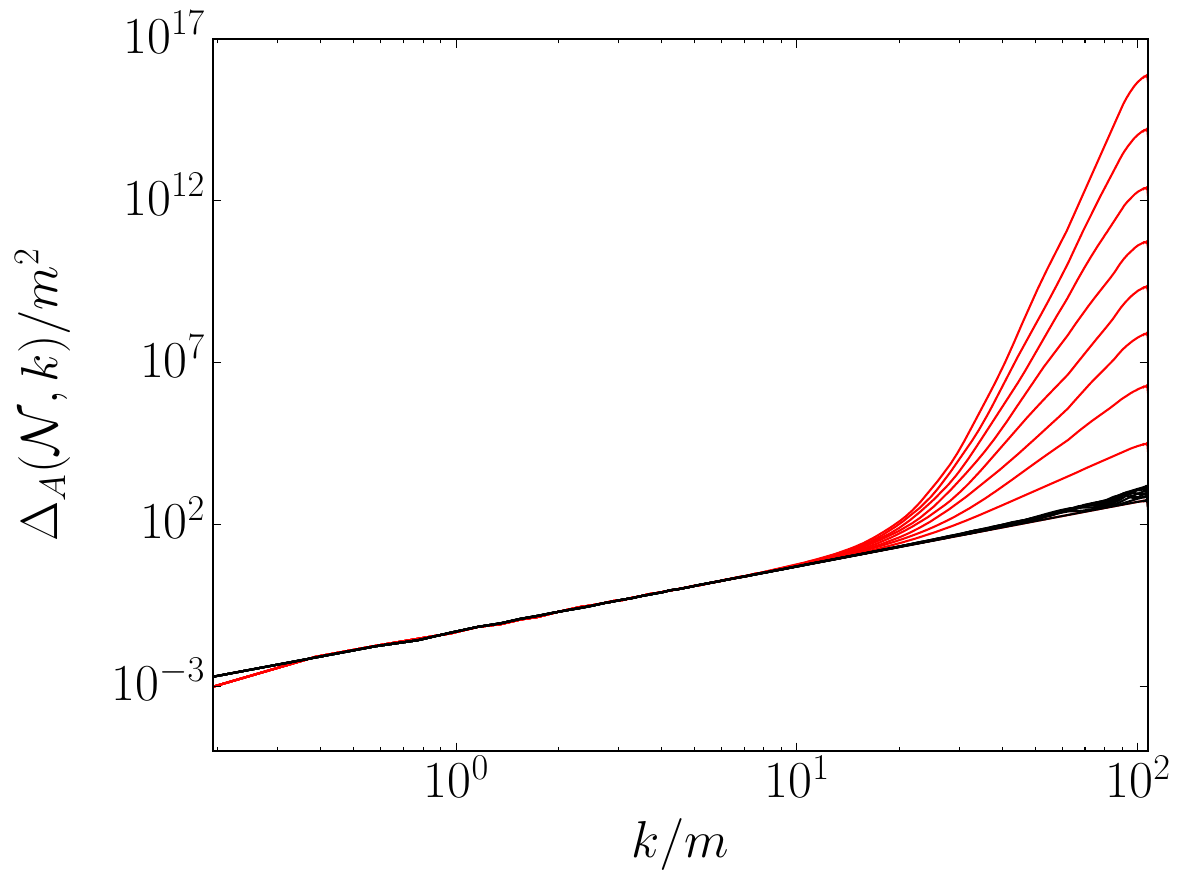}
\caption{Illustrative comparison between two simulations for the baseline setup, but with different time step: $\delta\mathcal{N}~=~10^{-3}$~(red) and $\delta\mathcal{N}=10^{-4}$ (black).}
\label{fig:UVproblem}
 \end{figure}

In order to ensure numerical stability, an appropriate time step ($\delta\mathcal{N}$) must be set for every resolution, \ie UV coverage. In Fig.~\ref{fig:UVproblem} we show the comparison of the evolution of the power spectrum of the gauge field for simulations with $\delta\mathcal{N}=10^{-3}$ (red) and $\delta\mathcal{N}=10^{-4}$ (black). The figure shows the evolution for $8 \times 10^{-3}$ e-folds. It clearly shows that the simulation with the worst time resolution (red lines) begins to grow uncontrollably in the most UV modes, and that the instability even propagates to more intermediate modes. For the simulation with a correct time step (black lines) we also observe a residual growth at the most UV modes, which is always present and cannot be avoided decreasing even more the time step. This residual growth, however, saturates after some time steps and do not alter the subsequent evolution.

A related effect is also present due to the exponential growth of the scale factor during inflation. The earlier we start the simulations, \ie further back in the inflationary period, the more rapidly the scale factor grows. The effect is similar to that shown in Fig.~\ref{fig:UVproblem} if the time step is not set correctly. We control this issue by adjusting the time step accordingly, so that we capture correctly the evolution of the scale factor and avoid non-physical instabilities in the evolution.

For the simulations with the largest couplings a combination of both aspects occurs, as they are required to start early during inflation and, simultaneously, greater extra inflationary periods emerge during the non-linear evolution. The latter implies that we need to include very UV scales on the lattice. Is in this cases where the use of the initial intermediate cutoff $k_{\rm BD}$ allows us to avoid excessively reducing the time step, as it becomes possible to eliminate the residual and the potential artificial growth shown in Fig.~\ref{fig:UVproblem}.

\subsection{Intermediate cutoff, $k_{\rm BD}$}

 \begin{figure}[t]
\includegraphics[width=8.8cm]{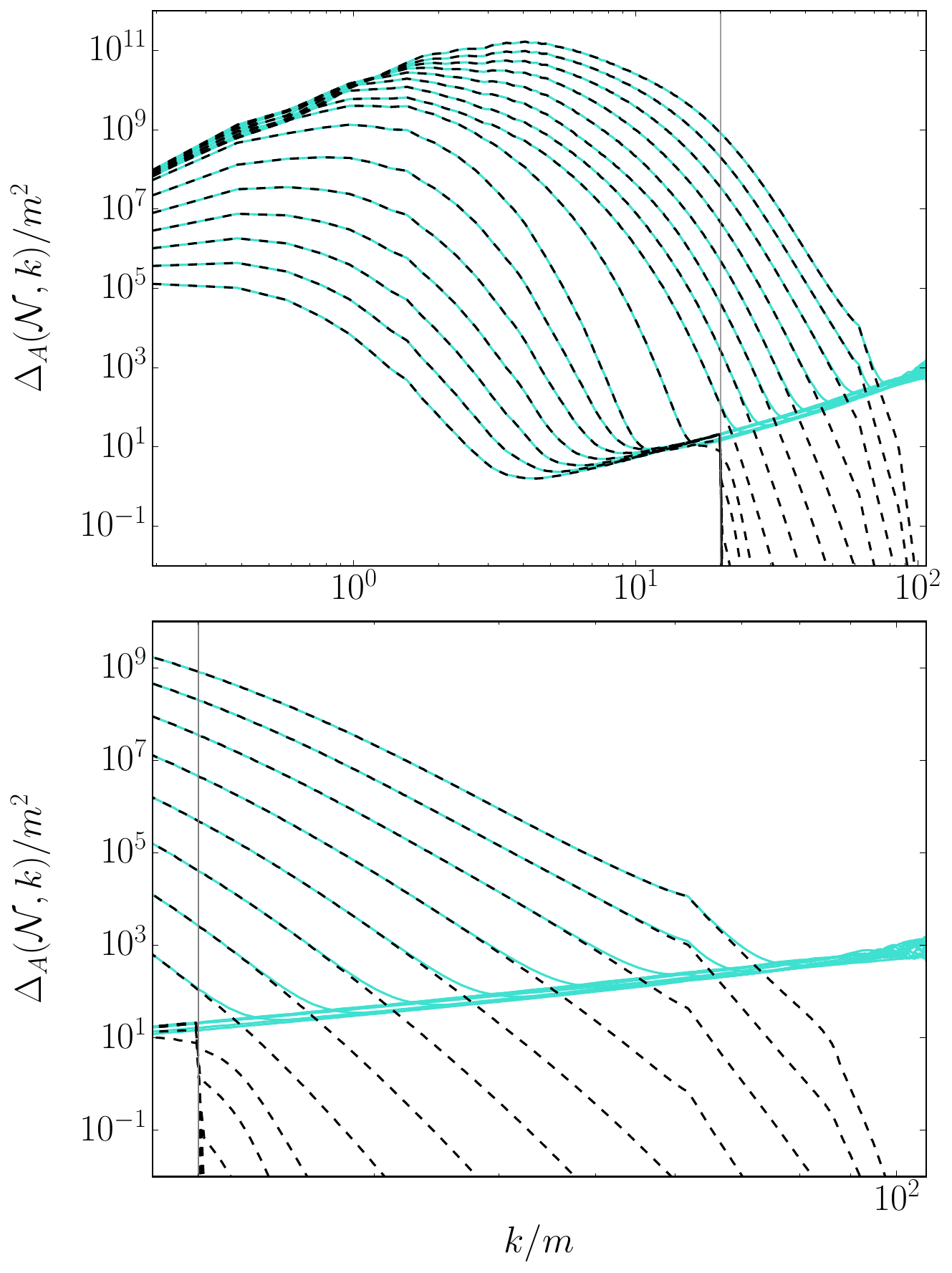}
\caption{Comparison of the evolution of the gauge field power spectrum where the BD vacuum initial condition is set up to $k_{\rm UV}$ (solid turquoise) and where a cutoff has been applied at $k_{\rm BD}$ (dashed black). The top panel shows the full range $[k_{\rm IR}, k_{\rm UV}]$, while the bottom panel provides a zoomed-in view of the cutoff region.
}
\label{fig:fromZeroVSfromBD}
 \end{figure}
The inclusion of an intermediate cutoff scale, $k_{\rm BD}$, in the initialization procedure eliminates the excessive influence of the quantum BD tail of the electromagnetic contributions on the expansion equations and the EOM, see Sec.~\ref{subsec:switch}. The cutoff is always set so that it provides sufficient support for modes to become excited in the linear regime. In order to verify the validity of this approach, it is necessary to ensure that modes in the range $[k_{\rm BD},k_{\rm UV}]$ grow properly in the non-linear regime, regardless they start from the vacuum solution or from vanishing amplitude. 

We show a comparison between different initialisation in Fig.~\ref{fig:fromZeroVSfromBD}, where we include the case with all modes in $[k_{\rm IR},k_{\rm UV}]$ starting in the BD solution (solid lines) and the case where BD is only set in the range $[k_{\rm IR},k_{\rm BD}]$, where $k_{\rm BD}/m = 20$ (dashed lines). We simulate starting from $\mathcal{N}_{\rm switch}$ until the end of non-linear inflation. We observe that the cutoff modes grow at the same rate as when starting from the Bunch-Davies vacuum solution and become excited due to the non-linear dynamics. We also show a zoomed-in view of the cutoff region in the lower panel to provide a detailed comparison between both cases.

\subsection{Initialization techniques}

Here we present a quantification on the differences on the initialisation techniques discussed in Sec.~\ref{subsec:excitedInit} and detailed in App.~\ref{App:BD}. The method where the gauge field already start excited at $\mathcal{N}_{\rm switch}$ allows to skip the evolution period between $\mathcal{N}_{\rm start}$ and $\mathcal{N}_{\rm switch}$, which lasts typically $\Delta\mathcal{N} \sim 2-3$. 

\begin{figure}[t]
\includegraphics[width=8.8cm]{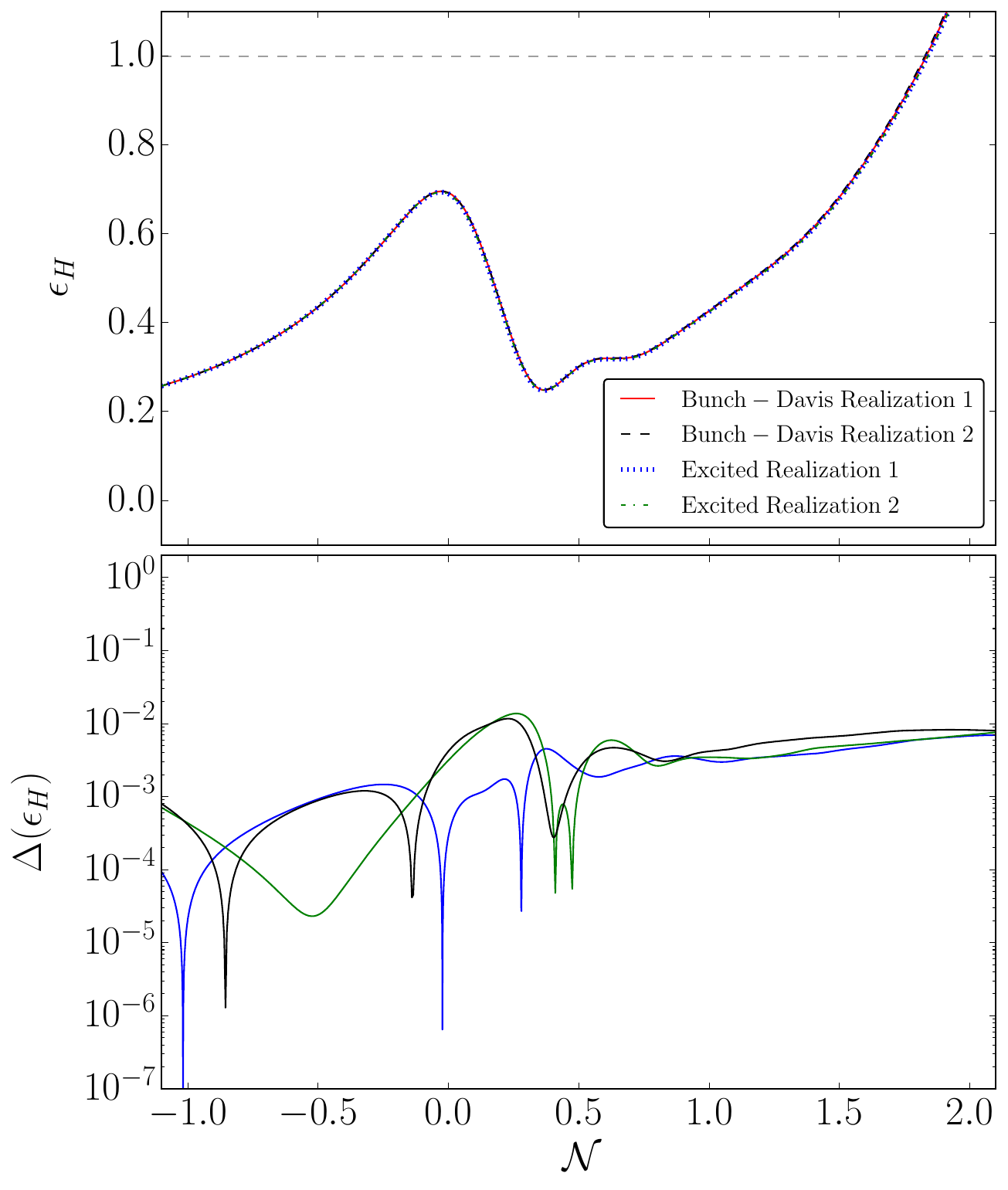}
\caption{\textit{Top}: Evolution of $\epsilon_H$ where we include two different realizations of simulations that start at $\mathcal{N}_{\rm start}$ with the BD vacuum solution (red and black lines) and two realizations of simulations starting at $\mathcal{N}_{\rm switch}$  with the gauge field already excited (blue and green lines). \textit{Bottom}: The relative difference in $\epsilon_H$ with respect to the first case in the top panel; we maintain the same colour scheme as in the top panel.
}
\label{fig:epsilonHfromBDfromExcited}
 \end{figure}

We already showed in Fig.~\ref{fig:SpectraIL15excitedVSbd} that both techniques lead to equivalent spectra. Here we provide additional evidence to support this equivalence. We include in Fig.~\ref{fig:epsilonHfromBDfromExcited} a comparison between both techniques for our baseline simulation. The top panel shows the evolution of $\epsilon_H$, using two different realizations per case. The bottom panel shows the relative difference between cases, using an equivalent version of (\ref{eq:reldiff_epsilon}) where we take the first realization of the case starting from Bunch-Davis (red) as a reference. We maintain the same colour scheme as in the upper panel. The relative differences between different realizations of the same technique and between different methods are of the same order and, remarkably, never greater than $1\%$. 

\subsection{Time integrators}

\begin{figure}[h]
\includegraphics[width=8.8cm]{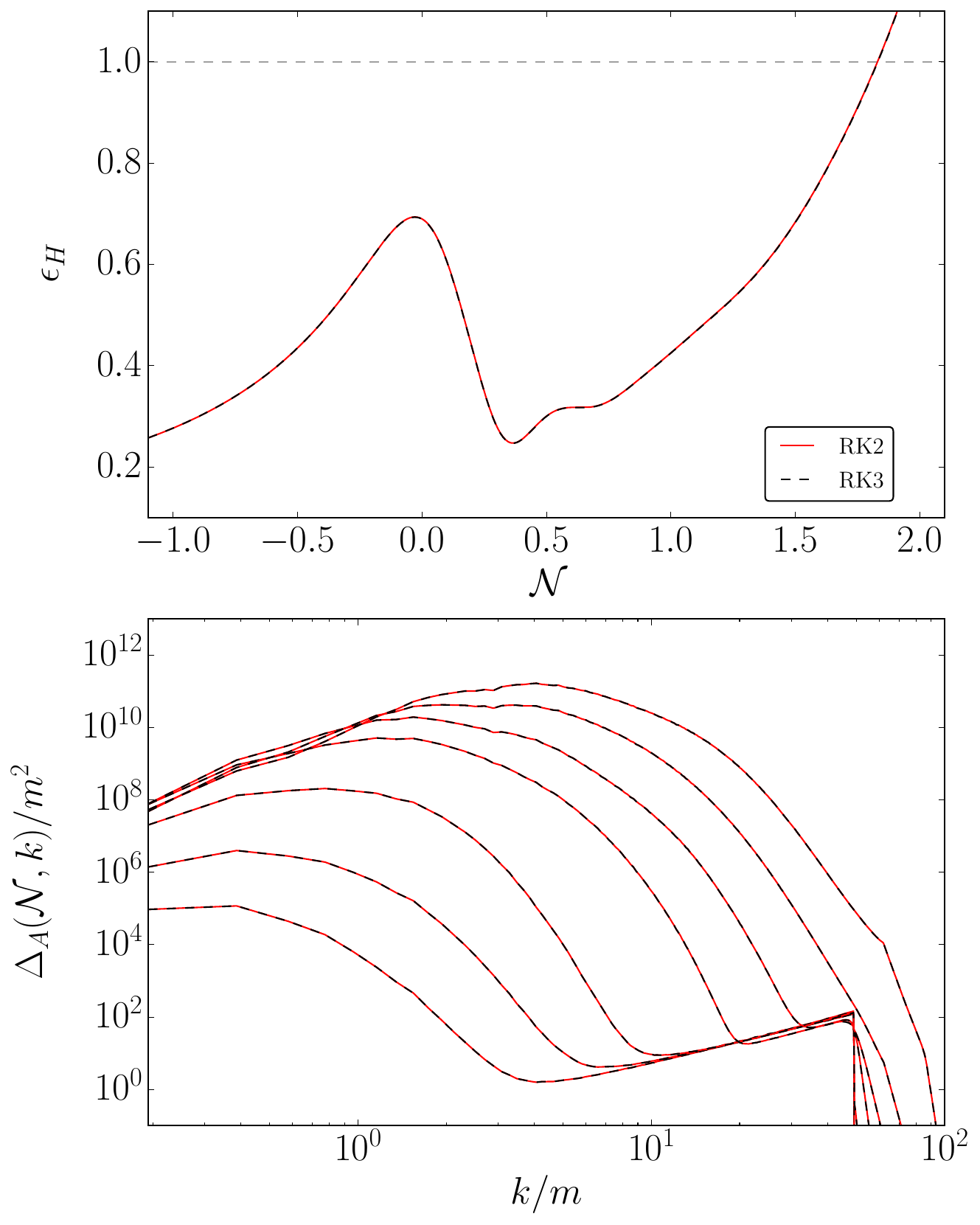}
\caption{Comparison of simulations for different time integrators: RK2 (solid red) and RK3 (dashed black). In the top panel, we show the evolution of the global variable $\epsilon_H$, and in the bottom panel, the growth of the total power spectrum of the gauge field.
}
\label{fig:epsilonHandSpectraRK2vsRK3}
 \end{figure}
 
Finally, we show the comparison between different time integrators. For the production runs we use a low storage Runge-Kutta 2nd order (RK2) time integrator, which minimizes memory usage \cite{Carpenter1994Thirdorder2R,Carpenter1994Fourthorder2R}. This integrator preserves numerical accuracy up to order $\mathcal{O}(d\mathcal{N}^2)$, which is the standard for PDE integrators. We contrast this with a higher order RK3 integrator and show the comparison in Fig.~\ref{fig:epsilonHandSpectraRK2vsRK3}. The outcome form RK2 is depicted in solid red, whereas the solution from RK3 is in black dashed. In the top panel, we see the overlap in the evolution of the global variable $\epsilon_H$, and in the bottom panel, how the gauge field spectra for both integrators match. 

Additionally, we also analyse the evolution of the energy conservation and Gauss constraints. As it is customary \cite{Figueroa:2020rrl, Figueroa:2023oxc} we measure this by using
\begin{equation}
    \Delta_H = \frac{|LHS-RHS|}{\sqrt{LHS^2+RHS^2}}\; , \label{eq:HubbleSch} 
\end{equation}
\begin{equation}
     \Delta_G = \frac{\left\langle|LHS-RHS|\right\rangle_{\rm V}}{\left\langle\sqrt{\sum_{i=1}^3(LHS_i)^2+RHS^2}\right\rangle_{\rm V}}\; ,
     \label{eq:GaussSch} 
\end{equation}
where $\langle ... \rangle_{\rm V} \equiv \frac{1}{N^3}\sum_{\mathbf{n}}(...)$ stands fo volume averaging. $LHS$ and $RHS$ refer to the {\it left-} and {\it right-hand sides} of Eq.~(\ref{eq:HubbleLatProgram}) and (\ref{eq:GaussLatProgram}) respectively, and $LHS_i=\Delta_{i}^{-}E_{i}$ (considering no sum over repeated indices).

\begin{figure}[h!]
\includegraphics[width=8.8cm]{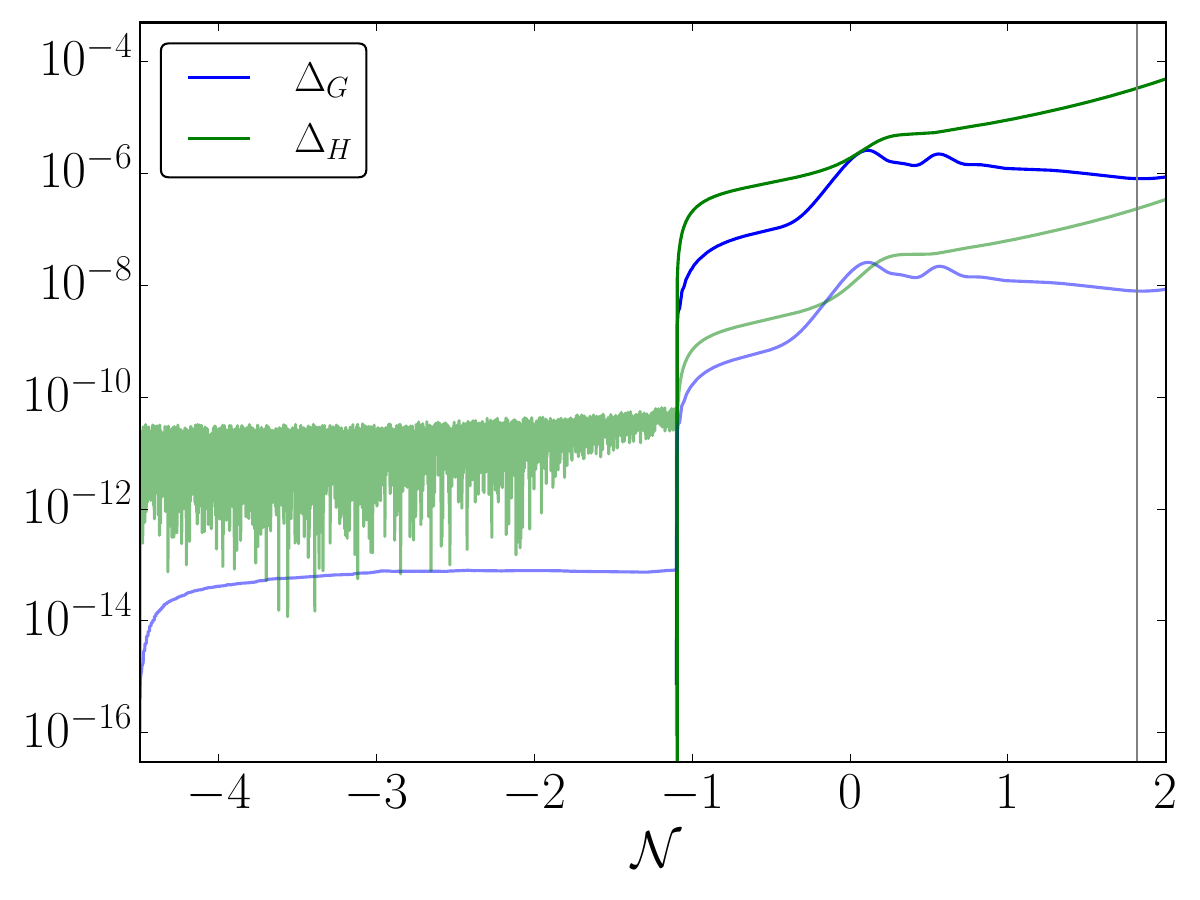}
\caption{Evolution of the energy conservation constraint, $\Delta_{H}$ (green), and the Gauss law constraint, $\Delta_{G}$ (blue) using the RK2 integrator. The lighter lines correspond to a simulation starting at $\mathcal{N}_{\rm start}$ with BD initial conditions, while the dark ones starts at $\mathcal{N}_{\rm switch}$ from an excited state. The vertical gray line indicates the end of inflation.
 }
\label{fig:GaussandFriedmann}
 \end{figure}
We show the evolution of $\Delta_{H}$ (green) and $\Delta_{G}$ (blue) in Fig.~\ref{fig:GaussandFriedmann}, for the RK2 integrator. The lighter lines correspond to the initialization method with an initial linear phase starting in $\mathcal{N}_{\rm start}$ with BD as a initial condition, while the dark ones correspond to the initialization from an excited state of the gauge field $\mathcal{N}_{\rm switch}$. Both constraints remain below $\mathcal{O}(10^{-4})$ during the non-linear regime until the end of inflation. The differences in order between initialization methods arise from the use of different time steps, as the method corresponding to the excited configuration permits a larger $\delta N$. Similar levels of constraint preservation are observed for other coupling values.

\bibliography{automatic,manual}

\end{document}